\documentclass[a4paper,11pt]{tkbook} 

\usepackage{amsmath,amssymb}
\usepackage{fancybox}
\usepackage{graphicx}
\usepackage{color}
\usepackage[vcentermath]{youngtab}

\usepackage{fancyheadings}
\renewcommand{\chaptermark}[1]%
  {\markboth{#1}{}}
\renewcommand{\sectionmark}[1]%
  {\markright{\thesection\ #1}}
\makeatletter
\renewcommand{\theequation}{%
   \thesection.\arabic{equation}}
 \@addtoreset{equation}{section}
\makeatother

\def\figurename#1#2{\bfseries Figure #2}
\def\tablename#1#2{\bfseries Table #2}

\newcommand{\ket}[1]{| \, {#1} \, \rangle}
\newcommand{\Ket}[1]{\Big| \; {#1} \; \Big\rangle}

\newcommand{\del}{\partial}

\newcommand{\half}{\frac{1}{2}}

\newcommand{\LS}{\ \ \ \ \ \ \ \ \ \ }
\newcommand{\ls}{\ \ \ \ \ }
\newcommand{\wt}{\widetilde}
\newcommand{\wh}{\widehat}
\newcommand{\ve}{\varepsilon}
\newcommand{\ol}{\overline}
\newcommand{\ul}{\underline}
\newcommand{\dps}{\displaystyle}

\newcommand{\bsubeq}{\begin{subequations}}
\newcommand{\esubeq}{\end{subequations}}

\newcommand{\tc}{\textcolor}
\newcommand{\noi}{\noindent}

\newcommand{\w}{\wedge}
\renewcommand{\d}{{\rm d}}
\newcommand{\christoffel}[3]{\genfrac{\{}{\}}{0pt}{}{#1}{#2\,#3}}
\newcommand{\Circ}[1]{\overset{\mbox{\tiny $\circ$}}{#1}}

\newcommand{\eps}{\epsilon}

\renewcommand{\l}{\ell}
\newcommand{\G}{\Gamma}
\newcommand{\nn}{\nonumber}

\newcommand{\Tr}{{\rm Tr}}

\newcommand{\e}{{\rm e}}

\newcommand{\hG}{{\wh{\Gamma}}}
\newcommand{\tG}{{\wt{\Gamma}}}
\newcommand{\nt}{{\natural}}
\newcommand{\Yng}{\tiny \yng}
\newcommand{\para}{\parallel}

\newcounter{Enumerate}

\definecolor{midnightblue}{named}{MidnightBlue}
\definecolor{myblue}{named}{NavyBlue}
\definecolor{blueviolet}{named}{BlueViolet}
\definecolor{mycyan}{named}{CornflowerBlue}
\definecolor{mypurple}{named}{Orchid}
\definecolor{mygreen}{named}{OliveGreen}
\definecolor{redorange}{named}{RedOrange}
\definecolor{orangered}{named}{OrangeRed}
\definecolor{myred}{named}{BrickRed}
\definecolor{brown}{named}{Brown}
\definecolor{maroon}{named}{Maroon}
\definecolor{mygray}{gray}{.45}

\renewcommand{\r}{\tc{myred}}

\begin{document}

\allowdisplaybreaks{

\thispagestyle{empty}


{\normalsize
\begin{flushright}
KEK-TH-941, OU-HET 466\\
hep-th/0402054\\
\end{flushright}
}

\vspace{10mm}

\begin{center}
\scalebox{1.1}{\huge\upshape Zero-mode Spectrum of Eleven-dimensional Theory}

\vspace{5mm}

\scalebox{1.1}{\huge\upshape on the Plane-wave Background}

\vspace{10mm}

{\Large Tetsuji {\sc Kimura}}

\vspace{7mm}

{\large\sl
Theory Division, Institute of Particle and Nuclear Studies,\\
High Energy Accelerator Research Organization (KEK) \\
Tsukuba, Ibaraki 305-0801, Japan

{\sf tetsuji@post.kek.jp}

{\normalsize\sl and}

Department of Physics,
Graduate School of Science, Osaka University\\
Toyonaka, Osaka 560-0043, Japan

{\sf t-kimura@het.phys.sci.osaka-u.ac.jp}

}
\vspace{25mm}

{\large\sc
A Dissertation Presented to the Faculty of Osaka University

in Candidacy for the Degree of Doctor of Science 

\vspace{5mm}

Recommended for Acceptance by the Department of Physics
}

\end{center}


\newpage

\thispagestyle{empty}

{\phantom{spacepage}}


\newpage

\thispagestyle{empty}

\vspace*{7.5cm}

\begin{quote}
\begin{center}
{\Large \sf This doctoral thesis is dedicated to my family.}
\end{center}
\end{quote}


\newpage

\thispagestyle{empty}

{\phantom{spacepage}}


\newpage

\thispagestyle{empty}

\begin{center}
\ul{\Large\sc Acknowledgements}
\end{center}

\begin{quote}
First I would like to thank my advisor, Professor Kiyoshi Higashijima,
for introducing me to various wonderful topics and problems in quantum
field theory, and for encouraging me even when I left Osaka University
and when I studied in the Theory Division, Institute of Particle and
Nuclear Studies, High Energy Accelerator Research Organization (KEK) 
for the last one year of my graduate studies. 
I thank him for everything.

I would like to thank Professor Satoshi Iso for his warm
hospitality and encouraging me during my stay in KEK.
I would like to also thank Dr. Kentaroh Yoshida, 
Dr. Machiko Hatsuda, 
Dr. Makoto Sakaguchi and Dr. Yasuhiro Sekino for advising me on various topics 
around supergravities and M-theory.

I would like to express my gratitude to
Professor Hiroshi Itoyama, 
Professor Takahiro Kubota, Professor Nobuyoshi Ohta, Professor
Norisuke Sakai, Professor Asato Tsuchiya,
Dr. Kazuyuki
Furuuchi, Dr. Koichi Murakami, 
Dr. Muneto Nitta, Dr. Dan Tomino, Dr. Hiroshi Umetsu and Dr. Naoto Yokoi
for introducing and discussing various topics and problems 
around string theories and field theories.
I also thank Naoyuki Kawahara, Toshiharu Maeda, 
Hidenori Terachi and 
Yastoshi Takayama for discussing in the superstring theory seminar in
KEK.

I am grateful for
Professor Tohru Eguchi, Professor Hiroaki Kanno, Professor Toshio Nakatsu, 
Professor Yuji Sugawara, Professor Yukinori Yasui,
Dr. Hiroyuki Fuji, Dr. Takahiro Masuda, Dr. Takao Suyama, 
Dr. Takashi Yokono, Tomoyuki Fujita, 
Yutaka Ookouchi, Kazuhiro Sakai and Kensuke Tsuda 
to introduce a lot of
interesting topics among geometry and topological strings.
Some unpublished studies are inspired by discussions and warm advice.

Thanks all members of high energy theory group in Osaka University
and all members of theory division in KEK for useful discussions,
comments and encouragement.

{Finally, I would like to thank all of my friends and my family for
their encouragement and powerful support.}

\vspace{10mm}

This thesis is supported in part by the Research Fellowships of the
Japan Society for the Promotion of Science for Young Scientists (No.15-03926).

\end{quote}

\newpage

\thispagestyle{empty}

{\phantom{spacepage}}


\newpage

\thispagestyle{empty}

\vspace*{2.5cm}

\begin{center}
\ul{\Large\sc Abstract}
\end{center}

\begin{quote}
In this doctoral thesis 
we study zero-mode spectra of Matrix theory and 
eleven-dimensional supergravity on the plane-wave background.
This background is obtained via 
the Penrose limit of $AdS_4 \times S^7$ and $AdS_7 \times S^4$. 
The plane-wave background is a maximally supersymmetric spacetime
supported by non-vanishing constant four-form flux 
in eleven-dimensional spacetime. 
%
%
First, we discuss 
the Matrix theory on the plane-wave background
suggested by Berenstein, Maldacena and Nastase.
We construct the Hamiltonian, 32 supercharges and their commutation
relations.
We discuss a spectrum of one specific supermultiplet which represents
the center of mass degrees of freedom of $N$ D0-branes.
This supermultiplet would also represent a superparticle of
the eleven-dimensional supergravity in the large-$N$ limit.
%
Second, 
we study the linearized supergravity on the plane-wave background in
eleven dimensions.
Fixing the bosonic and fermionic fields in the light-cone gauge, 
we obtain the spectrum of physical modes.
We obtain the fact that the energies of the states in Matrix theory 
completely correspond to those of fields in supergravity.
Thus, we find that the Matrix theory on the plane-wave background
contains the zero-mode spectrum of the eleven-dimensional supergravity
completely.
Through this result,
we can argue the Matrix theory on the plane-wave as a candidate of 
quantum extension of eleven-dimensional supergravity, or as a
candidate which describes M-theory.

\end{quote}

\newpage

\pagenumbering{roman}

\tableofcontents


\newpage

\thispagestyle{empty}

{\phantom{spacepage}}


\newpage

\thispagestyle{empty}

\vspace*{7cm}

\begin{quote}
\tc{blueviolet}{\bfseries\itshape
Eleven-dimensional supergravity remains an enigma. It is hard to
  believe that its existence is just an accident, but it is difficult
  at the present time to state a compelling conjecture for what its
  role may be in the scheme of things.
}
\begin{flushright}
--- M.B. Green, J.H. Schwarz and E. Witten, ``{\sl Superstring Theory}''.
\end{flushright}
\end{quote}


\newpage

\thispagestyle{empty}

{\phantom{spacepage}}

\newpage
\pagenumbering{arabic}


\makeatletter
\renewcommand{\theequation}{%
   \thechapter.\arabic{equation}}
 \@addtoreset{equation}{chapter}
\makeatother

\setcounter{section}{0}
\renewcommand{\thechapter}{\Roman{chapter}}

\chapter{Introduction} \label{intro}


\newpage

\section*{Supergravity}

Eleven-dimensional (Lorentzian) spacetime is the maximal spacetime
in which one can formulate a consistent 
supersymmetric multiplet including fields with spin less than 
two\footnote{If the spacetime metric has two negative signatures,
one could formally construct the supersymmetric theory in {\sl
  twelve}-dimensions in which the supermultiplet would contain the
fields ``spin'' less than two.
This is because the Majorana-Weyl spinor with 32 real degrees of
freedom is the irreducible representation of spinors in such
``spacetime''. 
As you know the ``F-theory'' will be formulated in such
twelve dimensions \cite{V96, Sen96}, 
but this theory may not have a field theory realization.
Bars has been studying the {\sl two-time physics} 
in order to understand the field theory in such a specific spacetime
\cite{BK9703, BK9705, BDM99, BDPZ0312}.}.
Nahm first recognized this fact in his classification and
representation of supersymmetry algebra \cite{N78}.
Not so long after this understanding, 
Cremmer, Julia and Scherk realized that supergravity not only permits
up to seven extra dimensions from four dimensions 
but in fact takes its simplest and most
elegant form \cite{CJS78}. 
The unique supergravity in eleven-dimensional spacetime
contains a graviton $g_{MN}$, a gravitino $\Psi_M$ and a three-form
gauge field $C_{MNP}$ with 44, 128 and 84 on-shell degrees of freedom,
respectively. 
The theory was regarded not only as a candidate for 
the fundamental theory including quantum gravity 
but also as a mathematically important 
tool to derive a 
four-dimensional supergravity with extended supersymmetries via
{\sl dimensional reduction}.
The research interests in those days were to find a (supersymmetric)
grand unified theory which gives gauge groups greater than $SU(3)
\times SU(2) \times U(1)$, and to analyze the hidden symmetries of
extended supergravities in four dimensions \cite{CJ79, dWN82}.
In this context, eleven-dimensional supergravities on 
some non-trivially curved spacetimes (in particular, the product
space of four-dimensional anti-de Sitter spaces $AdS_4$ and
seven-dimensional Einstein spaces such as round or squashed $S^7$, or the
product space of seven-dimensional anti-de Sitter space $AdS_7$ and
four-dimensional Einstein space) were also
investigated via {\sl Kaluza-Klein mechanism} \cite{W81, DP82, D83,
  DNPW84, Ohta85}. 
Now we can read a lot of important works of supergravities in diverse
dimensions in the book edited by Salam and Sezgin \cite{SS}.
We can also study the review of supergravity 
from the reports written by 
van Nieuwenhuizen \cite{PvN81} and by Duff, Nilsson and Pope \cite{DNP86}.

Although the eleven-dimensional supergravity is
intrinsically important theory as we introduced above,
this theory has some serious problems as the fundamental {\sl field} theory:
In eleven dimensions,
we cannot impose Weyl condition on the $SO(10,1)$ Dirac spinor because
of odd-dimensional spacetime.
So we cannot make four-dimensional {\sl chiral} field theory via Kaluza-Klein
mechanism, i.e., via the {\sl smooth} compactifications of
eleven-dimensional spacetime \cite{W81}\footnote{But, performing an
  {\sl orbifold} compactification one can obtain supersymmetric chiral
  field theories in four-dimensional spacetime \cite{HW96, A0011, AW01}.}. 
Moreover, the eleven-dimensional supergravity is non-renormalizable in
perturbation.
Although ten-dimensional supergravities are also non-renormalizable, 
they had barely survived 
because ten-dimensional supergravities could be regarded as the
low energy effective theory of ten-dimensional {\sl superstrings}, 
which are {\sl renormalizable} as perturbation theories.
As you know Salam also stated below
in the introduction of the proceedings of the Trieste
Spring School 1986 \cite{SUGRA86}:
\begin{center}
\fbox{\parbox{16cm}{\it
``Supergravity is dead. Long live supergravity in the context of
    superstrings''.
This seemed to be the motto of the Fourth Spring School on
Supergravity and Supersymmetry which was held at the International
Centre for Theoretical Physics at Trieste between 7 -- 15 April 1986.
}}
\end{center}
Through the above recognition,
the eleven-dimensional supergravity was abandoned in the middle eighties.

\section*{Super $p$-branes}

Theories of supersymmetric
extended objects in diverse dimensions are mysterious.
In the early eighties, Green and Schwarz constructed 
supersymmetric one-dimensional extended objects
(called the ``Green-Schwarz (GS) superstrings'') 
in ten-dimensional spacetime \cite{GS84-1}. 
Moreover it was shown that 
the GS superstrings also live {\sl classically} in $D=3,4$ and $6$ dimensions.
In the case of spatially two-dimensional objects (the membranes),
Bergshoeff, Sezgin and Townsend showed that the supermembrane can
classically propagate in $D=4,5,7$ and $11$ dimensions \cite{BST87, BST88}.
Thus people wondered which $p$-branes can exist in
$D$-dimensional spacetime ($p$ denotes the
spatial dimensions of extended objects).
A simple way to understand this question is to consider the numbers of
boson and fermion degrees of freedom {\sl on the $d$-dimensional 
worldvolume} of extended objects ($d = p +1$) \cite{AETW87}.
If the numbers of boson and fermion degrees of freedom are equal,
we can classically discuss the $p$-brane in $D$-dimensional spacetime.
Here let us explain the way of counting of the numbers of boson and
fermion degrees of freedom in the Green-Schwarz type theory \cite{D96}.
As a $p$-brane moves through $D$-dimensional spacetime,
its trajectory is described by the functions $X^M (\sigma^i)$, where
$X^M$ represent not only the spacetime coordinates but also the scalar
functions on the worldvolume ($M = 0, 1, \cdots, D-1$), 
and $\sigma^i$ denote the $d$-dimensional worldvolume coordinates ($i
= 0,1, \cdots, d-1$).
Choosing the static gauge $X^{\mu} (\sigma) = \sigma^{\mu}$ ($\mu
= 0,1,\cdots, d-1$),
we find that the number of on-shell bosonic degrees of freedom is 
\begin{align}
N^{\rm scalar}_{\rm B} \ &= \ D - d \; .
\label{scalar-dof}
\end{align}
In order to describe the super $p$-brane we should count the number of
fermionic degrees of freedom on the worldvolume.
Let us introduce anticommuting fermionic coordinates
$\theta^{\alpha} (\sigma)$ in the $D$-dimensional {\sl spacetime}.
We can impose the $\kappa$-symmetry on the fermionic coordinates,
which implies that half of the fermionic degrees of freedom are
redundant and may be gauged away from the physical degrees of freedom.
The net result is that the theory exhibits a $d$-dimensional {\sl
  worldvolume} supersymmetry whose number of fermionic generators is
half of the generators in the original {\sl spacetime} supersymmetry.
Let $M$ be the minimal number of real components of the minimal spinor 
and $N$ be the number of supersymmetry of $D$-dimensional spacetime,
and let $m$ and $n$ be the corresponding quantities in $d$-dimensional
worldvolume (see Table \ref{dof-fermion}).

\begin{table}[h]
\begin{center}
\begin{tabular}{c|ccc} \hline
dimension ($D$ or $d$) & irreducible spinor & minimal number ($M$ or $m$) &
supersymmetry ($N$ or $n$) \\ \hline \hline
2 & Majorana-Weyl & 1 & $1,2,\cdots, 32$ \\
3 & Majorana & 2 & $1,2, \cdots, 16$ \\
4 & Majorana or Weyl & 4 & $1,2, \cdots, 8$ \\
5 & Dirac & 8 & $1,2,3,4$ \\
6 & Weyl & 8 & $1,2,3,4$ \\
7 & Dirac & 16 & $1,2$ \\
8 & Majorana or Weyl & 16 & $1,2$ \\
9 & Majorana & 16 & $1,2$ \\
10 & Majorana-Weyl & 16 & $1,2$ \\
11 & Majorana & 32 & 1 \\ \hline
\end{tabular}
\caption{\sl The minimal number of fermion in $D$-dimensional
  (Lorentzian) spacetime and $d$-dimensional (Lorentzian)
  worldvolume. We also describe the number of supersymmetry.}
\label{dof-fermion}
\end{center}
\end{table}
\noi
Since the $\kappa$-symmetry always halves the number of fermionic
degrees of freedom and on-shell condition also halves it again,
we can write the number of on-shell fermionic degrees of freedom as
\begin{align}
N_{\rm F} \ &= \ \half m n \ = \ \frac{1}{4} MN 
\; . \label{spinor-dof}
\end{align}
{\sl Worldvolume} supersymmetry demands $N_{\rm B}^{\rm scalar} =
N_{\rm F}$, hence
\begin{align}
D - d \ &= \ \half mn \ = \ \frac{1}{4} MN
\; . \label{match-scalar}
\end{align}
Notice that this relation is satisfied except for the superstring
$d=2$, in which left- and right-moving modes should be treated 
independently.
In the case of the superstring, the following relation is obeyed:
\begin{align}
D - 2 \ &= \ n \ = \ \half MN
\; . \label{match-scalar-2}
\end{align}
On the worldvolume, bosons and fermions subject to
(\ref{match-scalar}) or (\ref{match-scalar-2}) belong to a scalar
supermultiplet of the worldvolume supersymmetry.
The solutions of scalar multiplets are categorized into four
compositions via division algebra ${\mathbb R}$, ${\mathbb C}$, ${\mathbb H}$
and ${\mathbb O}$ \cite{Sie87, AETW87, Ev88};
for example,
the GS superstrings in $D=3,4,6$ and $10$ dimensions belong to 
the ${\cal R}$-, ${\cal C}$-, ${\cal H}$- and
${\cal O}$-sequence, respectively \cite{BST88}.

We can consider other possibilities on the worldvolume supersymmetry.
If vectors also live on the worldvolume, the number of 
the on-shell bosonic degrees of freedom $N_{\rm B}^{\rm vector}$ is 
\begin{align}
N_{\rm B}^{\rm vector} \ &= \ D - d + (d-2) \ = \ D - 2
\; .
\label{vector-dof}
\end{align}
Thus the matching condition (\ref{match-scalar}) replaces
\begin{align}
D-2 \ &= \ \half mn \ = \ \frac{1}{4} M N
\; .
\label{match-vector}
\end{align}
In this case there lives a supersymmetric vector multiplet on the worldvolume.
The case of existence of an antisymmetric tensor field is also considerable.
We summarize the results of the possibilities of super $p$-branes in
various spacetime dimensions in Table \ref{branescan}, 
which is called the {\sl Brane Scan} \cite{D96}.
\begin{table}[h]
\begin{center}
\begin{tabular}{ccccccccccccccc}
~&$D\uparrow$ &
&&&&&&&&&~\\

~&11&$\cdot$&~
&&\tc{red}{\bf S}&&\phantom{{\bf S}/{\bf V}}&{\bf T}&\phantom{{\bf S}/{\bf V}}&\phantom{{\bf S}/{\bf V}}&\phantom{{\bf S}/{\bf V}}&\phantom{\bf ?}&~\\

~&10&$\cdot$&{\bf V}
&\tc{red}{\bf S}/{\bf V}&{\bf V}&{\bf V}&{\bf V}&\tc{mygreen}{\bf
    S}/{\bf V}&{\bf V}&{\bf V}&{\bf V}&{\bf V}&~\\ 

~&9&$\cdot$&\tc{red}{\bf S}
&&&&\tc{mygreen}{\bf S}&&&&&\phantom{{\bf S}/{\bf V}}&~\\

~&8&$\cdot$&~
&&&\tc{mygreen}{\bf S}&&&&&&&~\\

~&7&$\cdot$&~
&&\tc{mygreen}{\bf S}&&&{\bf T}&&&&&~\\

~&6&$\cdot$&{\bf V}
&\tc{mygreen}{\bf S}/{\bf V}&{\bf V}&\tc{blue}{\bf
    S}/{\bf V}&{\bf V}&{\bf V}&&&&&~\\

~&5&$\cdot$&\tc{mygreen}{\bf S}
&&\tc{blue}{\bf S}&&&&&&&&~\\

~&4&$\cdot$&{\bf V}
&\tc{blue}{\bf S}/{\bf V}&\tc{maroon}{\bf S}/{\bf V}&{\bf V}&&&&&&&~\\

~&3&$\cdot$&\tc{blue}{\bf S}/{\bf V}
&\tc{maroon}{\bf S}/{\bf V}&{\bf V}&&&&&&&&~\\

~&2&$\cdot$&\tc{maroon}{\bf S}
&&&&&&&&&&~\\


~&\phantom{0}&$\cdot$&$\cdot$
&$\cdot$&$\cdot$&$\cdot$&$\cdot$&$\cdot$&$\cdot$&$\cdot$&$\cdot$&$\cdot$&$\cdot$~\\ 
~&~&\phantom{0}&1
&2&3&4&5&6&7&8&9&10&11& $d$ $\rightarrow$
\end{tabular}
\caption{\sl The brane scan, 
where the spacetime dimensions $D$ are plotted vertically and the
worldvolume dimensions $d$ of $p$-branes ($d= p+1$) are plotted
horizontally. Note that {\bf S}, {\bf V} and {\bf T} 
denote scalar, vector and antisymmetric tensor multiplets.
The colored symbols of scalar multiplets 
such as \tc{maroon}{\bf S}, \tc{blue}{\bf S},
\tc{mygreen}{\bf S} and \tc{red}{\bf S} represent the solutions of
${\cal R}$-, ${\cal C}$-, ${\cal H}$- and ${\cal O}$-sequences,
respectively.}
\label{branescan}
\end{center}
\end{table} 

Most of the super $p$-branes in Table \ref{branescan} are
interpreted as solitons rather than fundamental extended objects.
Here we use the word {\sl solitons} to mean any such {\sl non-singular}
lumps of field energy which solve the (supergravity) field equations, 
which have finite mass per unit $p$-volume and which are prevented from
dissipating by some topological conservation law. 
We can understand that only the super $p$-branes in the 
${\cal O}$-sequences are {\sl fundamental}
objects, which are described by singular configurations with
$\delta$-function sources at the spacetime locations of $p$-branes.
Moreover we know that only the super $p$-branes in the ${\cal O}$-sequences 
are quantum consistent objects, which do not have Lorentz anomalies
in the light-cone gauge \cite{Bars88, BP88}.
The other super $p$-branes can be regarded as the solitons, for
example, super $p$-branes of vector multiplets in ten dimensions are
interpreted as Dirichlet $p$-branes (D$p$-branes), 
which carry the Ramond-Ramond
charges and which are solitonic non-perturbative objects in type
IIA/IIB string theories, etc \cite{Pol95}.  


\section*{Supermembrane}

In eleven-dimensional spacetime,
there exists a supergraviton (point particle) \cite{CJS78}, 
a supermembrane ($p=2$ in the ${\cal O}$-sequence) as a fundamental
object \cite{DS91},
and a super fivebrane as a solitonic, dual object of the supermembrane
\cite{G92}.
The supermembrane couples to a three-form gauge field
$C_3$ electrically via 
\begin{align*}
\int \! C_3 
\end{align*}
and the fivebrane couples to $C_3$ magnetically.
Supergraviton, supermembrane and super fivebrane 
appear in the eleven-dimensional supersymmetry algebra \cite{PKT95}.
The anticommutator of two supersymmetry generators $Q_{\alpha}$ is
schematically given by 
\begin{align*}
\{ Q_{\alpha} , Q_{\beta} \} \ &= \ 
(C \hG_M)_{\alpha \beta} P^M 
+ (C \hG_{MN})_{\alpha \beta} Z^{MN}
+ (C \hG_{MNPQR})_{\alpha \beta} Z^{MNPQR}
\; ,
\end{align*}
where $\hG_M$ is a Dirac gamma matrix in eleven-dimensional
  spacetime and $C$ is a charge conjugation matrix; $\hG_{M_1 M_2 \cdots
  M_n}$ are antisymmetrized products of Dirac gamma matrices.
We see that the right hand side involves not only the momentum $P^M$
of the superparticle but also the two-form central charge $Z^{MN}$ and
  five-form central charge $Z^{MNPQR}$, which are charges of
  supermembrane and super fivebrane, respectively.

Here we introduce a short review of the supermembrane \cite{BST87,
  BST88, PKT, dW99}. 
The supermembrane action is defined by
the Green-Schwarz type Lagrangian ${\cal L}_0$ and Wess-Zumino term ${\cal
  L}_{\rm WZ}$ as
\bsubeq \label{M2-L}
\begin{gather}
{\cal L} \ = \ {\cal L}_0 + {\cal L}_{\rm WZ}
\; , \\
{\cal L}_0 \ = \ - \sqrt{- g(Z)} 
\; , \ls
{\cal L}_{\rm WZ} \ = \ 
\frac{1}{6} \eps^{ijk} \, \Pi_i^{\r{\ul{A}}} \, \Pi_j^{\r{\ul{B}}} \,
\Pi_k^{\r{\ul{C}}} \, C_{\r{\ul{A}} \r{\ul{B}} \r{\ul{C}}} (Z)
\; , 
\end{gather}
\esubeq
where $Z^{\ul{M}} (\sigma) = \{ X^M (\sigma), \theta^{\alpha} (\sigma)
\}$ are eleven-dimensional 
superspace embedding coordinates ($\theta$ is a fermionic coordinate
denoted by $SO(10,1)$ Majorana spinor) and $\sigma^i$ ($i = 0,1,2$) are
worldvolume coordinates;
$\Pi_i^{\r{\ul{A}}} 
= \del Z^{\ul{M}}/ \del \sigma^i \wh{E}_{\ul{M}}{}^{\r{\ul{A}}}$ 
are pullbacks of the
superspace coordinates to the membrane worldvolume coordinates 
and $C_{\r{\ul{A}} \r{\ul{B}} \r{\ul{C}}}$ denotes the three-form
superfield\footnote{The convention about indices as follows. Curved
  space indices are denoted by $\ul{M} = \{ M , \alpha \}$, whereas
  tangent space indices are $\r{\ul{A}} = \{ \r{A} , \r{a} \}$. Here
  $M, \r{A}$ refer to commuting and $\alpha, \r{a}$ to anticommuting
  coordinates.}. 
Note that $g(Z)$ is a determinant of the worldvolume metric, 
and this is represented by the spacetime background metric $g_{MN}$ as
\begin{align*}
g (Z (\sigma)) \ &= \ 
\det \{ \Pi_i^{\r{A}} \, \Pi_j^{\r{B}} \, \eta_{\r{A} \r{B}} \}
\; ,
\ls
g_{MN} \ = \ \eta_{\r{A} \r{B}} \, e_M{}^{\r{A}} \, e_N{}^{\r{B}}
\; .
\end{align*}
Note that $\wh{E}_{\ul{M}}{}^{\r{\ul{A}}}$ is a supervielbein. 
In the flat superspace case,
the supervielbein and a three-form superfield $C_{\r{\ul{A}} \r{\ul{B}}
  \r{\ul{C}}}$ are given by 
\begin{align*}
\wh{E}_M{}^{\r{A}} \ &= \ \delta_M^{\r{A}} \; , 
&
E_M{}^{\r{a}} \ &= \ 0 \; , \\
E_{\alpha}{}^{\r{a}} \ &= \ \delta_{\alpha}^{\r{a}} \; , 
&
E_{\alpha}{}^{\r{A}} \ &= \ - (\ol{\theta} \hG^{\r{A}})_{\alpha} 
\; , \\
C_{MN \alpha} \ &= \ (\ol{\theta} \hG_{MN})_{\alpha} \; , 
&
C_{M \alpha \beta} \ &= \ 
(\ol{\theta} \hG_{MN})_{(\alpha} (\ol{\theta} \hG^N)_{\beta)}
\; , \\
C_{\alpha \beta \gamma} \ &= \ 
(\ol{\theta} \hG_{MN})_{(\alpha} (\ol{\theta} \hG^M)_{\beta} 
(\ol{\theta} \hG^N)_{\gamma)}
\; , 
&
C_{MNP} \ &= \ 0 \; .
\end{align*}
On general curved background \cite{BDPS89}, 
the supervielbeins and three-form gauge field become so complicated
that we have only a few solutions of curved spaces such as $AdS_4
\times S^7$, $AdS_7 \times S^4$ and their continuously deformed ones.

As in the case of Green-Schwarz superstring, 
the supermembrane action also has a reparametrization invariance and
fermionic $\kappa$-symmetry invariance.
In order to fix these local gauge symmetries
we can take the light-cone gauge 
\begin{align*}
X^+ (\tau) \ &= \ \tau
\; , \ls
\hG^{\r{+}} \theta \ = \ 0 
\; .
\end{align*}
Although we fix the above gauge symmetries in the supermembrane action,
there is a residual gauge symmetry such as diffeomorphism on the
membrane surface.
Thus we rewrite the supermembrane action (\ref{M2-L}) in the flat
spacetime background as a gauge theory action \cite{dWHN88}:
\begin{align}
w^{-1} {\cal L} \ &= \ 
\half D_{\tau} X^I D_{\tau} X^I + \frac{i}{2} \Psi^{\dagger} D_{\tau} \Psi
- \frac{1}{4} \{ X^I , X^J \}^2
+ \frac{i}{2} \Psi^{\dagger} \gamma^I \{ X^I , \Psi \}
\; , \label{APD-M2}
\end{align}
where $\Psi$ is an $SO(9)$ Majorana spinor satisfying the reality
condition $\Psi^{\dagger} = \Psi^T$ and 
$\gamma^I$ are $SO(9)$ Dirac's gamma matrix\footnote{Definitions are
described in section \ref{conv-membrane}.} with (flat) spacetime
indices $I = 1,2, \cdots, 9$;
the bracket $\{ * , * \}$ is the Lie bracket defined in terms of
an arbitrary function $w(\sigma^r)$ of worldvolume spatial coordinates
$\sigma^r$ ($r = 1,2$) as
\begin{align*}
\{ A , B \} \ &= \ \frac{1}{w} \eps^{rs} \del_r A \, \del_s B
\; , 
\end{align*}
with $\del_r = \del/\del \sigma^r$ and $\eps^{12} = 1$.
This system has, as mentioned above, 
a residual gauge symmetry called the 
``area preserving diffeomorphism'' (APD) and 
we define the covariant derivative of this gauge symmetry as
\begin{align*}
D_{\tau} X^I \ &= \ \del_{\tau} X^I - \{ \omega , X^I \}
\; ,
\end{align*}
where $\omega$ is a gauge field of this symmetry.
In 1988, de Wit, Hoppe and Nicolai argued that the supermembrane
Lagrangian (\ref{APD-M2}) might be written down as a supersymmetric
quantum mechanical theory in terms of the following ``matrix
regularization''
in order to analyze quantum properties of supermembrane: 
\begin{align*}
X^I (\sigma) \ &\to \ X^I (\tau)
\; , \ls
\Psi (\sigma) \ \to \ \Psi (\tau)
\; , \ls
\int \! \d^2 \sigma \, w (\sigma) \ \to \ \Tr
\; , \ls
\{ A , B \} \ \to \ - i [ A , B ]
\; . 
\end{align*}
Via this matrix regularization procedure,
the supermembrane action is written in terms of the $N \times N$ matrix
variables $X^I$ and $\Psi$ as
\begin{align}
L \ &= \ \Tr \Big\{ \half D_{\tau} X^I D_{\tau} X^I
+ \frac{i}{2} \Psi^{\dagger} D_{\tau} \Psi
+ \frac{1}{4} [ X^I , X^J ]^2
+ \half \Psi^{\dagger} \gamma^I [ X^I , \Psi ] \Big\}
\label{SQM}
\end{align}
with covariant derivative $D_{\tau} X^I = \del_{\tau} X^I + i [ \omega,
  X^I]$.

Type IIA superstring theory emerges via double dimensional reductions
of supermembrane theory in the eleven-dimensional spacetime \cite{DHIS87}.
Moreover, it is believed that all the D$p$-branes in type IIA string
theory emerge in various reductions from the extended objects such as
supermembrane and super fivebrane in the eleven-dimensional theory.
Thus one may think that the eleven-dimensional theory is the most
fundamental theory including gravity.
But, unfortunately,
we have not completely understood the supermembrane yet because of
a lot of problems: the difficulty of the classification of three
dimensional topologies, the interpretation of the Hilbert space \cite{dWLN89}, 
the zero mode spectrum of supermembranes \cite{dWN89, dW97, FH97}, etc.
In order to go beyond these difficulties,
a lot of scientists have been studying by using various methods.

Here let us introduce one of these difficulties; a supermembrane
instability problem.
When de Wit, Hoppe and Nicolai showed that the regularized
supermembrane could be described in terms of supersymmetric quantum
mechanics,
most people thought that the quantized supermembrane would have a
{\sl discrete spectrum} of states. 
In the case of string theory, the spectrum of states in the Hilbert
space of string can be put into one-to-one correspondence with
elementary particle states in the spacetime.
It is crucial that the massless spectrum contains a ``graviton'' and
that there is a mass gap separating the massive excitations from
massless states. 
However, for the supermembrane theory (and also for the super
$p$-brane theory as $p \geq 2$),
the spectrum does not seem to have these important properties.
We call this problem the {\sl membrane instability problem}.

This problem is explained simply at the classical level \cite{WT01}.
Consider a supermembrane 
whose energy is given by the area of the
membrane times a constant tension $T$.
Such a membrane can have a lot of long {\sl narrow} spikes at very low
cost in energy. 
If the spike is roughly cylindrical and has a radius $r$ and length
$L$, the energy of this spike is $2 \pi r L T$.
For a spike with large $L$ but a small $r \ll 1/TL$, the energy cost
is very small but the spike is very long.
This situation shows that {a membrane will tend to have
many fluctuations of this type}, making it difficult to conceive of the
membrane as single object which is well localized in spacetime.
Note that the string theory does not have this type of problems
because a long spike in a string always has energy {\sl proportional}
to the length of the string. 
In the quantum supermembrane theory 
the above process can also occur without energy loss because of the
existence of flat directions protected by the
supersymmetry (the quantum bosonic membrane theory is cured because
the flat directions rise via quantum corrections).
This phenomenon occurs in any {\sl quantum supersymmetric} $p$-brane
theories ($p \geq 2$).
By virtue of this phenomenon,
the supermembrane theory has a continuous spectrum and 
it is very difficult to distinguish the zero-modes from the other
excited states \cite{dWN89, dW97, FH97}.

Owing to the above serious problem,
the supermembrane theory has not been investigated more than the
superstring theories. 
On the other hand, the superstring theories have been well studied
since 1984, the ``first string revolution year'', in terms of of some keywords
such as the ``anomaly free'', ``mass gap'', ``derivation of GUTs'',
and so on.
Furthermore we have been re-investigating (super)string theories 
since 1995, the ``second revolution year'', with the keyword ``{duality}''.


\section*{Superstrings, Dualities and M-theory}

Since the first string revolution year,
five superstrings have been studied as
perturbatively consistent theories.
They are all anomaly free and live in ten-dimensional spacetime.
These five theories are introduced in the glossary of the
Polchinski's book \cite{Polchinski98} as:
\begin{quote}
\ul{Type IIA superstring theory}: 
{\sl a theory of closed oriented superstrings.
The right-movers and left-movers transform under separate spacetime
supersymmetries, which have opposite chiralities.} \\
\ul{Type IIB superstring theory}: 
{\sl a theory of closed oriented superstrings.
The right-movers and left-movers transform under separate spacetime
supersymmetries, which have the same chirality.} \\
\ul{Type I superstring theory}: 
{\sl the theory of open and closed unoriented superstrings, which is
  consistent only for the gauge group $SO(32)$. The right-movers and
  left-movers, being related by the open string boundary condition,
  transform under the same spacetime supersymmetry.} \\
\ul{Heterotic $E_8 \times E_8$ or $SO(32)$ superstring theory}: 
{\sl a string with different constraint algebras acting on the left-
  and right-moving fields. The case of phenomenological interest has a
$(0,1)$ superconformal constraint algebra, with spacetime
  supersymmetry acting only on the right-movers and with gauge group
  $E_8 \times E_8$ or $SO(32)$.}
\end{quote}
These superstring theories have ten-dimensional supergravities as the
massless excitation modes of superstring theory 
in the low energy limit, as mentioned by Salam.
The field contents of these superstring theories are summarized in
Table \ref{FC-II} and \ref{FC-IH}:
\begin{table}[h]
{\footnotesize
\begin{center}
\begin{tabular}{c||@{\ \ \ }c@{\ \ \ }|@{\ls}c@{\ls}|c} \hline 
 & sectors & fields & supersymmetry \\ \hline \hline
type IIA 
  & NS-NS & $g_{MN} (35)$, \ls $B_{MN} (28)$, \ls $\phi (1)$ & 32 \\
  & NS-R  & $\Psi_M (56)$, \ls $\psi (8)$ \\
  & R-NS  & $\wt{\Psi}_M (56)$, \ls $\wt{\psi} (8)$ \\
  & R-R   & $C_1 (8)$, \ls $C_3 (56)$ \\ \hline
type IIB
  & NS-NS & $g_{MN} (35)$, \ls $B_{MN} (28)$, \ls $\phi (1)$ & 32 \\
  & NS-R  & $\Psi_M (56)$, \ls $\psi (8)$ \\
  & R-NS  & $\Psi_M (56)$, \ls $\psi (8)$ \\
  & R-R   & $C_0 (1)$, \ls $C_2 (28)$, \ls $C_4^+ (35)$ \\ \hline
\end{tabular}
\end{center}
}
\caption{\sl Field contents of type IIA/IIB superstring theory in
  ten-dimensional spacetime.}
\label{FC-II}
\end{table}
\begin{table}[h]
{\footnotesize
\begin{center}
\begin{tabular}{c||@{\ \ \ }c@{\ \ \ }|@{\ls}c@{\ls}|c} \hline 
 & sectors & fields & supersymmetry \\ \hline \hline
type I
  & NS-NS & $g_{MN} (35)$, \ls $\phi (1)$ & 16 \\
  & NS-R  & $\Psi_M (56)$, \ls $\psi (8)$ \\
  & R-NS  & (reflections of NS-R)  \\
  & R-R   & $B_{MN} (28)$ \\ 
  & NS    & $A_M (8 \times 496)$ of $SO(32)$ gauge group \\
  & R     & $\lambda (8 \times 496)$ \\ \hline
heterotic 
  & boson & $g_{MN} (35)$, \ls $B_{MN} (28)$, \ls $\phi (1)$ & 16 \\
  & fermion & $\Psi_M (56)$, \ls $\psi (8)$ \\ 
  & gauge boson & $A_M (8 \times 496)$ of $SO(32)$ or $E_8 \times E_8$ \\ 
  & gauge fermion & $\lambda (8 \times 496)$ \\ \hline
\end{tabular}
\end{center}
}
\caption{\sl Field contents of type I/heterotic superstring theory in
  ten-dimensional spacetime.}
\label{FC-IH}
\end{table}

\noindent
Note that in all superstring theories 
there exists the supergravity multiplet which contains 
graviton $g_{MN}$, Kalb-Ramond field $B_{MN}$,
dilaton $\phi$, gravitino $\Psi_M$ and dilatino $\psi$. 
There exist various dimensional Ramond-Ramond fields $C_{p+1}$, which
couple to D$p$-branes in type IIA or type IIB string theory.
On the other hand, 
type I and heterotic string theory have gauge supermultiplets
containing gauge potential $A_M$ and gaugino $\lambda$ in the adjoint
representations.

In the first five years from 1984, 
the heterotic $E_8 \times E_8$ string theory was regarded 
as a candidate of the {\sl theory of everything}, i.e., a candidate
of the fundamental grand unified theory.
The heterotic $E_8 \times E_8$ theory has enough large gauge symmetry.
Via Calabi-Yau compactification mechanism \cite{CHSW85},
one could obtain four-dimensional quantum consistent field theory,
with $E_6$ gauge group and four supercharges.
Surprisingly, we could also obtain the generation numbers from the
geometric data of Calabi-Yau.
Since Maldacena have found that the AdS/CFT correspondence in 1997
\cite{M97, AMO},
people have studied some exact solutions for four-dimensional gauge
theories via gauge/gravity dualities \cite{GKP98, KW98, KS00, PT00,
  PZT01, IY01}.
In order to find new configurations and new phenomena in superstrings
or supergravities,
they engineered new (non-)compact manifolds with special holonomies
\cite{CGLP00, CGLP0102, CGLP0103, HKN0104, KN0104, AW01, KY0108, CGLP0108,
  HKN0110, KY0111, HKN0202}.
Unfortunately, however,
they found {\sl tremendously many vacua} from such compactifications
because we could compactify superstrings in terms of {\sl any}
Calabi-Yau manifolds, i.e., because 
we could not tell that some Calabi-Yau manifolds are more special than others.
Thus the string theorists wondered  whether string theories might or
might not predict any dynamics in four dimensions.
But they have studied around superstring theories in order to achieve 
the {\sl theory of everything}...

By virtue of the sting theorists' inexhaustible studies, 
one found some important properties among string theories:
the above five superstring theories are not distinct theories but 
they are closely related to one another via Dirichlet branes, which we
now regard as the
solitonic extended objects and as the sources of Ramond-Ramond fields 
in string theories,
and via perturbative and 
non-perturbative dualities such as T-duality, S-duality, and so on.
These observations leads to the postulate of an underlying fundamental
theory, called {\bf M-theory} \cite{GPR94, HT94, W95, Pol95, HW95, S96, T96}.
This situation is schematically represented by Figure \ref{dual-web:fig}.
\begin{figure}[h]
\begin{center}
\LS 
\includegraphics{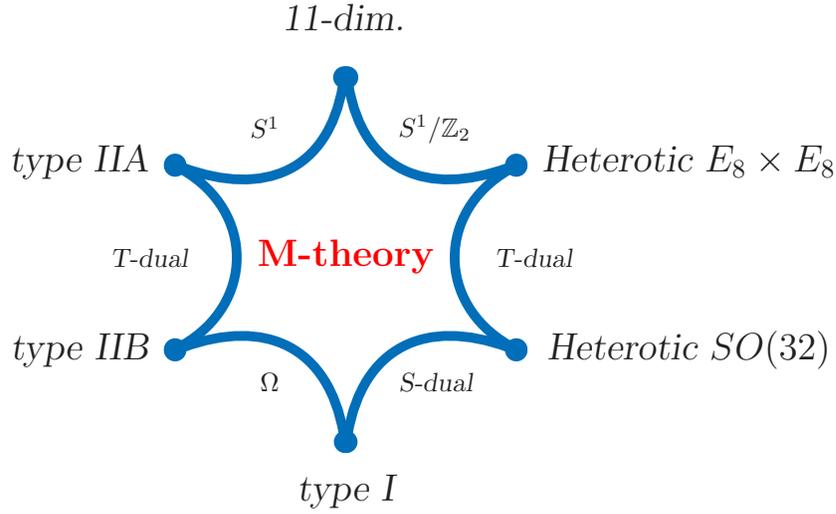}
\caption{\sl The duality web among five superstring theories and
  eleven-dimensional theory.}
\label{dual-web:fig}
\end{center}
\end{figure}

\noindent
We discuss a very rough explanation for the string duality web
described in Figure \ref{dual-web:fig}.
First, performing the worldsheet parity projection ($\Omega$ projection)
and introducing an appropriate orientifold plane in type IIB string
theory,
we obtain the closed string sector of type I string theory:
When we compactify one direction to a circle of radius $R$ and take 
T-duality to this circle in type IIA (or IIB) string theory,
we obtain type IIB (or type IIA) string theory on nine-dimensional
spacetime plus one circle of radius $\alpha'/R$. 
We also connect heterotic string theory with gauge group $E_8 \times
E_8$ to heterotic string with gauge group $SO(32)$ via T-duality:
S-duality is a duality under which the coupling constant of a quantum
theory changes non-trivially, including the case of strong-weak duality.
Via S-duality we can connect heterotic $SO(32)$ string theory 
to type I string theory.
Type IIB string theory is invariant under the S-duality
transformation.
Performing S-duality to type IIA string,
i.e., taking the strong coupling limit of type IIA string,
we may reach an unknown eleven-dimensional theory whose low energy
effective theory is the eleven-dimensional supergravity:
Performing compactification 
the eleven-dimensional theory on $S^1/{\mathbb Z}_2$,
we obtain heterotic $E_8 \times E_8$ string theory:
Furthermore, if we compactify some string theory on nontrivial compact
manifolds, for instance, $K3$ surface and Calabi-Yau three-fold,
we find deeper relations among these string theories.

There is a substantial piece of evidence that eleven-dimensional
quantum theory, i.e., M-theory, might underlie type IIA string theory
in the strong coupling limit.
The first evidence is the existence of the dilaton field in the low
energy action (see Table \ref{FC-II}).
When an eleven-dimensional gravitational theory is compactified on 
$x^{10}$-directions, 
the component of the metric $g_{10,10}$
behaves as a scalar field in the lower dimensional theory.
Furthermore this scalar field enters the lower dimensional action in
the same way that the dilaton does.
This suggests that the dilaton in type IIA string theory really
emerges via local compactification of higher dimensional theory,
say, via local compactification of eleven-dimensional theory.
The second piece of evidence is 
the existence of Ramond-Ramond one-form field in type IIA string theory.
Massless fields in type IIA string theory also appears via dimensional
reduction of eleven-dimensional supergravity.
This dimensional reduction keeps only the $p_{10} = 0$ states (where $p_{10}$
denotes the Kaluza-Klein momentum in the compactified direction),
but type IIA string theory has also states of $p_{10} \neq 0$ in the
form of $N$ {\sl D0-branes} and their bound states.
In this situation a D0-brane mass $m_0$ is given by
\begin{align}
m_0 \ &= \ \frac{1}{R} \ = \ \frac{1}{{\rm g} \l_s} \; , 
\label{D0-mass}
\end{align}
where $R$ is the radius of compactified direction, ${\rm g}$ is
the type IIA string coupling and $\l_s$ is the string length.
The Kaluza-Klein momentum $p_{10}$ is represented as $p_{10} = N m_0$.
Furthermore we know that 
the D0-branes couple to Ramond-Ramond one-form gauge field in type IIA
string theory via $\dps \int \! C_1$.
Thus the D0-brane analysis is much important to understand the
mysterious properties of eleven-dimensional theory, the M-theory properties.

Through the above string dualities, 
a simple and intriguing model was proposed in order to define a
microscopic description of M-theory.


\section*{Matrix Theory}

In 1996, Banks, Fischler, Shenker and Susskind proposed that 
the degrees of freedom of M-theory in the infinite momentum frame
could be described in terms of D0-branes and that all dynamics of
M-theory in this frame are described by the system of the low energy
effective theory of $N$ D0-branes in the large-$N$ limit \cite{BFSS96}.
Furthermore, in 1997, Susskind refined the proposal by conjecturing that for
all finite $N$ the quantum theory describes the sector of $N$ units of
momentum of M-theory with discrete light-cone quantization (DLCQ)
\cite{Suss97}. 
We refer the ideas of Banks, Fischler, Shenker and
Susskind and Susskind's refinement to the ``BFSS conjecture'' and the
model described by the matrix variables is called the ``Matrix
theory'' 
(for the review lectures, see, for instance, \cite{Bi9710, B9710,
  BS9712, WT98, B9911, WT00, WT01}.).

Matrix theory is defined in the framework of type IIA string theory.
In this framework the string coupling is weak and the D0-brane
mass becomes infinitely heavy as in (\ref{D0-mass}).
Thus the Lagrangian of this theory 
should be described by the {\sl non-relativistic limit} of $N$ D0-brane system.
The relativistic effective theory of D-brane system is described by
Dirac-Born-Infeld (DBI) action \cite{DLP89, L89}.
In the non-relativistic limit
this action reduces to 
$(0+1)$-dimensions of the ten-dimensional $U(N)$ super Yang-Mills theory:
\begin{align*}
L \ &= \ 
\frac{(2 \pi \alpha')^2}{2 {\rm g} \l_s} \Tr \Big\{
\dot{X}^I \dot{X}^I + \half [ X^I , X^J ]^2 
+ \Psi^{\dagger} \big( i \dot{\Psi} + \gamma^I [ X^I , \Psi ] \big)
\Big\}
\; , 
\end{align*}
where we set the gauge potential $A_0 = 0$.
The bosonic fields $X^I$, which have dimensions of $({\rm mass})^1$,
and fermionic fields $\Psi$, the mass dimensions $3/2$,
 are described as $N \times N$ matrix variables.
This Lagrangian gives the same Hamiltonian as the one of
matrix-regularized supermembrane theory via an appropriate field
rescaling\footnote{There is one relation in eleven-dimensional
  spacetime such as $R = {\rm g}^{2/3} \l_{11}$, 
where $\l_{11}$ is the Planck length of eleven dimensions.}!

While the BFSS conjecture is based on a different viewpoint from the
matrix-regularized supermembrane theory,
the Matrix theory provides us a lot of new interpretations for the
supermembrane in M-theory.
Here we introduce a few piece of important evidence.
One is that the Hilbert space of the matrix quantum mechanics contains
multiple particle states.
This observation resolves the problem of the continuous spectrum and
the membrane instability problem in
the supermembrane theory \cite{dWLN89}.
It is natural to think of the 
Matrix theory as a {\sl second quantized theory}
from the point of view of the target space.
Another evidence is the fact that quantum effects in the Matrix theory
give rise to long-range interactions between a pair of quanta, i.e., 
a pair of D0-branes.
These interactions have precisely the structure expected from 
the light-front supergravity.
There are lectures around this topic written by Taylor
\cite{WT98, WT00, WT01}.

Although the Matrix theory has been well studied in various works 
and there are many non-trivial results to check the above arguments,
there still exist serious question which have not been understood:
{\sl Can we formulate the Matrix theory on curved spacetime
  backgrounds without any inconsistency?}
Well-defined construction of 
Matrix theory on (arbitrary) curved spacetime
background is one of the most interesting and mysterious subjects
because we would like to understand whether the Matrix theory is a
fundamental description of M-theory
through various relations (or correspondences) 
between matrix model and supermembrane theory.
There are a lot of attempts around the Matrix theory on curved
background \cite{DOS9702, Sen97, S9710, DO9710}.
In particular, 
Taylor and Van Raamsdonk discussed the Matrix theory on weakly curved
spacetime background \cite{TvR98, TvR9904, TvR9910} but 
it is still difficult to analyze the Matrix theory on curved spaces.

In the end of the last century, 
one ten-dimensional spacetime background was discovered as a specific
limit of the product space of anti-de Sitter space and the Einstein
space which is a well-known background in supergravity \cite{G00,
FP0105}. 
This specific spacetime is the ``plane-wave background'' as the ``Penrose
limit'' of the $AdS_5 \times S^5$ spacetime which appears in the near
horizon limit of D3-brane in type IIB theory.
This plane-wave background is so useful that the study on the ``AdS/CFT
correspondence'' has been developed rapidly \cite{BMN02}.

There is also such a specific spacetime in eleven dimensions.
This eleven-dimensional spacetime background was first discovered by 
Kowalski-Glikman \cite{KG8401, KG8402} and was obtained 
as the Penrose limit of $AdS_4 \times S^7$ or $AdS_7 \times S^4$
backgrounds which appear in the near horizon limit of M2-brane or
M5-brane, respectively \cite{FP0105}.
This eleven-dimensional plane-wave background is also useful to analyze
Matrix theory on non-trivially curved background.
Although there is no tunable parameter in the flat background,
we can introduce one tunable mass parameter $\mu$ from the constant
four-form flux on the plane-wave.
Thus we can perform a Matrix perturbation theory for M-theory on such
a specific background!

In this doctoral thesis,
we will investigate a zero-mode spectrum included in Matrix theory on
the plane-wave background and will compare this to the massless
spectrum in the eleven-dimensional supergravity on the same
background.
This task should be an intrinsic work for Matrix theory on curved
background because Matrix theory on curved spacetime must also include
the superparticle subject to the eleven-dimensional supergravity as in
the case of flat spacetime background.


\section*{Organization}

The subjects of the doctoral thesis are organized as follows:

In {\bf section \ref{BMN}} we will review the Matrix theory on the plane-wave
background proposed by Berenstein, Maldacena and Nastase.
Introducing the construction procedure of this matrix model, 
we will construct the Hamiltonian and the supercharges 
of 32 local supersymmetry on the plane-wave. 
There we will discuss only the $U(1)$ part of the system, i.e., the
center of mass degrees of freedom of $N$ D0-branes which corresponds
to the superparticles. 
We will construct the supermultiplet including the ground state and
will read the energy spectrum of this multiplet.

In {\bf chapter \ref{MAIN}} 
we will analyze the (linearized) supergravity on the
same background in eleven dimensions. 
We will define the light-cone Hamiltonian in terms of the differential
operators and argue the Klein-Gordon type field equations.
Making bosonic and fermionic fields fluctuate we will obtain the field
equations for these fluctuation fields.
Since it is difficult to read the correct energies of them, we should
combine them in appropriate re-definitions. 
After these analyses we will obtain the zero-point energy spectrum of
these fluctuation and we will compare them with the result obtained in
chapter \ref{BMN}.

We devote {\bf chapter \ref{summary}} to the conclusion and discussions for
future problems.
We will discuss only the superparticles in both Matrix theory and
supergravity. In this chapter we will argue 
the possibilities to study some properties derived from extended
objects such as M2-brane and M5-brane in M-theory.

In {\bf appendix \ref{convention}} we will discuss the notation and
convention for some variables in the main chapters.
In particular we will write down the definitions of Dirac gamma matrices
and Majorana spinors in eleven-dimensional Minkowski spacetime.
The gamma matrices and spinors in $SO(9)$ Euclidean space and their
$SU(4) \times SU(2)$ decomposition rules are also introduced.

In {\bf appendix \ref{Lagrangians}}
we will discuss the dimensional reduction procedure of ten-dimensional
super Yang-Mills theory.
The nonabelian D-branes' effective action with non-vanishing background
fields will be also discussed.
Furthermore we will write down the eleven-dimensional supergravity Lagrangian.

In {\bf appendix \ref{geometry}}  
we will mention the Penrose limit of eleven-dimensional product spaces
such as $AdS_4 \times S^7$ and $AdS_7 \times S^4$.
We will also argue the geometrical 
properties of the plane-wave spacetime and its coset
construction via the Penrose limit of $AdS_{4(7)} \times S^{7(4)}$
spacetimes.


\makeatletter
\renewcommand{\theequation}{%
   \thesection.\arabic{equation}}
 \@addtoreset{equation}{section}
\makeatother

\chapter{Matrix Theory on the Plane-wave} \label{BMN}

\newpage

\setcounter{section}{0}
\renewcommand{\thesection}{\thechapter.\arabic{section}}

\indent
On 2002, 
Berenstein, Maldacena and Nastase proposed 
the Lagrangian of the DLCQ of Matrix theory
on the plane-wave background 
in a similar way of constructing type IIB superstring Lagrangian on the
ten-dimensional plane-wave background \cite{BMN02}.
This model is very useful to understand the properties of matrix model
on some specific curved spacetime and is now called the ``BMN matrix model''.
Not long after that,
Dasgupta, Sheikh-Jabbari and Van Raamsdonk found that 
the light-cone Hamiltonian of supermembrane on the plane-wave
background exactly corresponds to that of BMN matrix model
via matrix regularization \cite{DSvR0205}.
Furthermore Sugiyama and Yoshida explained the supersymmetric 
quantum mechanics of 
supermembrane theory on the plane-wave in the same way as the
quantum mechanics of supermembrane on flat background discussed by 
de Wit, Hoppe and Nicolai \cite{dWHN88, SY0206}.
They started the discussion from
the supermembrane Lagrangian as a gauge theory of area preserving
diffeomorphism and construct the light-cone Hamiltonian, 32
supercharges, their commutation relations, brane charges and their
matrix regularizations. Their results are consistent with the
BMN matrix model.

In this chapter we discuss the spectrum of the center of mass degrees
of freedom in the BMN matrix model.
We describe the Hamiltonian and supercharges in the $N \times N$
matrix representations and study their commutation relations.
We also define the ground state of this system and construct the 
supermultiplet of the $U(1)$ free part of the matrix model
in terms of the oscillator method as discussed by Dasgupta,
Sheikh-Jabbari and Van Raamsdonk \cite{DSvR0205, DSvR0207}, 
Kim and Plefka \cite{KP0207-1}, Kim and Park \cite{KP0207-2}, and
Nakayama, Sugiyama and Yoshida \cite{NSY02}.


\section{Derivation of Lagrangian}

In this section we construct the Lagrangian of the discrete light-cone
quantization (DLCQ) of Matrix theory on the plane-wave background
which was suggested by Berenstein, Maldacena and Nastase \cite{BMN02}. 
Let us first consider the action for single D0-brane on the plane-wave
and next expand this action to the non-abelian matrix model via
various techniques.

The single D0-brane action 
would be described as the superparticle action 
moving in the eleven-dimensional plane-wave background  
in the Green-Schwarz formalism, where
we use superspace coordinates and supervielbeins of spacetime background.
Here we write the superparticle action
\begin{align}
S \ &= \ \int \! \d t \, e^{-1} (t) \, \Big\{ \half 
\eta_{\r{A} \r{B}} \, \Pi_{t}^{\r{A}} \, \Pi_{t}^{\r{B}} 
\Big\}
\ = \ 
\int \! \d t \, \Big\{ - \Pi_{t}^{\r{+}} \, \Pi_{t}^{\r{-}}
+ \half \Pi_{t}^{\r{I}} \, \Pi_{t}^{\r{I}} \Big\}
\; . \label{super-P}
\end{align}
Note that $\Pi_t^{\r{A}} = \del_t Z^{\ul{M}} E_{\ul{M}}{}^{\r{A}}$ 
are pullbacks from the eleven-dimensional 
curved spacetime\footnote{The index $\r{I}$
  runs from 1 to 9 in the tangent space.} 
spanned by the superspace coordinates
$Z^{\ul{M}} = (X^M , \theta^{\alpha})$ to the worldline coordinate
$t$, and the supervielbeins are denoted by $E_{\ul{M}}{}^{\r{A}}$;
the einbein of the worldline ``metric'' is denoted by $e (t)$ and we
can choose $e (t) = 1$ because of the existence of
diffeomorphism of one-dimensional worldline.
As discussed in appendix \ref{PL},
the plane-wave background is the Penrose limit of the
$AdS_{4(7)} \times S^{7(4)}$ spacetime.
Thus we can describe the supervielbein on the plane-wave as the Penrose
limit of the $AdS_{4(7)} \times S^{7(4)}$ supervielbein 
and we obtain them by
substituting the geometrical variables of the plane-wave 
(\ref{KG-BG2-app}) into the supervielbein 
on the $AdS_{4(7)} \times S^{7(4)}$ background (\ref{TK-supervielbein-sol}).

The superparticle action (\ref{super-P}) has a fermionic gauge
symmetry called the $\kappa$-symmetry which the Green-Schwarz
superstring action also has.
This $\kappa$-symmetry should be gauge-fixed by choosing the fermionic
light-cone gauge (This procedure is adopted when we obtain the superstring in
$AdS_5 \times S^5$ and its Penrose limit \cite{M0112}).
Here we can choose the following fermionic gauge-fixing
\begin{align}
\hG^{\r{+}} \theta \ &= \ 0 \label{K-fix}
\end{align}
which is equivalent to the condition $\ol{\theta} \hG^{\r{+}} = 0$.
Under this condition the fermionic matrix ${\cal M}^2$ in the
supervielbein (\ref{TK-supervielbein-sol}) vanishes 
and we can simply write the components of supervielbein and the pullback
\begin{gather*}
\Pi^{\r{+}} 
\ = \ \d X^{+}
\; , \ls
\Pi^{\r{I}} 
\ = \ \d X^I
\; , \\
\Pi^{\r{-}} 
\ = \ \d X^{-} - \half G_{++} \, \d X^+ + \ol{\theta} \hG^{\r{-}} \d \theta
- \frac{\mu}{4} e^{\r{+}} \ol{\theta} \hG^{\r{-}} \hG^{\r{1} \r{2}
  \r{3}} \theta 
\; ,
\end{gather*}
where $\mu$ is a parameter included in the plane-wave metric discussed
in appendix \ref{PL}.
Thus the superparticle action is rewritten as\footnote{From now on we
  use the relation $\hG^{\r{1} \r{2} \r{3}} = \hG^{123}$ because
  these directions are flat on the plane-wave background 
(see the plane-wave metric in appendix \ref{PL}).}
\begin{align}
S \ &= \ \int \! \d t \Big\{
\half \sum_{I=1}^9 (\del_t X^I)^2 - \ol{\theta} \hG^{\r{-}} \del_t \theta
- \half \Big[ \Big( \frac{\mu}{3} \Big)^2 \sum_{\wt{I}=1}^3 (X^{\wt{I}})^2 
+ \Big( \frac{\mu}{6} \Big)^2 \sum_{I'=4}^9 (X^{I'})^2 \Big]
+ \frac{\mu}{4} \ol{\theta} \hG^{\r{-}} \hG^{123} \theta
\Big\}
\; , \label{S-0}
\end{align}
where we also choose the bosonic light-cone gauge fixing 
$X^+ = t$, $\del_t X^- = 0$\footnote{Here we do not mention the strict
definitions of variables.}.
Note that 
the $SO(10,1)$ Majorana spinor $\theta$ can be represented by the $SO(9)$
Majorana spinor $\Psi$ because of the fermionic light-cone gauge
fixing (\ref{K-fix})
\begin{align*}
\theta \ &\equiv \ \frac{1}{2^{3/4}} \left(
\begin{array}{c}
0 \\
\Psi
\end{array} \right) 
\; , \ls
\ol{\theta} \ = \ \theta^T C \ = \ 
\frac{1}{2^{3/4}} \Big( \, - \Psi^T \; , \; 0 \, \Big)
\; .
\end{align*}
Utilizing the $SO(9)$ Majorana spinor $\Psi$,
we reduce the $\theta$ bilinear terms in (\ref{S-0}) to the
following:
\begin{align*}
- \ol{\theta} \hG^{\r{-}} \del_t \theta
\ &= \ 
\frac{i}{2} \Psi^{\dagger} \del_t \Psi
\; , \ls
\frac{\mu}{4} \ol{\theta} \hG^{\r{-}} \hG^{123} \theta
\ = \ 
- \frac{i \, \mu}{8} \Psi^{\dagger} \gamma^{123} \Psi 
\; .
\end{align*}
Definitions of the $SO(9)$ gamma matrices $\gamma^{\r{I}}$ are 
summarized in appendix \ref{conv-membrane}.
Thus we write down the superparticle action as follows:
\begin{align}
S \ &= \ \int \! \d t \Big\{
\half \sum_{I=1}^9 (\del_t X^I)^2 
+ \frac{i}{2} \Psi^{\dagger} \del_t \Psi
- \half \Big[ \Big( \frac{\mu}{3} \Big)^2 \sum_{\wt{I}=1}^3 (X^{\wt{I}})^2 
+ \Big( \frac{\mu}{6} \Big)^2 \sum_{I'=4}^9 (X^{I'})^2 \Big]
- \frac{i \, \mu}{8} \Psi^{\dagger} \gamma^{123} \Psi
\Big\}
\; . \label{S-1}
\end{align}

Let us consider the supersymmetry invariance of the action (\ref{S-1})
and generalize it to the multi-superparticle action, i.e., $N$
D0-branes' action represented by non-abelian $U(N)$ gauge symmetry group.
First we look for the supersymmetry transformation of the type
\begin{align}
\delta X^I \ &\equiv \ \Psi^{\dagger} \gamma^I \, \eps (t)
\; , \nn \\
\delta \Psi \ &\equiv \ b \, \del_t X^I \gamma^I \, \eps (t)
+ \mu \, X^I \gamma^I M_{I}' \, \eps (t)
\; , \label{var-SUSY-BMN-1} \\
\eps (t) \ &= \ \exp \big( \mu M t) \, \eps_0
\; , \nn
\end{align}
where $b$ is a numerical constant and $\eps_0$ is a constant $SO(9)$
Majorana spinor; $M$ and $M_I'$ are matrix valued parameters. 
We will determine the values of these variables via properties of the
invariance of action $S$ under the supersymmetry of type
(\ref{var-SUSY-BMN-1}).
The invariance of the action under the supersymmetry transformations
(\ref{var-SUSY-BMN-1}) leads to the following equation: 
\begin{align}
\begin{split}
0 \ &= \ 
\int \! \d t \Big\{
\big( 1 - b i \big) \del_t X^I (\del_t \Psi)^{\dagger} \gamma^I \eps
\Big\}
\\
\ & \ \ \ \ 
+ \mu \int \! \d t \Big\{
\del_t X^I \Psi^{\dagger} \gamma^I M \eps
+ i \, \del_t X^I \, \Psi^{\dagger} \gamma^I M_I' \eps 
- b \frac{i}{4} 
\del_t X^I \, \Psi^{\dagger} \gamma^{123} \gamma^I \eps
\Big\}
\\
\ & \ \ \ \ 
+ \mu^2 \int \! \d t \Big\{
i \, X^I \, \Psi^{\dagger} \gamma^I \big( M_I' M \big) \eps
- \frac{1}{9} X^{\wt{I}} \Psi^{\dagger} \gamma^{\wt{I}} \eps
- \frac{1}{36} X^{I'} \Psi^{\dagger} \gamma^{I'} \eps
- \frac{i}{4} X^I \Psi^{\dagger} \gamma^{123} \gamma^I M_I' \eps
\Big\}
\; .
\end{split} \label{VS}
\end{align}
{}From now on we omit summation symbols with respect to the spacetime
coordinates.
We consider the invariance (\ref{VS}) 
order by order with respect to the parameter $\mu$.
The terms of order $\mu^0$ determine the constant as $b$ in the
supersymmetry transformation (\ref{var-SUSY-BMN-1}) as $b= -i
$. 
The terms of order $\mu^1$ in (\ref{VS}) give the equations
\begin{gather*}
M + i M_{\wt{I}}' - \frac{1}{4} \gamma^{123} \ = \ 0
\; , \ls
M + i M_{I'}' + \frac{1}{4} \gamma^{123} \ = \ 0
\; ,
\end{gather*}
and the terms of order $\mu^2$ in (\ref{VS}) leads to
\begin{gather*}
i M_{\wt{I}}' M - \frac{1}{9} - \frac{i}{4} \gamma^{123} M_{\wt{I}}' 
\ = \ 0 
\; , \ls
i M_{I'}' M - \frac{1}{36} + \frac{i}{4} \gamma^{123} M_{I'}' 
\ = \ 0 
\; .
\end{gather*}
Then we obtain the values of unknown parameters $M$ and $M_I'$ as
\begin{gather}
M \ = \ - \frac{1}{12} \gamma^{123} \; , \ls
i M_{\wt{I}}' \ = \ \frac{1}{3} \gamma^{123}
\; , \ls
i M_{I'}' \ = \ - \frac{1}{6} \gamma^{123}
\; .
\end{gather}

The extension to the non-abelian theory is obvious;
besides the usual commutator terms which are present in the Lagrangian
and supersymmetry transformation rules in flat spaces,
we have an extra coupling of order $\mu$.
Indeed, it was found that a term $F_{tIJK} \Tr(X^I X^J X^K)$ 
should be included
in the action for $N$ D0-branes in constant Ramond-Ramond field strength
\cite{T9701, M99, M0303}
(see also appendix \ref{nonabelian}).
In our case, the coupling is
\begin{align*}
F_{+\wt{I} \wt{J} \wt{K}} \Tr ( X^{\wt{I}} X^{\wt{J}} X^{\wt{K}} ) 
\ &= \ - \mu \, \eps_{\wt{I} \wt{J} \wt{K}}
\, \Tr ( X^{\wt{I}} X^{\wt{J}} X^{\wt{K}} ) 
\; .
\end{align*}
Thus the action is written in terms of $N \times N$ matrix valued
fields $X^I$ and $\Psi$ 
\begin{align}
\begin{split}
S \ &= \
\int \! \d t \, \Tr \Big\{ \half (\del_t X^I)^2
+ \frac{i}{2} \Psi^{\dagger} \del_t \Psi 
- \half \Big[
\Big( \frac{\mu}{3} \Big)^2 (X^{\wt{I}})^2
+ \Big( \frac{\mu}{6} \Big)^2 (X^{I'})^2 \Big]
\\
\ & \LS \ls
- \frac{i \, \mu}{8} \Psi^{\dagger} \gamma^{123} \Psi
+ d \, \mu \, g \, \eps_{\wt{I} \wt{J} \wt{K}}
\, ( X^{\wt{I}} X^{\wt{J}} X^{\wt{K}} ) 
+ \frac{1}{4} \, g^2 \, [ X^I, X^J ]^2 
+ \half \, g \, \Psi^{\dagger} \gamma^I [ X^I , \Psi ] 
\Big\}
\; .
\end{split} \label{S-3}
\end{align}
We explain newly introduced terms in the above action from the
viewpoint of the dimensional reduction of 
ten-dimensional $U(N)$ super Yang-Mills as in appendix \ref{SYM}.
The matrix valued fields $X^I$ and $\Psi$, whose mass dimensions are
$-1/2$ and $0$, are not only the 
the adjoint representations of $U(N)$ gauge group but also
the dynamical variables in ten-dimensional SYM.
The parameter $g$ is the Yang-Mills coupling with mass dimensions $3/2$. 
The quartic term $\frac{1}{4} g^2 [ X^I , X^J ]^2$ can be derived from
the reduction of the field strength of $U(N)$ gauge potential.
We obtain the three-point vertex term 
$\frac{1}{2} g \Psi^{\dagger} \gamma^{I} [ X^I , \Psi]$ 
from the covariant derivative of fermion 
$D_M \Psi = \del_M \Psi + i g [ A_M , \Psi ]$ in super Yang-Mills.
Notice that although the fermion $\Psi$ is the $SO(9)$ Majorana spinor
in our derivation, we can also regard this as 
the $SO(9,1)$ Majorana-Weyl spinor in ten dimensions.

The supersymmetry transformations of this system should be extended as
\begin{align}
\delta X^I \ &= \ \Psi^{\dagger} \gamma^I \eps (t)
\;, \nn \\
\delta \Psi \ &= \ 
- i \del_t X^I \gamma^I \eps (t)
- \frac{i \mu}{3} X^{\wt{I}} \gamma^{\wt{I}} \gamma^{123} \eps (t)
+ \frac{i \mu}{6} \, X^{I'} \gamma^{I'} \gamma^{123}
\eps (t)
+ \half \, g \, [ X^I , X^J ] \gamma^{IJ} \eps (t)
\; , \label{dy-SUSY-1} \\
\eps (t) \ &= \ \exp \big( - \frac{\mu}{12} \gamma^{123} t \big) \eps_0
\; . \nn
\end{align}
Last,
we introduce the gauge potential $A_t$ as 
an auxiliary matrix variable of this system
and rewrite the derivative $\del_t$ to 
the covariant derivative $D_{t} X^I = \del_t X^I + i g [ A_t , X^I ]$:
\begin{align}
\begin{split}
S \ &= \ \int \! \d t \, \Tr \Big\{
\half D_t X^I \, D_t X^I 
+ \frac{i}{2} \Psi^{\dagger} D_t \Psi
- \half \Big[ \Big( \frac{\mu}{3} \Big)^2 (X^{\wt{I}})^2 
+ \Big( \frac{\mu}{6} \Big)^2 (X^{I'})^2 \Big]
\\
\ & \LS \ls
- \frac{i \, \mu}{8} \Psi^{\dagger} \gamma^{123} \Psi
- \frac{i \, \mu}{3} g \, \eps_{\wt{I} \wt{J} \wt{K}} \, 
X^{\wt{I}} X^{\wt{J}} X^{\wt{K}}
+ \frac{1}{4} \, g^2 \, [ X^I , X^J ]^2 
+ \half \, g \, \Psi^{\dagger} \gamma^I [ X^I , \Psi ] \Big\}
\; . 
\end{split} \label{S-4'}
\end{align}
Here we can interpret that the covariant derivative $D_t X^I$ comes
from the dimensional reduction of field strength 
\begin{align*}
F_{MN} \ = \ \del_M A_N - \del_N A_M + i g [ A_M , A_N ]
\end{align*} 
in the ten-dimensional $U(N)$ super Yang-Mills Lagrangian.
This action (\ref{S-4'}) is also obtained by the matrix regularization of the
supermembrane on the plane-wave under the appropriate rescaling 
of some variables \cite{DSvR0205, SY0206}.

Now let us re-define the field variables in order for the
compatibility of the description of the nonabelian 
Dirac-Born-Infeld type Lagrangian discussed in appendix \ref{nonabelian}.
Combining the Yang-Mills coupling $g$ and field variables 
\begin{gather*}
g X^I \ \equiv \ X'{}^I \; , \ls
g A_t \ \equiv \ A_t' \; , \ls
g \Psi \ \equiv \ \Psi' \; ,
\end{gather*}
we rewrite the action (\ref{S-4'}) as
\begin{align}
\begin{split}
S \ &= \ \frac{1}{g^2} \int \! \d t \, \Tr \Big\{
\half D_t X'{}^I \, D_t X'{}^I 
+ \frac{i}{2} \Psi'{}^{\dagger} D_t \Psi'
- \half \Big[ \Big( \frac{\mu}{3} \Big)^2 (X'{}^{\wt{I}})^2 
+ \Big( \frac{\mu}{6} \Big)^2 (X'{}^{I'})^2 \Big]
\\
\ & \LS \ls
- \frac{i \, \mu}{8} \Psi'{}^{\dagger} \gamma^{123} \Psi'
- \frac{i \, \mu}{3} \eps_{\wt{I} \wt{J} \wt{K}} \, 
X'{}^{\wt{I}} X'{}^{\wt{J}} X'{}^{\wt{K}}
+ \frac{1}{4} \, [ X'{}^I , X'{}^J ]^2 
+ \half \, \Psi'{}^{\dagger} \gamma^I [ X'{}^I , \Psi' ] \Big\}
\; . 
\end{split} \label{S-4}
\end{align}
Note that the mass dimensions of $X'$ and $\Psi'$ are $1$ and $3/2$,
respectively. 
But, for simplicity, we omit the prime symbol in field variables. 
We also rewrite the Yang-Mills coupling $g$ 
in terms of the 
D0-brane mass (or tension) $m_0$ and the Regge constant $\alpha'$ as 
$g^{-2} = (2 \pi \alpha')^2 m_0$. 
(We will discuss this relation in appendix \ref{nonabelian}.)
{}From the viewpoint of DLCQ with compactification $x^- \sim x^- + 2
\pi R$, the D0-brane mass is represented in terms of $R$ as $m_0 = 1/R$.
Thus, we can write the overall factor of the action (\ref{S-4}) is 
\begin{align*}
\frac{1}{g^{2}} \ &= \ \frac{(2 \pi \alpha')^2}{R} \; .
\end{align*}
In the next section 
we will construct the Hamiltonian, supercharges and their commutation
relations in terms of the conventions adopted by Dasgupta,
Sheikh-Jabbari and Van Raamsdonk \cite{DSvR0205}.
We will also analyze one specific spectrum.


\section{Hamiltonian, Supercharges and their Commutation Relations}

We would like to study the zero-mode spectrum of this matrix model.
Before starting a discussion, we must prepare some operators such as 
Hamiltonian, supercharges, and
the commutation relations between them.
Here we review such preliminary discussed by Dasgupta,
Sheikh-Jabbari and Van Raamsdonk \cite{DSvR0205}.
Now we rewrite the matrix model Lagrangian (\ref{S-4}) via the
following rescaling\footnote{The re-definition
  (\ref{rescale-SM-to-M}) is somewhat complicated and looks like
  strange. Of course we can discuss the same investigation without this
re-definition. 
But we will analyze the system described by the action (\ref{BMN-L}),
the same representation as \cite{DSvR0205}, where Dasgupta,
Sheikh-Jabbari and Van Raamsdonk suggested the perturbation 
of the BMN matrix model.}:
\begin{align}
\begin{split}
t \ &= \ R^{2/3} \tau
\; , \ls
A_t \ = \ R^{-2/3} A_{\tau}
\; , \ls 
\mu \ = \ R^{-2/3} \wt{\mu}
\; , \\ 
X^I \ &= \ R^{1/3} \wt{X}^I
\; , \ls
\Psi \ = \ R^{1/2} \wt{\Psi}
\end{split} \label{rescale-SM-to-M} 
\end{align}
and $2 \pi \alpha' \equiv 1$.
Under the above rescaling, 
the Matrix Theory Lagrangian describing the DLCQ of M-theory 
on the plane-wave background \cite{DSvR0205} is given by
\begin{align}
S \ &= \ \int \! \d \tau \, {\cal L} \; , \nn \\
\begin{split}
{\cal L} \ &= \ 
\Tr \Big\{
\frac{1}{2R} \wt{D}_{\tau} \wt{X}^I \, \wt{D}_{\tau} \wt{X}^I 
+ \frac{i}{2} \wt{\Psi}^{\dagger} \wt{D}_{\tau} \wt{\Psi} 
+ \frac{R}{2} \wt{\Psi}^{\dagger} \gamma^{I} [ \wt{X}^{I}, \wt{\Psi} ] 
+ \frac{R}{4} [ \wt{X}^{I} , \wt{X}^{J} ]^2 
\Big\}
\\
\ & \ \ \ \ 
+ R \, \Tr \Big\{
- \half \Big[ \Big( \frac{\wt{\mu}}{3R} \Big)^2 (\wt{X}^{\wt{I}})^2
+ \Big( \frac{\wt{\mu}}{6R} \Big)^2 (\wt{X}^{I'})^2 \Big]
- \frac{i \, \wt{\mu}}{3R} \, \eps_{\wt{I} \wt{J} \wt{K}} \, 
\wt{X}^{\wt{I}} \wt{X}^{\wt{J}} \wt{X}^{\wt{K}} 
- \frac{i \, \wt{\mu}}{8R} \wt{\Psi}^{\dagger} \gamma^{123} \wt{\Psi}
\Big\}
\; , 
\end{split} \label{BMN-L}
\end{align}
where the covariant derivative $\wt{D}_{\tau} \wt{X}^I$ is given by
$\wt{D}_{\tau} \wt{X}^I = \del_{\tau} \wt{X}^{I} +i [ A_{\tau} , \wt{X}^I ]$.
For simplicity, we omit the tildes written above the rescaled variables.
Performing Legendre transformation, we obtain the Hamiltonian of this
system. 
We define the canonical momenta of $X^I$ and $\Psi$ in terms of the 
right-derivative:
\begin{align*}
(P_I)_{kl} \ &= \ 
\frac{\del}{\del (\del_{\tau} X^I)_{lk}} {\cal L}
\ = \ 
\frac{1}{R} (D_{\tau} X^I)_{kl}
\; , \ls
(S)_{kl} \ = \ 
\frac{\del}{\del (\del_{\tau} \Psi)_{lk}} {\cal L}
\ = \ 
\frac{i}{2} (\Psi^{\dagger})_{kl}
\; ,
\end{align*}
where $k$ and $l$ are indices of $N \times N$ matrices.
Thus the Hamiltonian is described as
\begin{align}
\begin{split}
H \ &= \ \Tr \{ P_I \del_{\tau} X^I \} 
+ \Tr \{ S \del_{\tau} \Psi \} - {\cal L} 
\\
\ &= \ 
R \, \Tr \Big\{
\half (P_I)^2 
- \half \Psi^{\dagger} \gamma^I [ X^I , \Psi ] 
- \frac{1}{4} [ X^I , X^J ]^2 
\\
\ & \LS \ \ \  
+ \half \Big[ \Big( \frac{\mu}{3R} \Big)^2 (X^{\wt{I}})^2 
+ \Big( \frac{\mu}{6R} \Big)^2 (X^{I'})^2 \Big]
+ \frac{i \, \mu}{3R} \, \eps_{\wt{I} \wt{J} \wt{K}} \, 
X^{\wt{I}} X^{\wt{J}} X^{\wt{K}}
+ \frac{i \, \mu}{8R} \Psi^{\dagger} \gamma^{123} \Psi
\Big\}
\; , 
\end{split} \label{BMN-H}
\end{align}
where we solved some Dirac constraints and substituted them 
into the Hamiltonian, or simply, 
wrote down this Hamiltonian under the gauge $A_{\tau} = 0$.

As for the case of flat spacetime,
the $U(1)$ part of the theory (i.e., the free part describing the center of
mass degrees of freedom) decouples from the $SU(N)$ part (the
interaction part of the theory).
On the plane-wave background,
the $U(1)$ sector is described by the harmonic oscillator Hamiltonian 
with bosonic oscillators in the $SO(3)$ directions of mass $\mu/3$ and
in the $SO(6)$ directions of mass $\mu/6$ as well as $8$ fermionic
oscillators of mass $\mu/4$.
Thus unlike the flat spacetime case,
the different polarization states have different masses.

Here we pick up the symmetry algebra of this Matrix theory and
provide explicit expressions for the bosonic generators in terms of
the matrix variables $X^I$ and $P_I$.
The bosonic generators include the harmonic oscillators $a^I$,
the Hamiltonian $H$, the light-cone
momentum $P^+$ (a central terms of the algebra) and the rotation
generators of $SO(3) \times SO(6)$ symmetry $\Sigma^{\wt{I} \wt{J}}$ and
$\Sigma^{I'J'}$, respectively.
These variables satisfy the following algebra \cite{BMN02, DSvR0205}
\begin{gather}
[ a^{\wt{I}} , a^{\dagger \wt{J}} ] \ = \
P^+ \, \delta^{\wt{I} \wt{J}} \; , \LS
[ a^{I'} , a^{\dagger J'} ] \ = \
P^+ \, \delta^{I'J'} \; , 
\nn \\
[ H , a^{\wt{I}} ] \ = \ - \frac{\mu}{3} a^{\wt{I}} 
\; , \LS \ \ \
[ H , a^{I'} ] \ = \ - \frac{\mu}{6} a^{I'} \; , 
\nn \\
[ \Sigma^{\wt{I} \wt{J}} , a^{\wt{K}} ] 
\ = \ 
i \big( \delta^{\wt{J} \wt{K}} \, a^{\wt{I}} 
- \delta^{\wt{I} \wt{K}} \, a^{\wt{J}}
\big)
\; , \LS
[ \Sigma^{I'J'} , a^{K'} ] 
\ = \ 
i \big( \delta^{J'K'} \, a^{I'} 
- \delta^{I'K'} \, a^{J'}
\big)
\; , \label{alg-U1} \\
\begin{split}
i [ \Sigma^{\wt{I} \wt{J}} , \Sigma^{\wt{K} \wt{L}} ] 
\ &= \ 
  \delta^{\wt{I} \wt{K}} \Sigma^{\wt{J} \wt{L}}
+ \delta^{\wt{J} \wt{L}} \Sigma^{\wt{I} \wt{K}}
- \delta^{\wt{I} \wt{L}} \Sigma^{\wt{J} \wt{K}}
- \delta^{\wt{J} \wt{K}} \Sigma^{\wt{I} \wt{L}}
\; , \\
i [ \Sigma^{I'J'} , \Sigma^{K'L'} ] 
\ &= \ 
  \delta^{I'K'} \Sigma^{J'L'} + \delta^{J'L'} \Sigma^{I'K'}
- \delta^{I'L'} \Sigma^{J'K'} - \delta^{J'K'} \Sigma^{I'L'}
\; .
\end{split} \nn 
\end{gather}
Note that the harmonic oscillators $a^{\dagger I}$ 
and $a^I$ are creation and annihilation
operators corresponding to the decoupled $U(1)$ part of the theory
which describes the center of mass degrees of freedom (a particle) in
a harmonic potential.
These generators are realized by the Matrix theory variables $X^I$,
$P^I$ and $\psi_{i \alpha}$:
\begin{gather*}
P^+ \ = \ \frac{1}{R} \Tr ({\bf 1})
\; , \\
a^{\wt{I}} \ = \ 
\frac{1}{\sqrt{R}}
\, \Tr \Big(
\sqrt{\frac{\mu}{6R}} X^{\wt{I}} + i \sqrt{\frac{3R}{2\mu}} P^{\wt{I}} \Big)
\; , \ls
a^{I'} \ = \ 
\frac{1}{\sqrt{R}}
\, \Tr \Big(
\sqrt{\frac{\mu}{12R}} X^{I'} + i \sqrt{\frac{3R}{\mu}} P^{I'} \Big)
\; , \\
\begin{split}
\Sigma^{\wt{I} \wt{J}} \ &= \ 
\Tr \Big( P^{\wt{I}} X^{\wt{J}} - P^{\wt{J}} X^{\wt{I}}
- i \eps^{\wt{I} \wt{J} \wt{K}} 
\, \psi^{\dagger i \alpha} 
\, (\sigma^{\wt{K}})_{\alpha}{}^{\beta} 
\, \psi_{i \beta} \Big)
\; , \\
\Sigma^{I'J'} \ &= \ 
\Tr \Big( P^{I'} X^{J'} - P^{J'} X^{I'} 
- \half \psi^{\dagger i \alpha} 
\, ({\sf g}^{I'J'})_{i}{}^{j} 
\, \psi_{j \alpha} \Big)
\; .
\end{split}
\end{gather*}
Notice that we have already used 
the $SU(4) \times SU(2)$ decomposition rule with respect to the
fermionic variables $\psi_{i \alpha}$ discussed in
appendix \ref{4-2-repre}; 
the gamma matrix in the last equation is defined as
${\sf g}^{I'J'} = \half \{{\sf g}^{I'} ({\sf g}^{J'})^{\dagger} -
{\sf g}^{J'} ({\sf g}^{I'})^{\dagger}\}$.
These generators expressed by the matrix variables 
satisfy the algebra (\ref{alg-U1}) 
via the (anti-)commutation relations
\begin{gather*}
[ X^{\wt{I}}_{kl} , P^{\wt{J}}_{mn} ] \ = \ 
i \, \delta^{\wt{I} \wt{J}} \, \delta_{kn} \, \delta_{lm}
\; , \ls
[ X^{I'}_{kl} , P^{J'}_{mn} ] \ = \ 
i \, \delta^{I'J'} \, \delta_{kn} \, \delta_{lm}
\; , \\
\{ (\psi^{\dagger i \alpha})_{kl} , (\psi_{j \beta})_{mn} \}
\ = \ \delta_j^i \, \delta_{\beta}^{\alpha} \, \delta_{kn} \, \delta_{lm}
\; .
\end{gather*}
These (anti-)commutation relations are also introduced when one
discuss the quantum mechanics of regularized supermembrane theory
in the light-cone gauge 
\cite{SY0206}.

The 32 components of the $SO(10,1)$ spacetime supersymmetry decompose
into two 16 components supersymmetry in the light-cone gauge.
One supersymmetry is linearly realized and the other 
nonlinearly realized as we shall discuss now. 
As discussed in the previous section,
the Matrix theory Lagrangian (\ref{BMN-L}) has the invariance of
nonlinearly realized supersymmetry transformation.
We rewrite the rescaled transformation rule of (\ref{dy-SUSY-1}):
\begin{align*}
\delta_{\eps} X^I \ &= \ \sqrt{R} \, \Psi^{\dagger} \gamma^I \eps (\tau)
\; , 
\ls
\delta_{\eps} \omega \ = \ \sqrt{R} \, \Psi^{\dagger} \eps (\tau) 
\; , \nn \\
\delta_{\eps} \Psi \ &= \ 
\sqrt{R} \, \Big( - \frac{i}{R} D_{\tau} X^I \gamma^I \eps (\tau)
+ \frac{1}{2} [ X^I , X^J ] \gamma^{IJ} \eps (\tau)
- \frac{i \, \mu}{3 R} X^{\wt{I}} \gamma^{\wt{I}} \gamma^{123} \eps (\tau)
+ \frac{i \, \mu}{6 R} X^{I'} \gamma^{I'} \gamma^{123} \eps (\tau)
\Big) 
\; , 
\nn \\
\eps (\tau) \ &= \ 
\exp \Big( - \frac{\mu}{12} \, \gamma^{123} \, \tau \Big) \, \eps_0
\; .
\end{align*}
We call this symmetry the ``dynamical supersymmetry'' whose
supercharges are written by
\begin{gather}
Q \ = \ \sqrt{R} \, \Tr \Big\{ 
P^I \gamma^I \Psi
- \frac{i}{2} [ X^I , X^J ] \gamma^{IJ} \Psi 
- \frac{\mu}{3R} X^{\wt{I}} \gamma^{\wt{I}} \gamma^{123} \Psi
- \frac{\mu}{6 R} X^{I'} \gamma^{I'} \gamma^{123} \Psi \Big\}
\; . \label{super-Q-9}
\end{gather}
The Lagrangian (\ref{BMN-L}) also has a linearly realized supersymmetry
whose transformation rule is 
\begin{gather*}
\delta_{\eta} X^I \ = \ 0
\; , \ls
\delta_{\eta} \omega \ = \ 0 
\; , \ls
\delta_{\eta} \Psi \ = \ \frac{1}{\sqrt{R}} \eta (\tau)
\; , \\
\eta (\tau) \ = \ \exp \Big( \frac{\mu}{4} \gamma^{123} \tau \Big) \,
\eta_0
\; ,
\end{gather*}
where the $SO(9)$ Majorana spinor $\eta_0$ is constant. This
supersymmetry is called the ``kinematical supersymmetry'' whose
supercharge is realized as
\begin{align}
q \ &= \ \frac{1}{\sqrt{R}} \Tr (\Psi)
\; .
\label{super-q-9}
\end{align}
Note that the dynamical supersymmetry acts on the $SU(N)$ interaction
part of theory whereas the kinematical supersymmetry acts only on the
free $U(1)$ part. 
In addition, the kinematical supercharges generate the overall
polarization states.
Between the dynamical and kinematical 
supersymmetries there are some nontrivial relation as follows \cite{DSvR0205}:
\begin{align}
\begin{split}
\{ Q_{\alpha} , Q_{\beta} \} 
\ &= \ 
2 \delta_{\alpha \beta} H 
+ \frac{\mu}{3} \big( \gamma^{\wt{I} \wt{J}} \gamma^{123}
\big)_{\alpha \beta} \, \Sigma^{\wt{I} \wt{J}} 
- \frac{\mu}{3} \big( \gamma^{I'J'} \gamma^{123}
\big)_{\alpha \beta} \, \Sigma^{I'J'} 
\; , \\
\{ Q_{\alpha} , q_{\beta} \}
\ &= \ 
- \sqrt{\frac{2 \mu}{3}} \Big(
\Big\{ \half \big( 1 - i \gamma^{123} \big) \gamma^{\wt{I}}
\Big\}_{\alpha \beta} \, a^{\wt{I}}
- \Big\{ \half \big( 1 + i \gamma^{123} \big) \gamma^{\wt{I}}
\Big\}_{\alpha \beta} \, a^{\dagger \wt{I}}
\Big)
\\
\ & \ \ \ \ 
+ \sqrt{\frac{\mu}{3}} \Big(
\Big\{ \half \big( 1 - i \gamma^{123} \big) \gamma^{I'}
\Big\}_{\alpha \beta} \, a^{\dagger I'}
- \Big\{ \half \big( 1 + i \gamma^{123} \big) \gamma^{I'}
\Big\}_{\alpha \beta} \, a^{\dagger I'}
\Big)
\; , \\
\{ q_{\alpha} , q_{\beta} \} \ &= \ \delta_{\alpha \beta} P^+
\; .
\end{split} \label{algebra-Qq}
\end{align}
Unlike the flat spacetime case,
the commutation relations between the Hamiltonian and supercharges do
not vanish:
\begin{align}
[ H , Q_{\alpha} ] \ &= \ \frac{\mu}{12} \big( i \gamma^{123} Q \big)_{\alpha}
\;, \ls
[ H , q_{\alpha} ] \ = \ - \frac{\mu}{4} \big( i \gamma^{123} q \big)_{\alpha}
\; . \label{com-HQ}
\end{align}
Thus different members of a multiplet of supersymmetric states
generated by acting with supercharges will have different energies,
although the energy differences will still be exactly determined by
the supersymmetry algebra (\ref{com-HQ}).


\section{Spectrum of the Ground State Supermultiplet}

In this section we discuss a supermultiplet generated by the
kinematical supercharges, which is the $U(1)$ part of the theory 
including the ground state.
We would like to compare the supermultiplet of the $U(1)$ free sector 
in the Matrix theory on the plane-wave background 
with the massless spectrum of eleven-dimensional
linearized supergravity on the plane-wave background \cite{KY03} which will be
discussed in chapter \ref{MAIN}.
For later convenience,
we express the $SO(9)$ Majorana spinor supercharge $q$ 
in terms of the $SU(4) \times SU(2)$ representation
(for the decomposition rule, see appendix \ref{4-2-repre}).
And we construct the states labeled by
the $SU(4)$ indices $i = 1,2,\cdots,4$ and the $SU(2)$ indices $\alpha = 1,2$.
Under the decomposition rules the supercharges are represented as follows:
\begin{align*}
\begin{split}
Q_{i \alpha} \ &= \ 
\sqrt{R} \, \Tr \Big\{
- \Big( P^{\wt{I}} + \frac{i \, \mu}{3 R} X^{\wt{I}} \Big) 
(\sigma^{\wt{I}})_{\alpha}{}^{\beta} \, \psi_{i \beta}
+ \Big( P^{I'} - \frac{i \, \mu}{6 R} X^{I'} \Big)
({\sf g}^{I'})_{ij} \, \eps_{\alpha \beta} \, \psi^{\dagger j \beta}
\\
\ & \LS \ls
+ \half [ X^{\wt{I}} , X^{\wt{J}} ] \eps^{\wt{I} \wt{J} \wt{K}} \,
(\sigma^{\wt{K}})_{\alpha}{}^{\beta} \, \psi_{i \beta}
- \frac{i}{2} [ X^{\wt{I}} , X^{\wt{J}} ] ({\sf g}^{I'J'})_i{}^j \,
\psi_{j \alpha}
\\
\ & \LS \ls
+ i [ X^{\wt{I}} , X^{J'} ] (\sigma^{\wt{I}})_{\alpha}{}^{\beta}
({\sf g}^{I'})_{ij} \, \eps_{\beta \gamma} \, \psi^{\dagger j \gamma}
\Big\}
\; , \\
q_{i \alpha} \ &= \ \frac{1}{\sqrt{R}} \Tr (\psi_{i \alpha})
\; .
\end{split}
\end{align*}
The algebras (\ref{algebra-Qq}) and (\ref{com-HQ}) 
are also rewritten as 
\bsubeq
\begin{gather}
\{ Q^{\dagger i \alpha} \, , \, Q_{j \beta} \}
\ = \ 
2 \delta_j^i \, \delta_{\beta}^{\alpha} \, H
+ \frac{\mu}{3} \eps^{\wt{I} \wt{J} \wt{K}} \,
(\sigma^{\wt{K}})_{\beta}{}^{\alpha} \, \delta_j^i 
\, \Sigma^{\wt{I} \wt{J}}
+ \frac{i \, \mu}{6} \delta_{\beta}^{\alpha} \, ({\sf g}^{I'J'})_j{}^i
\, \Sigma^{I'J'}
\; , \\
\{ q_{i \alpha} \, , \, Q_{j \beta} \} 
\ = \ 
- i \sqrt{\frac{\mu}{3}} \, ({\sf g}^{I'})_{ij} \, \eps_{\alpha \beta} \,
a^{I'}
\; , \ls
\{ q^{\dagger i \alpha} \, , \, Q_{j \beta} \} 
\ = \ 
- i \sqrt{\frac{2 \mu}{3}} \, (\sigma^{\wt{I}})_{\beta}{}^{\alpha} \, 
\delta_j^i \, a^{\wt{I} \dagger}
\; , \\
\{ q^{\dagger i \alpha} \, , \, q_{j \beta} \} 
\ = \ 
\delta_{\beta}^{\alpha} \, \delta_j^i \, P^+
\; , \\
[ H \, , \, Q_{i \alpha} ] \ = \ 
\frac{\mu}{12} Q_{i \alpha} 
\; , \ls
[ H \, , \, q_{i \alpha} ] \ = \ 
- \frac{\mu}{4} q_{i \alpha} 
\; . \label{com-HQ-2}
\end{gather}
\esubeq

Here we define the ground state $\ket{\Lambda}$ annihilated by
supercharges of the kinematical supersymmetry with arbitrary indices $i$
and $\alpha$:
\begin{align*}
q_{i \alpha} \ket{\Lambda} \ &= \ 0
\ls \ \ \mbox{for all \ $i,\alpha$} \; . 
\end{align*}
Starting from this ground state we construct the bosonic and fermionic
states generated by the kinematical supercharges $q^{\dagger i
  \alpha}$.
These states are also eigenstates of the Hamiltonian because of the
commutation relation $[H, q^{\dagger i \alpha}] = \frac{\mu}{4}
q^{\dagger i \alpha}$.
Since it is somewhat difficult to display the supersymmetric
states in terms of the supercharges themselves,
we introduce the Young Tableaux
\begin{align*}
q^{\dagger i \alpha} \ \sim \ \big( \, \Yng(1) , \Yng(1) \, \big)
\; .
\end{align*}
The first and second boxes $\Yng(1)$ in the right hand side indicate
the Young Tableaux of $SU(4)$ and $SU(2)$ fundamental representations,
respectively. 
Since we define the ground state $\ket{\Lambda}$ as a singlet with
respect to the action on the supercharge $q_{i \alpha}$, 
we label this state as
\begin{align}
\ket{\Lambda} \ &= \ \Ket{1 \; , \; 1}
\; .
\end{align}
We find that the energy of this state is zero by using the commutation
relation (\ref{com-HQ-2})\footnote{Notice that the parameter $\mu$ is
  rescaled in (\ref{rescale-SM-to-M}). 
But since the ``time'' variable $\tau$ is also rescaled, the Hamiltonian 
(\ref{BMN-H}) is defined the rescaled-time-evolution operator. Thus we
can obtain the energy eigenvalues of the states with correct mass dimensions.}.
The ``first floor'' is generated by acting the kinematical supercharge
$q^{\dagger i \alpha}$:
\begin{align}
\big( \, \Yng(1) , \Yng(1) \, \big) \otimes \Ket{1 \; , \; 1} \ &= \ 
\Ket{ \Yng(1) \; , \; \Yng(1) } \; .
\end{align}
The energy of the first floor is evaluated to $\mu/4$.
The ``second floor'' is also generated by the supercharge acting on
the first floor:
\begin{align}
\big( \, \Yng(1) , \Yng(1) \, \big) \otimes \Ket{\Yng(1) \; , \;
  \Yng(1)}
\ &= \ 
\tc{mygray}{
\Ket{\Yng(1,1) \; , \; \Yng(1,1)}
\oplus }
\Ket{\Yng(1,1) \; , \; \Yng(2)}
\oplus \Ket{\Yng(2) \; , \; \Yng(1,1)}
\tc{mygray}{
\oplus \Ket{\Yng(2) \; , \; \Yng(2)}
}
\; .
\end{align}
Notice that the generators $q^{\dagger i \alpha}$ is a fermionic
charge.
Thus the states symmetric with respect to the supercharges, are
forbidden as a member of the supermultiplet and these terms are
written by gray color.
In the same way we obtain the ``third floor'' as 
\begin{align}
&\big( \, \Yng(1) , \Yng(1) \, \big) \otimes 
\Big\{ \Ket{\Yng(1,1) \; , \; \Yng(2)} \oplus \Ket{\Yng(2) \; , \;
  \Yng(1,1)} \Big\}
\nn \\
& \LS \ = \
\tc{mygray}{
\Ket{\Yng(1,1,1) \; , \; \Yng(2,1)}
\oplus}
\Ket{\Yng(1,1,1) \; , \; \Yng(3)}
\oplus \Ket{\Yng(2,1) \; , \; \Yng(2,1)}
\tc{mygray}{
\oplus \Ket{\Yng(2,1) \; , \; \Yng(3)}
}
\oplus \Ket{\Yng(2,1) \; , \; \Yng(2,1)}
\tc{mygray}{
\oplus \Ket{\Yng(3) \; , \; \Yng(2,1)}
}
\; .
\end{align}
Here the two $\Ket{\Yng(2,1) \; , \; \Yng(2,1)}$ states are generated
from different states in the second floor.
In this case these states are linearly combined and 
only the antisymmetrized combination is chosen as a member of
supermultiplet (because of the fermionic generators).

The states in the higher ``floors'' are also described in terms of the
Young Tableaux. 
Since we generate the states by using fermionic supercharges
$q^{\dagger i \alpha}$,
the highest state is generated 
when we act eight supercharges on the ground state and 
the process will stop. 
The ninth supercharge annihilate the highest state.
Here we continue to generate the other states:

\noi
\ul{Fourth floor}:
\begin{align}
\begin{split}
&\big( \, \Yng(1) , \Yng(1) \, \big) \otimes 
\Big\{ \Ket{\Yng(1,1,1) \; , \; \Yng(3)} \oplus \Ket{\Yng(2,1) \; , \;
\Yng(2,1)} \Big\} 
\\
& \LS \ = \ 
\tc{mygray}{
\Ket{\Yng(1,1,1,1) \; , \; \Yng(3,1)}
\oplus}
\Ket{\Yng(1,1,1,1) \; , \; \Yng(4)}
\oplus \Ket{\Yng(2,1,1) \; , \; \Yng(3,1)}
\tc{mygray}{
\oplus \Ket{\Yng(2,1,1) \; , \; \Yng(4)}
}
\\
& \LS \ls \
\tc{mygray}{
\oplus \Ket{\Yng(2,1,1) \; , \; \Yng(2,2)}
}
\oplus \Ket{\Yng(2,1,1) \; , \; \Yng(3,1)}
\oplus \Ket{\Yng(2,2) \; , \; \Yng(2,2)}
\tc{mygray}{
\oplus \Ket{\Yng(2,2) \; , \; \Yng(3,1)}
}
\\
& \LS \ls \
\tc{mygray}{
\oplus \Ket{\Yng(3,1) \; , \; \Yng(2,2)}
\oplus \Ket{\Yng(3,1) \; , \; \Yng(3,1)}
}
\end{split}
\end{align}
\ul{Fifth floor}:
\begin{align}
\begin{split}
&\big( \, \Yng(1) , \Yng(1) \, \big) \otimes 
\Big\{ \Ket{\Yng(1,1,1,1) \; , \; \Yng(4)} 
\oplus \Ket{\Yng(2,1,1) \; , \; \Yng(3,1)} 
\oplus \Ket{\Yng(2,2) \; , \; \Yng(2,2)} 
\Big\} 
\\
& \LS \ = \ 
\Ket{\Yng(2,1,1,1) \; , \; \Yng(4,1)}
\tc{mygray}{
\oplus \Ket{\Yng(2,1,1,1) \; , \; \Yng(5)} 
\oplus \Ket{\Yng(2,1,1,1) \; , \; \Yng(3,2)} 
}
\oplus \Ket{\Yng(2,1,1,1) \; , \; \Yng(4,1)} 
\\
& \LS \ls \
\oplus \Ket{\Yng(2,2,1) \; , \; \Yng(3,2)}
\tc{mygray}{
\oplus \Ket{\Yng(2,2,1) \; , \; \Yng(4,1)}
\oplus \Ket{\Yng(3,1,1) \; , \; \Yng(3,2)}
\oplus \Ket{\Yng(3,1,1) \; , \; \Yng(4,1)}
}
\\
& \LS \ls \
\oplus \Ket{\Yng(2,2,1) \; , \; \Yng(3,2)}
\tc{mygray}{
\oplus \Ket{\Yng(3,2) \; , \; \Yng(3,2)}
}
\end{split}
\end{align}
\ul{Sixth floor}:
\begin{align}
\begin{split}
&\big( \, \Yng(1) , \Yng(1) \, \big) \otimes 
\Big\{ \Ket{\Yng(2,1,1,1) \; , \; \Yng(4,1)}
\oplus \Ket{\Yng(2,2,1) \; , \; \Yng(3,2)}
\Big\}
\\
& \LS \ = \ 
\tc{mygray}{
\Ket{\Yng(2,2,1,1) \; , \; \Yng(4,2)}
\oplus \Ket{\Yng(2,2,1,1) \; , \; \Yng(5,1)}
\oplus \Ket{\Yng(3,1,1,1) \; , \; \Yng(4,2)}
\oplus \Ket{\Yng(3,1,1,1) \; , \; \Yng(5,1)}
}
\\
& \LS \ls \
\tc{mygray}{
\oplus \Ket{\Yng(2,2,1,1) \; , \; \Yng(3,3)}
\oplus} 
\Ket{\Yng(2,2,1,1) \; , \; \Yng(4,2)}
\oplus \Ket{\Yng(2,2,2) \; , \; \Yng(3,3)}
\tc{mygray}{
\oplus \Ket{\Yng(2,2,2) \; , \; \Yng(4,2)}
} 
\\
& \LS \ls \
\tc{mygray}{
\oplus \Ket{\Yng(3,2,1) \; , \; \Yng(3,3)}
\oplus \Ket{\Yng(3,2,1) \; , \; \Yng(4,2)}
}
\end{split}
\end{align}
\ul{Seventh floor}:
\begin{align}
\begin{split}
&\big( \, \Yng(1) , \Yng(1) \, \big) \otimes 
\Big\{ \Ket{\Yng(2,2,1,1) \; , \; \Yng(4,2)}
\oplus \Ket{\Yng(2,2,2) \; , \; \Yng(3,3)}
\Big\}
\\
& \LS \ = \ 
\tc{mygray}{
\Ket{\Yng(2,2,2,1) \; , \; \Yng(4,3)}
\oplus \Ket{\Yng(2,2,2,1) \; , \; \Yng(5,2)}
\oplus \Ket{\Yng(3,2,1,1) \; , \; \Yng(4,3)}
\oplus \Ket{\Yng(3,2,1,1) \; , \; \Yng(5,2)}
}
\\
& \LS \ls \ 
\tc{mygray}{\oplus} \Ket{\Yng(2,2,2,1) \; , \; \Yng(4,3)}
\tc{mygray}{
\oplus \Ket{\Yng(3,2,2) \; , \; \Yng(4,3)}
}
\end{split}
\end{align}
\ul{Eighth floor}:
\begin{align}
\begin{split}
&\big( \, \Yng(1) , \Yng(1) \, \big) \otimes 
\Ket{\Yng(2,2,2,1) \; , \; \Yng(4,3)}
\\
& \LS \ = \  
\Ket{\Yng(2,2,2,2) \; , \; \Yng(4,4)}
\tc{mygray}{
\oplus \Ket{\Yng(2,2,2,2) \; , \; \Yng(5,3)}
\oplus \Ket{\Yng(3,2,2,1) \; , \; \Yng(4,4)}
\oplus \Ket{\Yng(3,2,2,1) \; , \; \Yng(5,3)}
}
\end{split}
\end{align}
The state in the ``eighth floor'' is the highest state which is
annihilated by ninth supercharge.
Thus we find that the supermultiplet contains the above states from
the ground state $\ket{\Lambda}$ to the highest state
$\Ket{\Yng(2,2,2,2)\; , \; \Yng(4,4)}$.
The energy eigenvalues of the above states are also obtained by the
commutation relations (\ref{com-HQ}).
We summarize the members of supermultiplet in Table \ref{KP0207-1:table1}.
\begin{table}[h]
\begin{align*}
\begin{array}{c||@{\ls}c@{\ls}|c} \hline
\mbox{$N$-th Floor} & \mbox{$SU(4) \times SU(2)$ Representations} 
& \mbox{Energy Eigenvalues} \\ \hline \hline
8 & ({\bf 1}, {\bf 1}) & 2 \mu  \\
7 & (\ol{\bf 4}, {\bf 2}) & 7 \mu/4 \\
6 & (\ol{\bf 6}, {\bf 3}) \LS (\ol{\bf 10}, {\bf 1}) & 3 \mu / 2 \\ 
5 & ({\bf 4}, {\bf 4}) \LS (\ol{\bf 20}, {\bf 2}) & 5\mu/4 \\ 
4 & ({\bf 1}, {\bf 5}) \LS ({\bf 15} ,{\bf 3}) \LS({\bf 20'}, {\bf 1}) 
  &  \mu \\ 
3 & (\ol{\bf 4}, {\bf 4}) \LS ({\bf 20}, {\bf 2}) & 3\mu/4 \\ 
2 & ({\bf 6}, {\bf 3}) \LS ({\bf 10}, {\bf 1}) & \mu/2 \\
1 & ({\bf 4}, {\bf 2}) & \mu/4 \\
\mbox{ground state} & ({\bf 1}, {\bf 1}) & 0 \\ \hline
\end{array}
\end{align*}
\caption{\sl The simplest multiplet grouped into irreducible
  representations of $SU(4) \times SU(2)$ on each Floor of equal energies.}
\label{KP0207-1:table1}
\end{table}

If the Matrix theory conjecture \cite{BFSS96} is correct 
(and if M-theory conjecture \cite{W95} is also correct) 
even on curved spacetime background,
the resulting spectrum should correspond to the massless spectrum of
eleven-dimensional supergravity,
because the Matrix theory is proposed as a candidate of 
the well-defined description of M-theory, whose low energy effective
theory is eleven-dimensional supergravity. 
Thus, in the next chapter, 
we will construct the supermultiplet of the ground state
in eleven-dimensional supergravity and compare it to the result
obtained here.

Note that we have considered
  only this $U(1)$ free sector of the Matrix theory.
The remaining $SU(N)$ sector, 
which describes the interactions among $N$ D0-branes from the
viewpoint of type IIA string theory, would also describe 
the M-branes dynamics from the M-theory point of view \cite{BMN02,
  DSvR0205, MSvR0211}. 
It is quite interesting to investigate these dynamics in the
  supergravity side.
But since this topic is beyond the scope of this doctoral thesis,
we would like to consider this in the future.


\chapter{Eleven-dimensional Supergravity Revisited} \label{MAIN}


\newpage

\setcounter{section}{0}
\renewcommand{\thesection}{\thechapter.\arabic{section}}

\indent
In this chapter,
we discuss the eleven-dimensional supergravity on the plane-wave
background. 
Eleven-dimensional supergravity Lagrangian with full interaction terms
was discovered
by Cremmer, Julia and Scherk \cite{CJS78}.
But, for simplicity and later convenience,
we describe the Lagrangian and classical field equations
without the terms derived from spacetime torsion.
With this formulation
we make all the bosonic/fermionic fields fluctuate 
around classical field equations
and construct linearized field equations.
{}From the linearized field equations   
we study the zero point energy spectrum 
on the plane-wave background and compare 
with the zero-mode spectrum of the Matrix theory on the plane-wave.


\section{Supergravity Lagrangian}

As mentioned in chapter \ref{intro}, the eleven-dimensional supergravity
is one of the simplest model in supersymmetric field theories because
there are a few number of bosonic and fermionic fields 
\begin{align*}
e_M{}^{\r{A}} \ &: \ \ \mbox{vielbein} \; , \ls 
E_{\r{A}}{}^M \ : \ \ \mbox{inverse vielbein}  \\
\Psi_M \ &: \ \ \mbox{gravitino (vectorial Majorana spinor)}  \\
C_{MNP} \ &: \ \ \mbox{three-form gauge field} \\
\omega_M{}^{\r{AB}} \ &: \ \ \mbox{spin connection}
\end{align*}
The number of on-shell degrees of freedom of the vielbein (graviton), gravitino
and three-form gauge field are 44, 128 and 84, respectively.
Notice that 
the spin connection is independent of the vielbein 
in the first order formalism, 
but it is expressed by the vielbein in the second order formalism.
By using these fields we describe the on-shell Lagrangian (up to
torsion) \cite{CJS78}
\begin{align}
S \ &= \ \frac{1}{2 \kappa^2} \int \! \d^{11} x \, {\cal L} \; , 
\nn \\
\begin{split}
{\cal L} \ &= \ 
e \, {\cal R} 
- \half e \, \ol{\Psi}{}_M \hG{}^{MNP} D_N (\omega) \Psi_P 
- \frac{1}{48} e \, F_{MNPQ} \, F^{MNPQ} \\
\ & \ \ \ \
- \frac{1}{192} e \, \ol{\Psi}{}_M \, \tG{}^{MNPQRS} \, \Psi_N
F_{PQRS} 
- \frac{1}{(144)^2} \, \ve^{MNPQRSUVWXY} \, 
F_{MNPQ} \, F_{RSUV} \, C_{WXY}
\; , 
\end{split} \label{Lagrangian}
\end{align}
where $e = \det (e_M{}^{\r{A}}) = \sqrt{-\det g_{MN}}$
and $\hG_M$ is the gamma matrix defined in appendix \ref{SO11};
the eleven-dimensional gravitational constant is $\kappa$;
the rank six matrix $\tG^{MNPQRS}$ is defined by
\begin{align*}
\tG{}^{MNPQRS} \ &= \ 
\hG{}^{MNPQRS} + 12 g^{M[P} \hG{}^{QR} g^{S]N} \; .
\end{align*}
The Lagrangian (\ref{Lagrangian}) contains two types of covariant
derivatives explicitly or implicitly. 
One is the covariant derivative for general
coordinate transformations denoted by $\nabla_M$, and the other is
the covariant derivative for local Lorentz transformations denoted by $D_M$.  
They are defined by the affine connection $\Gamma^R_{MN}$ and the
spin connection $\omega_M{}^{\r{A} \r{B}}$, for instance, as
\begin{align*}
\nabla_M A_N \ &= \ \del_M A_N - \Gamma^P_{NM} A_P \; , \ls
D_N \Psi_P \ = \ 
\del_N \Psi_P - \frac{i}{2} \omega_N{}^{\r{A} \r{B}} 
\Sigma_{\r{A} \r{B}} \Psi_P
\; . 
\end{align*}
Note that $\Sigma_{\r{A} \r{B}}$ are the generators of the Lorentz algebra
in the tangent space.
The covariant derivative $\nabla_M$ does not appear in the Lagrangian
explicitly but the Einstein-Hilbert term (the scalar curvature) is the
contraction of Riemann tensor, which is defined by the commutator of
the covariant derivative $\nabla_M$. 
The precise definitions are described in appendix \ref{connections}.
We mention that the Lagrangian (\ref{Lagrangian}) is defined up to
torsion contributions 
because the terms derived from the torsion do not contribute to the
analysis of the linearized supergravity in this thesis.
We should consider such a contribution to the Lagrangian
in order to discuss full nonlinear supergravity.
(We will prepare the full supergravity Lagrangian in appendix \ref{11SUGRA}.)
In addition, this Lagrangian is invariant under the local
supersymmetry transformation with fermionic parameter $\ve (x)$: 
\begin{gather*}
\delta e_M{}^{\r{A}} \ = \ \half \ol{\ve} \hG^{\r{A}} \Psi_M
\; , \ls 
\delta C_{MNP} \ = \ - \frac{3}{2} \ol{\ve} \hG_{[MN} \Psi_{P]}
\; , \\
\delta \Psi_M \ = \ 
2 D_M \ve + 2 F_{NPQR} T_M{}^{NPQR} \ve 
\; , \ls
T_M{}^{NPQR} \ = \ \frac{1}{288} 
\Big( \hG_M{}^{NPQR} - 8 \delta_M^{[N} \hG_{\phantom{M}}^{PQR]} \Big) 
\; ,
\end{gather*}
where we also neglect the higher order contribution with respect to
torsion.


\subsection*{Classical Field Equations}

Varying $g_{MN}$, $\Psi_M$ and $C_{MNP}$,
we obtain classical field equations 
from the Lagrangian (\ref{Lagrangian}):
\bsubeq \label{eom-class}
\begin{align}
0 \ &= \half g_{MN} {\cal R} - {\cal R}_{MN} 
- \frac{1}{96} g_{MN} F_{PQRS} F^{PQRS}
+ \frac{1}{12} F_{MPQR} F_{N}{}^{PQR} 
\; , \label{eom-g} \\
0 \ &= \ 
\hG{}^{MNP} D_N \Psi_P 
+ \frac{1}{96} \tG{}^{MNPQRS} \Psi_N F_{PQRS} \; , 
\label{eom-psi} \\
0 \ &= \ 
\nabla^Q \big\{ e F_{QMNP} \big\}
- \frac{18}{(144)^2} \, g_{MZ} \, g_{NK} \, g_{PL} \, \ve^{ZKLQRSUVWXY} 
F_{QRSU} F_{VWXY}
\; . \label{eom-C} 
\end{align}
\esubeq
Note that we neglect the gravitino quadratic term $\ol{\Psi}_M
\tG^{MNPQRS} \Psi_N$ which does not contribute to the linearized field
equations for fluctuation fields which we will calculate in later discussions.
{}From the classical field equation for the metric $g_{MN}$, 
we find that some relations among curvatures and four-form flux $F_{MNPQ}$.  
Contracting curved spacetime indices in (\ref{eom-g}),
we obtain the equations
\bsubeq \label{eom-class-g-2}
\begin{gather}
{\cal R}_{MN} \ = \ 
- \frac{1}{144} g_{MN} \, F_{PQRS} \, F^{PQRS}
+ \frac{1}{12} F_{MPQR} \, F_N{}^{PQR}
\; , \label{eom-class-g-2-2} \\
{\cal R} \ = \ 
\frac{1}{144} F_{PQRS} \, F^{PQRS} 
\; . \label{eom-class-g-2-1} 
\end{gather}
\esubeq
Thus we find that if there are some non-vanishing constant four-form flux in
eleven-dimensional spacetime, 
the spacetime have nontrivial curvature and will be compactified. 
This mechanism derived from the existence of constant flux 
is called ``spontaneous compactification'', and the 
assumption of a constant flux is called 
the ``Freund-Rubin Ansatz'' \cite{FR80, DP82}.

{}From now on we consider field equations for fluctuation fields
in terms of equations (\ref{eom-class-g-2-2}), (\ref{eom-psi}) and
(\ref{eom-C}).


\subsection*{Fluctuations}

Let us consider equations of motion of fluctuation fields in
eleven-dimensional spacetime.
We make fields fluctuate around the eleven-dimensional spacetime
background: 
\begin{gather}
g_{MN} \ = \ \Circ{g}{}_{MN} + h_{MN} \; , \ls
g^{MN} \ = \ \Circ{g}{}^{MN} + \wt{h}{}^{MN} \;, \nn \\ 
\Psi_M \ = \ 0 + \psi_M \; , \label{fluc} \\
F_{MNPQ} \ = \ \Circ{F}{}_{MNPQ} + {\cal F}{}_{MNPQ} \; , \ls
{\cal F}{}_{MNPQ} \ = \ 4 \del_{[M} {\cal C}{}_{NPQ]} \; . \nn
\end{gather}
In order to preserve the Lorentz invariance in the tangent space,
we assume that the gravitino background field vanishes. 
The fluctuation of the inverse metric $\wt{h}^{MN}$ is represented 
by $\wt{h}{}^{MN} = - \Circ{g}{}^{MP} \, \Circ{g}{}^{NQ} \, h_{PQ} = - h^{MN}$.
Under the above expansions, 
we calculate fluctuations of the determinant of vielbein $e$, 
affine connection $\Gamma^P_{MN}$, Ricci tensor ${\cal R}_{MN}$ and
scalar curvature ${\cal R}$ as follows\footnote{From now on we omit
  the circle in (\ref{fluc}), which is the symbol of classical background.}:
\begin{align*}
\delta e \ &= \ 
\half e \, g^{MN} \, h_{MN} 
\; , \\
\delta \Gamma^{M}_{NP} \ &= \ 
\half g^{MR} 
\big( \nabla_N h_{PR} + \nabla_P h_{NR} - \nabla_R h_{NP} \big) \; , \\
\delta {\cal R}_{MN} 
\ &= \
- \half \Big\{ 
\nabla_N \nabla_M h_P{}^P 
- \nabla_N \nabla^P h_{MP} - \nabla_M \nabla^P h_{NP}
\Big\} 
+ \half \wh{\Delta} h_{MN}
\; , \\
\delta {\cal R} \ &= \ 
- h_{PQ} \, g^{MP} \, g^{NQ} \, {\cal R}_{MN} 
+ \Circ{g}{}^{MN} \, \delta {\cal R}_{MN} \; .
\end{align*}
Note that the above covariant derivative $\nabla_M$ is written in
terms of the classical affine connection $\Gamma^P_{MN} =
\half {g}{}^{PR} (\del_M {g}{}_{NR} + \del_N {g}{}_{MR}
- \del_R {g}{}_{MN})$ because the plane-wave background, on which we
analyze the physical modes, is torsion free;
the operator $\wh{\Delta}$ is called the Lichnerowicz operator
which acts on the rank two symmetric tensor $h_{MN}$ below \cite{DP82}: 
\begin{align*}
\wh{\Delta} h_{MN} \ &= \ 
- \nabla_P \nabla^P h_{MN}  
- 2 R_{MPNQ} \, h^{PQ} + {\cal R}_M{}^P \, h_{PN} + {\cal R}_N{}^P \,
h_{PM}
\; .
\end{align*}
In terms of these expressions 
we derive the following linearized field equations for fluctuation fields 
from the classical field equations (\ref{eom-class}):
\bsubeq \label{general-fluctuate}
\begin{align}
\delta {\cal R}_{MN} \ &= \
- \half \Big\{ 
\nabla_N \nabla_M h_P{}^P - \nabla_N \nabla^P h_{MP}
- \nabla_M \nabla^P h_{NP}
\Big\}
+ \half \wh{\Delta} h_{MN}
\nn \\
\ &= \ 
- \frac{1}{144} h_{MN} \, F_{PQRS} \, F^{PQRS}
- \frac{1}{72} g_{MN} \, F^{PQRS} \, {\cal F}_{PQRS}
+ \frac{1}{36} h^{PU} \, g_{MN} \, F_{PQRS} \, F_U{}^{QRS}
\nn \\
\ & \ \ \ \ 
+ \frac{1}{12} \Big( 
{\cal F}_{MPQR} \, F_N{}^{PQR} + {\cal F}_{NPQR} \, F_M{}^{PQR}  
\Big)
- \frac{1}{4} h^{PU} \, F_{MPQR} \, F_{NU}{}^{QR}
\; , 
\label{fluc-g'}
\end{align}
\begin{align}
0 \ &= \ \hG{}^{MNP} \, D_N \psi_P 
+ \frac{1}{96} \tG{}^{MNPQRS} \, F_{PQRS} \, \psi_N 
\; , \label{fluc-psi}
\end{align}
\begin{align}
0 \ &= \  
e \, \Big\{ \half h_U{}^U \, g^{QR} - h^{QR} \Big\} 
\tc{mygray}{
\nabla_R F_{QMNP}
}
+ e \, \nabla^Q {\cal F}_{QMNP}
\nn \\
\ & \ \ \ \ 
- e \, \Big\{
F_{SMNP} \Big( \nabla^Q h_Q{}^S - \half \del^S h_{Q}{}^Q \Big)
+ F_{QSNP} \nabla^Q h_M{}^S 
+ F_{QMSP} \nabla^Q h_N{}^S 
+ F_{QMNS} \nabla^Q h_P{}^S 
\Big\}
\nn \\
\ & \ \ \ \ 
- \frac{1}{576} \, \ve^{ZKLQRSUVWXY} 
{\cal F}_{QRSU} \, g_{MZ} \, g_{NK} \, g_{PL} \, F_{VWXY}
\nn \\
\ & \ \ \ \ 
\tc{mygray}{
- \frac{18}{(144)^2} \, \ve^{ZKLQRSUVWXY}
\big( 
h_{MZ} \, g_{NK} \, g_{PL}
+ h_{NK} \, g_{MZ} \, g_{PL}
+ h_{PL} \, g_{MZ} \, g_{NK}
\big) F_{QRSU} \, F_{VWXY}
}
\; . 
\label{fluc-C}
\end{align}
\esubeq
Notice that the gray-colored terms do not contribute to the equations under
the Freund-Rubin ansatz which we will assume on the plane-wave
background in the next section. 


\section{Plane-wave Background}

In the previous section 
we defined the supergravity Lagrangian and
derived the classical field equations from it.
Furthermore we made fields fluctuate around general classical
backgrounds.
Since the main theme of this section is to investigate the spectrum of
fluctuation fields around the plane-wave background,
we introduce the geometrical variables on this specific spacetime
\begin{align}
\begin{split}
\d s^2 \ &= \ - 2 \d x^+ \d x^- + G_{++} \cdot (\d x^+)^2 + \sum_{I=1}^9 (
\d x^{I})^2 \; , 
\\
G_{++} \ &= \ - \Big[ \Big( \frac{\mu}{3} \Big)^2 \sum_{\wt{I}=1}^3
  (x^{\wt{I}})^2 + \Big( \frac{\mu}{6} \Big)^2 \sum_{I'=4}^9
  (x^{I'})^2 \Big] \; . 
\end{split} \label{KG-BG1}
\end{align}
Under this background 
we can set the constant four-form flux as the Freund-Rubin ansatz
\begin{align*}
F_{123+} \ &= \ \mu \ \neq \ 0 \; . 
\end{align*}
In our consideration, no contributions from torsion are included, i.e.,
the affine connection is symmetric under lower indices: $\Gamma^P_{MN} =
\Gamma^P_{NM}$. 
The components of vielbein, affine connection, spin connection and
their curvature tensors are described below (see also appendix
\ref{PL}): 
\begin{gather}
e_+{}^{\r{+}} \ = \ e_-{}^{\r{-}} \ = \ 1
\; , \ls
e_+{}^{\r{-}} \ = \ - \half G_{++} 
\; , \nn \\
E_{\r{+}}{}^+ \ = \ E_{\r{-}}{}^- \ = \ 1
\; , \ls
E_{\r{+}}{}^- \ = \ \half G_{++} 
\; , \nn \\
\omega_+{}^{\r{I} \r{-}} 
\ = \ 
\half \del_I G_{++} \; , \label{KG-BG2}
\\
\Gamma^{I}_{++} \ = \ 
\Gamma^-_{+ I} \ = \ 
- \half \del_{I} G_{++}
\; , \nn \\
R^{I}{}_{+J+} \ = \ 
- \half \del_I \del_J G_{++}
\; ,
\ls
{\cal R}_{++} \ = \ 
\half \mu^2 
\;, \ls 
{\cal R} \ = \  0 
\; . \nn
\end{gather}
Note that this background is almost flat but non-trivial curvature
tensor which is proportional to the constant parameter $\mu$.
This constant comes from the non-vanishing constant flux $F_{123+}$.
Substituting (\ref{KG-BG2}) into the field equations for fluctuations
(\ref{general-fluctuate}), 
we will discuss the linearized supergravity and its spectrum on the
plane-wave background.


\section{Light-cone Hamiltonian on the Plane-wave}

Now let us discuss the Hamiltonian and its energy eigenvalue. 
We need to calculate and solve field equations for fluctuation modes 
around the plane-wave background in the next section. 
Then we will encounter Klein-Gordon type equations of motion and 
have to evaluate its energy spectrum.

We shall consider a Klein-Gordon type equation of motion 
for a field $\phi (x)$:
\begin{align}
\big( \Box + \alpha \, \mu \, i \del_- \big) \phi (x^+, x^-, x^I) \ &= \ 0 
\; , \label{Klein-Gordon}
\end{align}
where $\alpha$ is an arbitrary numerical constant and $x^+$ is an evolution
parameter. The d'Alembertian $\Box$ on the plane-wave background is given by 
\begin{align*}
\Box \ &= \ - \nabla^P \nabla_P \ = \ - \del^P \del_P
\nn \\
\ &= \ 
- \frac{1}{\sqrt{-g}} \del_M \big( \sqrt{-g} g^{MN} \del_N \big)
\ = \ 
2 \del_+ \del_- + G_{++} \cdot (\del_-)^2 - (\del_K)^2 
\; .
\end{align*}
The above Klein-Gordon type field equation 
will appear later as equations of motion of fluctuation fields. 
Fourier transformed expression of $\phi (x)$
\begin{align*}
\phi (x^+, x^-, x^I) \ &= \ 
\int \! \frac{\d p_- \d^9 p_I}{\sqrt{(2 \pi)^{10}}}
\, \e^{i (p_- x^- + p_I x^I)} \, \wt{\phi} (x^+, p_-, p_I)
\end{align*}
leads to the following expression:
\begin{align*}
0 \ &= \ 2 p_- \, i \del_+ - \wt{G}{}_{++} \cdot (p_-)^2 + (p_I)^2 
- \alpha \, \mu \, p_-
\; ,
\end{align*}
where $\wt{G}{}_{++}$ is defined by
\begin{align*}
\wt{G}{}_{++} \ &\equiv \ 
\Big( \frac{\mu}{3} \Big)^2 \sum_{\wt{I}=1}^3 (\del_{p_{\wt{I}}})^2
+ \Big( \frac{\mu}{6} \Big)^2 \sum_{I'=4}^9 (\del_{p_{I'}})^2
\; .
\end{align*}
By rewriting the above equation and defining the Hamiltonian $H = i \del_+$, 
we obtain the explicit expression for the Hamiltonian 
\begin{align*}
H \ &= \ \frac{1}{-2 p_-} \big\{
(p_I)^2 - \wt{G}{}_{++} \cdot (p_-)^2 - \alpha \, \mu \, p_- \big\}
\; .
\end{align*}
The energy spectrum of this Hamiltonian can be derived via the
standard technique of harmonic oscillators. 
Now we define ``creation/annihilation'' operators  
\begin{align*}
a^{\wt{I}} \ &\equiv \ 
\frac{1}{\sqrt{2 \wt{m}}} 
\big\{ p_{\wt{I}} + \wt{m} \del_{p_{\wt{I}}} \big\}
\; , &
\ol{a}{}^{\wt{I}} \ &\equiv \ 
\frac{1}{\sqrt{2 \wt{m}}} 
\big\{ p_{\wt{I}} - \wt{m} \del_{p_{\wt{I}}} \big\}
\; , &
\wt{m} \ &\equiv \ - \frac{1}{3} \mu \, p_- 
\; , \\
a^{I'} \ &\equiv \ 
\frac{1}{\sqrt{2 m'}} 
\big\{ p_{I'} + m' \del_{p_{I'}} \big\}
\; , &
\ol{a}{}^{I'} \ &\equiv \ 
\frac{1}{\sqrt{2 m'}} 
\big\{ p_{I'} - m' \del_{p_{I'}} \big\}
\; , &
m' \ &\equiv \ - \frac{1}{6} \mu \, p_- 
\; ,
\end{align*}
whose commutation relations are represented by 
\begin{align*}
[ a^{\wt{I}} , \ol{a}{}^{\wt{J}} ] \ &= \ \delta^{\wt{I} \wt{J}}
\; , \ls
[ a^{I'} , \ol{a}{}^{J'} ] \ = \ \delta^{I'J'}
\; , \ls 
[ a^{\wt{I}} , \ol{a}{}^{J'} ] \ = \ 
[ a^{I'} , \ol{a}{}^{\wt{J}} ] \ = \ 0 
\; .
\end{align*}
Thus we express the Hamiltonian in terms of the above oscillators:
\begin{align*}
H \ &= \ 
\frac{1}{3} \mu \sum_{\wt{I}} \ol{a}{}^{\wt{I}} a^{\wt{I}}
+ \frac{1}{6} \mu \sum_{I'} \ol{a}{}^{I'} a^{I'}
+ \frac{1}{2}\mu\left(2 + \alpha \right)
\; .
\end{align*}
Note that the last term implies the zero point energy $E_0$ 
of the system, which is represented by
\begin{align}
E_0 \ &= \ \half \mu \, {\cal E}{}_0 (\phi) \; , 
\ls
{\cal E}{}_0 (\phi) \ = \ 2 + \alpha 
\; . \label{zero-energy}
\end{align}
In the next section, we will use ${\cal E}{}_0$ 
to evaluate the energy of the zero-modes of fluctuation fields. 

After the above setup,
we will discuss the physical spectrum of fluctuation fields around the
plane-wave background.
First we will take the light-cone gauge for fluctuation fields and reduce
field equations of them.
After field re-definition we will discuss the zero-point energy and
the number of physical degrees of freedom.


\section{Field Equations for Fluctuations on the Plane-wave Background}

We discuss the spectrum of fluctuation fields on the plane-wave
background. 
In order to consider the spectrum of the physical fields 
we take the light-cone gauge fixing as follows:
\begin{align}
h_{-M} \ &= \ 0 \, \ls 
h^{+M} \ = \ 0 \, \ls
{\cal C}_{-MN} \ = \ 0 \, \ls
\psi_- \ = \ 0 \; .
\label{LC-GF}
\end{align}
We write all the field equations for fluctuation fields $h_{MN}$,
$\psi_M$ and ${\cal C}_{MNP}$ on the plane-wave background (\ref{KG-BG1}) and 
(\ref{KG-BG2}) under the light-cone gauge-fixing condition
(\ref{LC-GF}).
First, the following field equations are derived from (\ref{fluc-g'}):
\bsubeq \label{KG-fluc-g3}
\begin{align}
0 \ &= \ 
\half \Big\{ 
\nabla_+ \nabla_+ h_P{}^P 
- \nabla_+ \nabla^P h_{+P}
- \nabla_+ \nabla^P h_{+P}
- \Box \, h_{++}
\Big\}
- \Big( \frac{\mu}{3} \Big)^2 h_{\wt{K} \wt{K}}
- \Big( \frac{\mu}{6} \Big)^2 h_{L'L'}
\nn \\
\ & \ \ \ \ 
- \frac{1}{3} \mu \, G_{++} \, \del_- {\cal C}_{123}
- \mu \, {\cal F}_{+123}
- \frac{1}{2} \mu^2 \, h_{\wt{L}\wt{L}}
\; ,
\label{KG-fluc-g3++}
\\
0 \ &= \ 
\Big\{ 
\del_- \del_+ h_P{}^P 
- \del_- \del^P h_{+P}
\Big\}
- \frac{1}{3} \mu \, \del_- {\cal C}_{123}
\; ,
\label{KG-fluc-g3+-}
\\
0 \ &= \ 
\Big\{ 
\nabla_{\wt{I}} \nabla_+ h_P{}^P 
- \del_{\wt{I}} \del^P h_{+P}
- \del_+ \del^P h_{\wt{I}P}
- \Box \, h_{+\wt{I}}
\Big\}
+ \frac{1}{2} \mu \, \eps_{\wt{I} \wt{J} \wt{K}} \,
\del_- {\cal C}_{+\wt{J} \wt{K}} 
\; ,
\label{KG-fluc-g3+I}
\\
0 \ &= \ 
\Big\{ 
\nabla_{I'} \nabla_+ h_P{}^P 
- \del_{I'} \del^P h_{+P}
- \del_+ \del^P h_{I'P}
- \Box \, h_{+I'}
\Big\}
- \frac{1}{6} \mu \, \eps_{\wt{J}\wt{K}\wt{L}}
{\cal F}_{I'\wt{J}\wt{K}\wt{L}} 
\; ,
\label{KG-fluc-g3+I'}
\\
0 \ &= \ 
\del_- \del_- h_P{}^P 
\; ,
\label{KG-fluc-g3--}
\\
0 \ &= \ 
\del_{I} \del_- h_P{}^P 
- \del_- \del^P h_{IP}
\; ,
\label{KG-fluc-g3-I}
\\
0 \ &= \ 
\Big\{ 
\del_{\wt{J}} \del_{\wt{I}} h_P{}^P 
- \del_{\wt{J}} \del^P h_{\wt{I} P}
- \del_{\wt{I}} \del^P h_{\wt{J} P}
- \Box \, h_{\wt{I} \wt{J}}
\Big\}
+ \frac{4}{3} \mu \, \delta_{\wt{I} \wt{J}} \, \del_- {\cal C}_{123}
\; ,
\label{KG-fluc-g3IJ}
\\
0 \ &= \ 
\Big\{ 
\del_{J'} \del_{\wt{I}} h_P{}^P 
- \del_{J'} \del^P h_{\wt{I} P}
- \del_{\wt{I}} \del^P h_{J' P}
- \Box \, h_{\wt{I} J'}
\Big\}
+ \frac{1}{2} \mu \, \eps_{\wt{I} \wt{K} \wt{L}} \,
\del_- {\cal C}_{J' \wt{K} \wt{L}} 
\; ,
\label{KG-fluc-g3IJ'}
\\
0 \ &= \ 
\Big\{ 
\del_{J'} \del_{I'} h_P{}^P 
- \del_{J'} \del^P h_{I' P}
- \del_{I'} \del^P h_{J' P}
- \Box \, h_{I'J'}
\Big\}
- \frac{2}{3} \mu \, \delta_{I'J'} \, \del_- {\cal C}_{123}
\; .
\label{KG-fluc-g3I'J'}
\end{align}
\esubeq
The next four equations are the components of field equations (\ref{fluc-psi}):
\bsubeq \label{KG-fluc-psi3}
\begin{align}
0 \ &= \ 
\hG{}^{+NP} D_N \psi_P
\; ,
\label{KG-fluc-psi3+}
\\
0 \ &= \ 
\hG^{-NP} D_N \psi_P
+ \frac{1}{4} \mu \, \hG{}^{+-123I'} \, \psi_{I'}
+ \frac{1}{8} \mu \, 
\eps_{\wt{I} \wt{J} \wt{K}} \hG_{\wt{I} \wt{J}} \, \psi_{\wt{K}}
\; ,
\label{KG-fluc-psi3-}
\\
0 \ &= \ 
\hG^{\wt{I}NP} D_N \psi_P
- \frac{1}{4} \mu \, \hG^{+123} 
\big( \delta_{\wt{I} \wt{J}} - \hG_{\wt{I}} \hG_{\wt{J}} \big) \psi_{\wt{J}}
\; ,
\label{KG-fluc-psi3I}
\\
0 \ &= \ 
\hG{}^{I'NP} D_N \psi_P
+ \frac{1}{4} \mu \,
\hG^{+123} \big( \delta_{I'J'} - \hG_{I'} \hG_{J'} \big) \psi_{J'}
\; .
\label{KG-fluc-psi3I'}
\end{align}
\esubeq
Finally, we write the components of field equations (\ref{fluc-C})
under the light-cone gauge fixing:
\bsubeq \label{KG-fluc-C3}
\begin{align}
0 \ &= \  
\del_- \del^Q {\cal C}_{Q+I}
\; ,
\label{KG-fluc-C3+-I}
\\
0 \ &= \  
\del^Q {\cal F}_{Q+\wt{I} \wt{J}}
- \del_{K} G_{++} \del_- {\cal C}_{K \wt{I} \wt{J}}
\nn \\
\ & \ \ \ \ 
- \mu \, \eps_{\wt{I} \wt{J} \wt{L}} 
\Big( \del^Q h_{Q \wt{L}} - \half \del_{\wt{L}} h_{KK} \Big)
+ \mu \, \eps_{\wt{I} \wt{J} \wt{L}} \del^+ h_{+\wt{L}}
- \mu \, \eps_{\wt{J} \wt{K} \wt{L}} \del_{\wt{K}} h_{\wt{I} \wt{L}}
+ \mu \, \eps_{\wt{I} \wt{K} \wt{L}} \del_{\wt{K}} h_{\wt{J} \wt{L}}
\; ,
\label{KG-fluc-C3+IJ}
\\
0 \ &= \  
\del^Q {\cal F}_{Q+\wt{I} J'}
- \del_K G_{++} \del_- {\cal C}_{K \wt{I} J'}
+ \mu \, \eps_{\wt{I} \wt{K} \wt{L}} \del_{\wt{K}} h_{J' \wt{L}}
\; ,
\label{KG-fluc-C3+IJ'}
\\
0 \ &= \  
\del^Q {\cal F}_{Q+I'J'}
- \del_K G_{++} \del_- {\cal C}_{K I'J'}
+ \frac{1}{24} \mu \, \ve^{I'J' Q'R'S'U'} 
{\cal F}_{Q'R'S'U'} 
\; ,
\label{KG-fluc-C3+I'J'}
\\
0 \ &= \  
- \del_- \del^Q {\cal C}_{QIJ}
\; ,
\label{KG-fluc-C3-IJ}
\\
0 \ &= \  
\del^Q {\cal F}_{Q\wt{I}\wt{J} \wt{K}}
- \half \mu \, \eps_{\wt{I}\wt{J}\wt{K}} \del^+ h_{LL}
+ \mu \, \eps_{\wt{J} \wt{K} \wt{L}} \del^+ h_{\wt{I} \wt{L}}
- \mu \, \eps_{\wt{I} \wt{K} \wt{L}} \del^+ h_{\wt{J} \wt{L}}
+ \mu \, \eps_{\wt{I} \wt{J} \wt{L}} \del^+ h_{\wt{K} \wt{L}} 
\; ,
\label{KG-fluc-C3IJK}
\\
0 \ &= \  
\del^Q {\cal F}_{Q\wt{I}\wt{J}K'}
+ \mu \, \eps_{\wt{I}\wt{J}\wt{L}} \del^+ h_{K' \wt{L}} 
\; ,
\label{KG-fluc-C3IJK'}
\\
0 \ &= \  
\del^Q {\cal F}_{Q\wt{I}J'K'}
\; ,
\label{KG-fluc-C3IJ'K'}
\\
0 \ &= \  
\del^Q {\cal F}_{QI'J'K'}
+ \frac{1}{6} \mu \, \ve^{I'J'K' R'S'U'} 
\del_- {\cal C}_{R'S'U'} 
\; .
\label{KG-fluc-C3I'J'K'}
\end{align}
\esubeq
These equations are somewhat complicated and one might wonder whether
these equations can be solved explicitly.
But, we can obtain some constraints from the above equations, and
we will be able to solve the other equations completely when 
we substitute the constraints into the equations! 


\subsection*{Physical Modes of Bosonic Fields}

Now let us derive a physical spectrum of the bosonic fields under the
light-cone gauge-fixing: $h_{-M} = {\cal C}{}_{-MN} = 0$. 
All we have to do is to 
analyze physical modes in linearized field equations.
First, we find a traceless condition 
\begin{align}
0 \ &= \ h_{M}{}^M \ = \ h_{II}  \label{h---1.5}
\end{align}
from the field equation (\ref{KG-fluc-g3--}).
This condition, which the graviton $h_{MN}$ should satisfy,
is derived from the light-cone gauge-fixing $h_{-M} = 0$.
Substituting (\ref{h---1.5}) into (\ref{KG-fluc-g3-I})
leads to the divergence free condition for the graviton field $\del^M
h_{IM} = 0$ and we can rewrite $h_{I+}$ as
\begin{align*}
h_{I+} \ = \ \frac{1}{\del_-} \del_J h_{IJ} \; .
\end{align*}
Thus we find that $h_{I+}$ is non-dynamical.
Moreover, we obtain another constraint 
\begin{align*}
\del^M h_{+M} \ &= \ \frac{1}{3} \mu {\cal C}_{123}
\; ,
\end{align*}
which leads to the expression for $h_{++}$ from the equation
(\ref{KG-fluc-g3+-}) as
\begin{align*}
h_{++} \ &= \ 
\frac{1}{(\del_-)^2} \del_I \del_J h_{IJ}
+ \frac{1}{3 \del_-} \mu \, {\cal C}{}_{123} 
\; . 
\end{align*}
In contrast to the IIB supergravity case \cite{MT02}, 
$h_{++}$ includes the term
proportional to $\mu$. The appearance of this term is characteristic of 
our case\footnote{The spectrum of type IIA string theory and linearized
supergravity is studied in \cite{KS03}. In this case $h_{++}$ contains 
the additional term proportional to $\mu$. }. 
In the same way, we can read off the following condition 
from the equation (\ref{KG-fluc-C3+-I}):
$\del_J {\cal C}{}_{+IJ} = 0$. 
The equation (\ref{KG-fluc-C3-IJ}) leads to
the divergence free condition $\del^M {\cal C}_{MIJ} = 0$ 
and the expression for the field ${\cal C}{}_{+IJ}$ is
\begin{align*}
{\cal C}{}_{+IJ} \ = \ \frac{1}{\del_-} \del_K {\cal C}{}_{IJK} 
\; .
\end{align*}
We find that ${\cal C}{}_{+IJ}$ is also non-dynamical.

Under the light-cone gauge-fixing conditions and 
the above mentioned conditions for the non-dynamical modes, 
we can reduce field equations for $h_{MN}$ and ${\cal C}_{MNP}$ as follows:
\bsubeq 
\begin{align}
\mbox{equation (\ref{KG-fluc-g3IJ})} : &&
0 \ &= \ 
\Box \, h_{{\wt{I}}{\wt{J}}} 
- \frac{2}{3} \mu \, \delta_{{\wt{I}}{\wt{J}}} \, \del_{-} {\cal C}
\; ,
\label{h-II2} \\
\mbox{equation (\ref{KG-fluc-g3IJ'})} : &&
0 \ &= \ 
\Box \, h_{{\wt{I}}{J'}} 
- \mu \, \del_{-} {\cal C}{}_{\wt{I} J'} 
\; ,
\label{h-II'2} \\
\mbox{equation (\ref{KG-fluc-g3I'J'})} : &&
0 \ &= \ 
\Box \, h_{{I'}{J'}} 
+ \frac{1}{3} \mu \, \delta_{{I'}{J'}} \del_{-} {\cal C}
\; ,
\label{h-I'I'2} \\
\mbox{equation (\ref{KG-fluc-C3IJK})} : &&
0 \ &= \ 
\Box \, {\cal C}
+ 2 \mu \, \del_- h_{\wt{I} \wt{I}} 
\; , 
\label{c-III2} \\
\mbox{equation (\ref{KG-fluc-C3IJK'})} : &&
0 \ &= \ 
\Box \, {\cal C}{}_{\wt{I} J'} 
+ \mu \, \del_- h_{\wt{I} J'} 
\; , 
\label{c-III'2} \\
\mbox{equation (\ref{KG-fluc-C3IJ'K'})} : &&
0 \ &= \ 
\Box \, {\cal C}{}_{{\wt{I}}{J'}{K'}}
\; ,
\label{c-II'I'2} \\
\mbox{equation (\ref{KG-fluc-C3I'J'K'})} : &&
0 \ &= \ 
\Box \, {\cal C}{}_{I' J' K'}
- \frac{1}{6} \mu \, 
\ve^{I' J'K' W'X'Y'} \del_- {\cal C}{}_{W'X'Y'}
\; ,
\label{c-I'I'I'2}
\end{align}
\esubeq
where $\ve^{I'J'K'W'X'Y'}$ is the $SO(6)$ invariant tensor density 
(or equivalently, the Levi-Civita symbol) whose
normalization is $\ve^{456789} = \ve_{456789} =1$.
Note that 
we wrote the above equations in terms of the following 
two quantities defined by 
\begin{align*}
{\cal C}{}_{\wt{I} J'} \ &\equiv \ 
\half \eps_{\wt{I} \wt{K} \wt{L}} {\cal C}{}_{\wt{K} \wt{L} J'}
\; , \ls
{\cal C} \ \equiv \ 2 {\cal C}{}_{123} \; ,
\end{align*}
where we introduced the $SO(3)$ invariant tensor density (or
equivalently, Levi-Civita
symbol) $\eps_{\wt{I} \wt{J} \wt{K}}$ ($\eps_{123} = \eps^{123} = 1$).


Now let us solve the above reduced equations of motion for fluctuation
modes, and derive the zero-mode energy spectrum and degrees of freedom of
bosonic fields.
We consider the field ${\cal C}{}_{\wt{I} J'K'}$. 
{}From the above equation (\ref{c-II'I'2}),
we find that this field does not couple to the other fields. 
So the zero point energy ${\cal E}{}_0 ({\cal C}{}_{\wt{I} J' K'})$ 
and degrees of freedom ${\cal D} ({\cal C}{}_{\wt{I} J'K'})$ are given by
\begin{align}
{\cal E}_0 ({\cal C}{}_{\wt{I} J' K'}) \ &= \ 2
\; , \ls
{\cal D} ({\cal C}{}_{\wt{I} J' K'}) \ = \ 45 \; .
\label{energy-dof-C-II'I'}
\end{align}

Next, we consider $SO(3) \times SO(6)$ tensor fields $h_{\wt{I} J'}$
and ${\cal C}{}_{\wt{I} J'}$ coupled to each other. 
In order to diagonalize these coupled fields, we define
two complex fields $H_{\wt{I} J'}$ and $\ol{H}{}_{\wt{I} J'}$ as 
\begin{align*}
H_{\wt{I} J'} \ &= \ h_{\wt{I} J'} + i {\cal C}{}_{\wt{I} J'} 
\; , \ls
\ol{H}{}_{\wt{I} J'} \ = \ h_{\wt{I} J'} - i {\cal C}{}_{\wt{I} J'} 
\; .
\end{align*}
By using these fields, (\ref{h-II'2}) and (\ref{c-III'2}) can be
rewritten as 
\begin{align*}
0 \ &= \ \big( \Box + \mu \, i \del_- \big) H_{\wt{I} J'}
\; , \ls
0 \ = \ \big( \Box - \mu \, i \del_- \big) \ol{H}{}_{\wt{I} J'}
\; .
\end{align*}
Thus the zero point energies and degrees of freedom 
of $H_{\wt{I} J'}$ and $\ol{H}{}_{\wt{I} J'}$ are given by
\begin{align}
{\cal E}{}_0 (H_{\wt{I} J'}) \ &= \ 3
\; , \ls
{\cal E}{}_0 (\ol{H}{}_{\wt{I} J'}) \ = \ 1
\; , \ls
{\cal D} (H_{\wt{I} J'}) \ = \ {\cal D} (\ol{H}{}_{\wt{I} J'}) 
\ = \ 18
\; . \label{energy-dof-H-II'}
\end{align}
Then we will solve the field equations (\ref{h-II2}), (\ref{h-I'I'2}) and
(\ref{c-III2}) concerning $h_{\wt{I} \wt{J}}$, $h_{I'J'}$ and
${\cal C}$. Since these fields are coupled to one another, 
we have to diagonalize these fields in order to solve the equations. 
Hence let us introduce the following fields: 
\begin{align*}
\ls &&
h_{\wt{I} \wt{J}}^{\perp}
\ &\equiv \ 
h_{\wt{I} \wt{J}} - \frac{1}{3} \delta_{\wt{I} \wt{J}} \, h_{\wt{K} \wt{K}}
\; , &
h_{I' J'}^{\perp}
\ &\equiv \ 
h_{I' J'} - \frac{1}{6} \delta_{I' J'} \, h_{K' K'}
\; , && \ls \\
&&
h \ &\equiv \ h_{\wt{K} \wt{K}} + i {\cal C} 
\; , &
\ol{h} \ &\equiv \ h_{\wt{K} \wt{K}} - i {\cal C} \; .
\end{align*}
Note that 
$h_{\wt{I} \wt{J}}^{\perp}$ and $h_{I' J'}^{\perp}$ are transverse modes
and two complex scalar fields $h$ and $\ol{h}$ are trace modes. 
In this re-definition we find $\Box h_{\wt{I} \wt{J}}^{\perp} = 0$, and
so its energy and degrees of freedom are given by 
\begin{align}
{\cal E}{}_0 (h_{\wt{I} \wt{J}}^{\perp}) \ &= \ 2 \; , \ls
{\cal D} (h_{\wt{I} \wt{J}}^{\perp}) \ = \ 5 
\; . \label{energy-dof-h-perp}
\end{align}
Since we also find $\Box h_{I' J'}^{\perp} = 0$,
we obtain the energy and degrees of freedom of $h_{I' J'}^{\perp}$:
\begin{align}
{\cal E}{}_0 (h_{I' J'}^{\perp}) \ &= \ 2 \; , \ls
{\cal D} (h_{I' J'}^{\perp}) \ = \ 20
\; . \label{energy-dof-h'-perp}
\end{align}
Similarly the field equations for $h$ and $\ol{h}$ are described by
\begin{align*}
\big( \Box + 2 \mu \, i \del_- \big) h \ &= \ 0 
\; , \ls 
\big( \Box - 2 \mu \, i \del_- \big) \ol{h} \ = \ 0 
\; .
\end{align*}
Thus the energies and degrees of freedom of them are
\begin{align}
{\cal E}{}_0 (h) \ &= \ 4
\; , \ls
{\cal E}{}_0 (\ol{h}) \ = \ 0
\; , \ls
{\cal D} (h) \ = \ 
{\cal D} (\ol{h}) \ = \ 1 
\; . \label{energy-dof-h-hbar}
\end{align}

Finally we consider (\ref{c-I'I'I'2}) by decomposing ${\cal
  C}{}_{I'J'K'}$ into self-dual part and anti-self-dual part as follows:
${\cal C}{}_{I'J'K'} \equiv 
{\cal C}{}_{I'J'K'}^{\oplus} + {\cal C}{}_{I'J'K'}^{\ominus}$,
where 
${\cal C}{}_{I'J'K'}^{\oplus}$ is a self-dual part and 
${\cal C}{}_{I'J'K'}^{\ominus}$ is an anti-self-dual part. 
These are defined by, respectively, 
\begin{align*}
{\cal C}{}_{I'J'K'}^{\oplus} \ &= \ 
\frac{i}{3!} \ve^{I'J'K'W'X'Y'} {\cal C}{}_{W'X'Y'}^{\oplus}
\; , \ls
{\cal C}{}_{I'J'K'}^{\ominus} \ = \ 
- \frac{i}{3!} \ve^{I'J'K'W'X'Y'} {\cal C}{}_{W'X'Y'}^{\ominus}
\; .
\end{align*}
Due to this decomposition,
the field equations of them are expressed as
\begin{align*}
\big( \Box + \mu \, i \del_- \big) {\cal C}{}_{I'J'K'}^{\oplus}
\ &= \ 0 
\; , \ls
\big( \Box - \mu \, i \del_- \big) {\cal C}{}_{I'J'K'}^{\ominus}
\ = \ 0 
\; ,
\end{align*}
and hence we find the energies and degrees of freedom of 
${\cal C}{}_{I'J'K'}^{\oplus}$ and ${\cal C}{}_{I'J'K'}^{\ominus}$:
\begin{align}
{\cal E}{}_0 ({\cal C}{}_{I'J'K'}^{\oplus}) 
\ &= \ 3
\; , \ls
{\cal E}{}_0 ({\cal C}{}_{I'J'K'}^{\ominus})
\ = \ 1
\; , \ls
{\cal D} ({\cal C}{}_{I'J'K'}^{\oplus}) \ = \ 
{\cal D} ({\cal C}{}_{I'J'K'}^{\ominus}) \ = \ 
10
\; . \label{energy-dof-C-I'I'I'}
\end{align}

Now we have fully solved the field equations for bosonic fluctuations
and have derived the spectrum of $h_{MN}$ and ${\cal C}_{MNP}$.
The resulting spectrum is splitting with a certain energy difference 
in contrast to the flat case. 
We summarize the spectrum of bosonic fields in Table
\ref{boson-KG}:
\begin{table}[htbp]
\begin{center}
\begin{tabular}{c|@{\ls}c@{\ls}|c} \hline
energy ${\cal E}{}_0$ 
& bosonic fields & degrees of freedom
\\ \hline \hline
$4$ & $h$ & $1$ \\ 
$3$ & $H_{\wt{I} J'}$ \LS ${\cal C}{}_{I'J'K'}^{\oplus}$ & $18+10$ \\
$2$ & $h_{\wt{I} \wt{J}}^{\perp}$ \LS ${\cal C}{}_{\wt{I} J'K'}$ 
      \LS $h_{I'J'}^{\perp}$ & $5+45+20$ \\
$1$ & $\ol{H}{}_{\wt{I} J'}$ \LS ${\cal C}{}_{I'J'K'}^{\ominus}$ & $18+10$ \\
$0$ & $\ol{h}$ & $1$ \\ \hline
\end{tabular}
\caption{\sl Zero point energy spectrum of the bosonic fields 
in eleven-dimensional supergravity on the plane-wave background.}
\label{boson-KG}
\end{center}
\end{table}


\subsection*{Physical Modes of Fermionic Fields}

Let us solve the field equations of the fluctuations of gravitino
imposed the light-cone gauge-fixing condition $\psi_- = 0$.
First,
we consider the equation (\ref{KG-fluc-psi3-}), which is rewritten as
\begin{align}
\hG^N D_N \psi_- - \hG^N D_- \psi_N \ &= \ 
J_- - \frac{1}{9} \hG_- \hG_N J^n \; .
\label{KG-fluc-psi4-}
\end{align}
Note that we represent the field equations (\ref{KG-fluc-psi3}) as
\begin{align*}
\hG^{MNP} D_N \psi_P \ &= \ J^M \; ,
\end{align*}
where $J^M$ in the right hand side of the above equation is described by
\begin{align*}
J^+ \ &= \ - J_- \ = \ 0 \; ,
&
J^- \ &= \ 
- \frac{1}{4} \mu \, \hG^{+-123I'} \psi_{I'}
-  \frac{1}{8} \mu \, \eps_{\wt{I} \wt{J} \wt{K}}
\, \hG_{\wt{I} \wt{J}} \, \psi_{\wt{K}} \; ,
\\
J^{\wt{I}} \ &= \ 
\frac{1}{4} \mu \, \hG^{+123} 
\big( \delta_{\wt{I} \wt{J}} - \hG_{\wt{I}} \hG_{\wt{J}} \big)
\psi_{\wt{J}} \; ,
&
J^{I'} \ &= \ 
- \frac{1}{4} \mu \, \hG^{+123} 
\big( \delta_{I'J'} - \hG_{I'} \hG_{J'} \big) \psi_{J'}
\; .
\end{align*}
Using the above variables and the properties of the plane-wave background
(\ref{KG-BG2}), 
we simplify the equation (\ref{KG-fluc-psi4-}) as
\begin{align}
\hG^M \psi_M \ &= \ 0 \; .
\label{KG-cond-6} 
\end{align}
This constraint is the condition which the on-shell gravitino should obey.
Next we consider the equation (\ref{KG-fluc-psi3+}), which is rewritten as
\begin{align}
0 \ &= \ g^{P+} \hG{}^N D_N \psi_P 
- g^{PN} \hG{}^+ D_N \psi_P
+ \half \big( 
\hG{}^+ \hG{}^N  - \hG{}^N \hG{}^+ 
\big) \hG{}^P D_N \psi_P
\; . \label{KG-fluc-psi3+2}
\end{align}
We find that the first and third term are deleted by light-cone
gauge-fixing and (\ref{KG-cond-6}). 
Thus we can reduce (\ref{KG-fluc-psi3+2}) to
$0 = \hG^+ ( - \del_- \psi_+ + \del_I \psi_I )$.
So we obtain the divergence free condition for the gravitino such as
$\del^M \psi_M = 0$,
which is also the condition that the on-shell gravitino should satisfy.
Thus we see that $\psi_+$ is expressed by the other fields
\begin{align*}
\psi_+ \ &= \ \frac{1}{\del_-} \del_I \psi_I
\end{align*}
and we find that this component of the gravitino is a non-dynamical field.

Here we shall reduce (\ref{KG-fluc-psi3I}) to
\begin{align}
0 \ &= \ 
\hG{}^{\r{+}} \Big( \del_+ + \half G_{++} \del_- \Big)
\psi_{\wt{I}}^{\oplus} 
+ \hG{}^{\r{-}} \del_- \psi_{\wt{I}}^{\ominus}
+ \hG{}^{\r{K}} \del_K 
(\psi_{\wt{I}}^{\oplus} + \psi_{\wt{I}}^{\ominus})
- \frac{1}{4} \mu \hG{}^{\r{+}\r{1}\r{2}\r{3}} 
\big( \delta_{\r{\wt{I}} \r{\wt{J}}} 
- \hG{}_{\r{\wt{I}}} \hG{}_{\r{\wt{J}}} \big) 
\psi_{\wt{J}}^{\oplus} 
\; , \label{KG-fluc-psi4I}
\end{align}
where 
we decomposed gravitino as $\psi_{\wt{I}} \equiv \psi_{\wt{I}}^{\oplus} +
\psi_{\wt{I}}^{\ominus}$. The $\psi_{\wt{I}}^{\oplus}$ and 
$\psi_{\wt{I}}^{\ominus}$ are defined as 
\begin{align*}
\psi_{\wt{I}}^{\oplus} \ &\equiv \ - \half \hG{}^{\r{-}}
\hG{}^{\r{+}} \psi_{\wt{I}} \; , \ls 
\psi_{\wt{I}}^{\ominus} \ \equiv \ - \half \hG{}^{\r{+}}
\hG{}^{\r{-}} \psi_{\wt{I}} \; , 
\end{align*}
which satisfy the projection conditions: 
$\hG{}^-\psi_{\wt{I}}^{\oplus} = \hG{}^+ \psi_{\wt{I}}^{\ominus} = 0$.
When we act $\hG{}^{\r{+}}$ on (\ref{KG-fluc-psi4I}) from the left, 
$\psi_{\wt{I}}^{\ominus}$ can be expressed in terms
$\psi_{\wt{I}}^{\oplus}$ as follows:
\begin{align}
\psi_{\wt{I}}^{\ominus} \ &= \ 
\frac{1}{2 \del_-} \hG{}^{\r{+}} \hG{}^{\r{K}}
\del_K \psi_{\wt{I}}^{\oplus} 
\; . \label{non-dyn--fermion}
\end{align}
Thus $\psi_{\wt{I}}^{\ominus}$ is not independent of $\psi_{\wt{I}}^{\oplus}$.
Similarly, when we act $\hG{}^{\r{-}}$ on (\ref{KG-fluc-psi4I}) from the left 
and utilize (\ref{non-dyn--fermion}), 
we obtain the following equation:
\begin{align}
0 \ &= \ 
\Box \, \psi_{\wt{I}}^{\oplus} 
- \half \mu \hG{}^{\r{1}\r{2}\r{3}} 
\big( \delta_{\r{\wt{I}} \r{\wt{J}}} 
- \hG{}_{\r{\wt{I}}} \hG{}_{\r{\wt{J}}} \big) 
\del_- \psi_{\wt{J}}^{\oplus}
\; . \label{pre-eom-psiI}
\end{align}
In order to solve this equation, 
let us decompose the gravitino fields into the traceless part and the
``trace'' part with respect to the spacetime indices as follows:
\begin{align*}
\psi_{\wt{I}}^{\oplus \perp} \ &\equiv \ 
\Big( \delta_{\r{\wt{I}} \r{\wt{J}}} 
- \frac{1}{3} \hG{}_{\r{\wt{I}}} \hG{}_{\r{\wt{J}}}
\Big) \psi_{\wt{J}}^{\oplus} 
\; , \ls
\psi_1^{\oplus \para} \ \equiv \ 
\hG{}^{\r{\wt{I}}} \psi_{\r{\wt{I}}}^{\oplus} \ = \ 
\hG{}^{\r{\wt{I}}} \psi_{\wt{I}}^{\oplus}
\; . 
\end{align*}
We denote the traceless part and the ``trace'' part 
to $\psi_{\wt{I}}^{\oplus \perp}$ and $\psi_1^{\oplus \para}$ 
and call them the $\hG$-transverse mode and the $\hG$-parallel mode,
respectively. 
Acting $\hG{}^{\r{\wt{I}}}$ on (\ref{pre-eom-psiI}) from the
left and contracting the index ${\r{\wt{I}}}$,
we obtain a equation with respect to the $\hG$-parallel mode
$\psi_1^{\oplus \para}$ 
\begin{align}
0 \ &= \ 
\Box \, \psi_1^{\oplus \para} 
- \mu \hG{}^{\r{1}\r{2}\r{3}} \del_- \psi_1^{\oplus \para} 
\; . \label{para-eom-psiI}
\end{align}
We also obtain a non-trivial equation for the $\hG$-transverse mode
$\psi_{\wt{I}}^{\oplus \perp}$ 
when we act $(\delta_{\r{\wt{K}} \r{\wt{I}}} - \frac{1}{3}
\hG{}_{\r{\wt{K}}} \hG{}_{\r{\wt{I}}})$ on
(\ref{pre-eom-psiI}):
\begin{align}
0 \ &= \ 
\Box \, \psi_{\wt{K}}^{\oplus \perp}
- \half \mu \hG{}^{\r{1}\r{2}\r{3}} \del_- \psi_{\wt{K}}^{\oplus \perp} 
\; . \label{perp-eom-psiI}
\end{align}
The field equations (\ref{para-eom-psiI}) and (\ref{perp-eom-psiI}) 
contain extra factors given by the gamma matrices $\hG^{\r{1} \r{2}
  \r{3}}$ which prevent us from our obtaining the Klein-Gordon type field
equations (\ref{Klein-Gordon}) for the gravitinos. 
Thus we decompose $\psi_{\wt{I}}^{\oplus \perp}$ 
and $\psi_1^{\oplus \para}$ in terms of the ``chiral projection operator''
$\half ( 1 \pm i\hG{}^{\r{1}\r{2}\r{3}})$ as follows: 
\begin{align*}
\psi_{\wt{I} {\rm R}}^{\oplus \perp} 
\ &\equiv \ 
\frac{1 + i \hG{}^{\r{1}\r{2}\r{3}}}{2} \psi_{\wt{I}}^{\oplus \perp}
\; , &
\psi_{\wt{I} {\rm L}}^{\oplus \perp} 
\ &\equiv \ 
\frac{1 - i \hG{}^{\r{1}\r{2}\r{3}}}{2} \psi_{\wt{I}}^{\oplus \perp}
\; , \\
\psi_{1 {\rm R}}^{\oplus \para} 
\ &\equiv \ 
\frac{1 + i \hG{}^{\r{1}\r{2}\r{3}}}{2} \psi_1^{\oplus \para}
\; , &
\psi_{1 {\rm L}}^{\oplus \para} 
\ &\equiv \ 
\frac{1 - i \hG{}^{\r{1}\r{2}\r{3}}}{2} \psi_1^{\oplus \para}
\; .
\end{align*}
These variables satisfy the following ``chirality'' conditions
\begin{align*}
i \hG{}^{\r{1}\r{2}\r{3}} \psi_{\wt{I} {\rm R}}^{\oplus \perp}
\ &= \ 
+ \psi_{\wt{I} {\rm R}}^{\oplus \perp} 
\; , &
i \hG{}^{\r{1}\r{2}\r{3}} \psi_{\wt{I} {\rm L}}^{\oplus \perp}
\ &= \ 
- \psi_{\wt{I} {\rm L}}^{\oplus \perp} 
\; , \\
i \hG{}^{\r{1}\r{2}\r{3}} \psi_{1 {\rm R}}^{\oplus \para}
\ &= \ 
+ \psi_{1 {\rm R}}^{\oplus \para} 
\; , &
i \hG{}^{\r{1}\r{2}\r{3}} \psi_{1 {\rm L}}^{\oplus \para}
\ &= \ 
- \psi_{1 {\rm L}}^{\oplus \para} 
\; .
\end{align*}
One can of course write down the above gravitino spinor fields 
in the $SO(9)$ Majorana spinor representation argued in appendix
\ref{conv-membrane}.
But we continue the discussion with the $SO(10,1)$ Majorana spinor
representation. 
Multiplying the ``chiral projection operators'' 
$\half (1 \pm i \hG{}^{\r{1}\r{2}\r{3}})$ to
the field equation for 
$\hG$-parallel mode (\ref{para-eom-psiI}) on the left, we obtain 
\begin{align}
0 \ &= \ 
\Big( \Box + \mu \, i \del_- \Big) 
\psi_{1 {\rm R}}^{\oplus \para}
\; , \ls
0 \ = \ 
\Big( \Box - \mu \, i \del_- \Big)
\psi_{1 {\rm L}}^{\oplus \para}
\; . \label{eom-psiI-1}
\end{align}
It appears that the equations (\ref{eom-psiI-1}) 
are the correct field equations for the $\hG$-parallel modes.
But it is impossible to read the zero-point energy from
only these equations.
The reason is that the {\sl correct} 
$\hG$-parallel mode is defined by the ``trace'' part of only the
$\psi_{\wt{I}}$ mode, which does not include the $\psi_{I'}$ mode.
Thus if we would like to obtain the correct informations of this
parallel mode,
we must also look at the $\hG$-parallel mode (i.e., the ``trace'' part)
of $\psi_{I'}$ and combine the field equations for these two 
$\hG$-parallel modes.
But we have not look at the field equations for the gravitino
$\psi_{I'}$ yet.
Thus we will discuss the energies of the $\hG$-parallel modes later.  

Let us discuss the $\hG$-transverse mode here. 
In the similar way to the $\hG$-parallel modes,
we obtain the Klein-Gordon type field equations 
when we perform the ``chiral projection'' to the field equation for
the $\hG$-transverse mode (\ref{perp-eom-psiI}) on the left:
\begin{align*}
0 \ &= \ 
\Big( \Box + \half \mu \, i \del_- \Big)
\psi_{\wt{I} {\rm R}}^{\oplus \perp}
\; , \ls
0 \ = \ 
\Big( \Box - \half \mu \, i \del_- \Big)
\psi_{\wt{I} {\rm L}}^{\oplus \perp}
\; . 
\end{align*}
In the case of these mode 
we can analyze the energies and the number of degrees of freedom
from these equations. 
We can read off the zero point energies and degrees of freedom of
$\psi_{\wt{I} {\rm R}}^{\oplus \perp}$ and $\psi_{\wt{I} {\rm
    L}}^{\oplus \perp}$ from the above equations:
\begin{align}
{\cal E}{}_0 (\psi_{\wt{I} {\rm R}}^{\oplus \perp}) 
\ &= \ \frac{5}{2} 
\; , \ls
{\cal E}{}_0 (\psi_{\wt{I} {\rm L}}^{\oplus \perp}) 
\ = \ \frac{3}{2}
\; , \ls
{\cal D} (\psi_{\wt{I} {\rm R}}^{\oplus \perp}) 
\ = \ {\cal D} (\psi_{\wt{I} {\rm L}}^{\oplus \perp}) 
\ = \ 
8 \times (3-1) \ = \ 16
\; . \label{energy-dof-psiI-perp}
\end{align}
We will discuss these quantities of $\psi_{1 {\rm R}}^{\oplus \para}$
and $\psi_{1 {\rm L}}^{\oplus \para}$ later. 


Now we argue the other gravitino fields labeled by curved indices $I'
= 4, 5, \cdots, 9$.
The decomposition rules for the gravitino fields $\psi_{I'}$ 
are quite similar to the previous discussions for the $\psi_{\wt{I}}$.
Let us rewrite the equation (\ref{KG-fluc-psi3I'}):
\begin{align*}
0 \ &= \ 
\Big\{ \hG{}^{\r{+}} \Big( \del_+ + \half G_{++} \del_- \Big)
+ \hG{}^{\r{-}} \del_- + \hG{}^{\r{K}} \del_K \Big\}
\psi_{I'}
+ \frac{1}{4} \mu \hG{}^{\r{+}\r{1}\r{2}\r{3}} 
\Big( \delta_{\r{I'}\r{J'}} -
\hG{}_{\r{I'}} \hG{}_{\r{J'}} \Big) \psi_{J'}
\; . 
\end{align*}
In the same way as the case of $\psi_{\wt{I}}$, 
we decompose the gravitino $\psi_{I'}$ into the $\hG$-transverse modes
and the the $\hG$-parallel modes,  
and we decompose them further in terms of the ``chiral projection
operators''. 
After these processes we obtain the following field equations:
\bsubeq
\begin{align}
0 \ &= \ 
\Big( \Box - \frac{5}{2} \mu \, i \del_- \Big)
\psi_{2 {\rm R}}^{\oplus \para}
\; , &
0 \ &= \ 
\Big( \Box + \frac{5}{2} \mu \, i \del_- \Big)
\psi_{2 {\rm L}}^{\oplus \para}
\; , \label{eom-psiI'-1} \\
0 \ &= \ 
\Big( \Box - \half \mu \, i \del_- \Big) 
\psi_{I' {\rm R}}^{\oplus \perp}
\; , &
0 \ &= \ 
\Big( \Box + \half \mu \, i \del_- \Big)
\psi_{I' {\rm L}}^{\oplus \perp}
\; , \label{eom-psiI'-2}
\end{align}
\esubeq
where the $\hG$-transverse mode and $\hG$-parallel mode 
are defined as
\begin{align*}
\psi_{I'}^{\oplus} \ &= \ 
- \half \hG{}^{\r{-}} \hG{}^{\r{+}} \psi_{I'} 
\; , \\
\psi_{I' {\rm R}}^{\oplus \perp} 
\ &= \ 
\frac{1 + i \hG{}^{\r{1}\r{2}\r{3}}}{2} \psi_{I'}^{\oplus \perp}
\; , &
\psi_{I' {\rm L}}^{\oplus \perp} 
\ &= \ 
\frac{1 - i \hG{}^{\r{1}\r{2}\r{3}}}{2} \psi_{I'}^{\oplus \perp}
\; , \\
\psi_{2 {\rm R}}^{\oplus \para} 
\ &= \ 
\frac{1 + i \hG{}^{\r{1}\r{2}\r{3}}}{2} \psi_{2}^{\oplus \para}
\; , &
\psi_{2 {\rm L}}^{\oplus \para} 
\ &= \ 
\frac{1 - i \hG{}^{\r{1}\r{2}\r{3}}}{2} \psi_{2}^{\oplus \para}
\; .
\end{align*}
We find the energy and the number of degrees of freedom for 
the $\hG$-transverse modes from (\ref{eom-psiI'-2}): 
\begin{align}
{\cal E}{}_0 (\psi_{I' {\rm R}}^{\oplus \perp}) 
\ &= \ \frac{3}{2} 
\; , \ls
{\cal E}{}_0 (\psi_{I' {\rm L}}^{\oplus \perp}) 
\ = \ \frac{5}{2}
\; , \ls
{\cal D} (\psi_{I' {\rm R}}^{\oplus \perp}) 
\ = \ {\cal D} (\psi_{I' {\rm L}}^{\oplus \perp}) 
\ = \ 
8 \times (6-1) \ = \ 40
\; . \label{energy-dof-psiI'-perp}
\end{align}
By the same discussion on $\psi_{1 {\rm
    R}}^{\oplus \para}$ and $\psi_{1 {\rm L}}^{\oplus \para}$ in the
    previous analysis,
it is also impossible to read the correct energies and the number
    of degrees of freedom for the $\hG$-parallel modes 
from only the equation (\ref{eom-psiI'-1}).
But when we summarize the equations for the $\hG$-parallel modes for
    $\psi_{\wt{I}}$ (\ref{eom-psiI-1}) 
and the equations for the $\hG$-parallel modes for $\psi_{I'}$
    (\ref{eom-psiI'-1}), 
we can obtain the correct field equations for them.
Thus we perform a linear combination of (\ref{eom-psiI-1}) and
(\ref{eom-psiI'-1}), and define new $\hG$-parallel modes as 
\begin{align*}
\psi_{\rm R}^{\oplus \para} \ &\equiv \ 
\frac{2}{5} \psi_{1 {\rm R}}^{\oplus \para} 
- \psi_{2 {\rm R}}^{\oplus \para}
\; , \ls
\psi_{\rm L}^{\oplus \para} \ \equiv \ 
\frac{2}{5} \psi_{1 {\rm L}}^{\oplus \para} 
- \psi_{2 {\rm L}}^{\oplus \para}
\; . 
\end{align*}
Then, by the on-shell gravitino condition (\ref{KG-cond-6}), 
 we find that the re-defined fermions satisfy the equations
\begin{align*}
0 \ &= \ 
\Big( \Box - \frac{3}{2} \mu \, i \del_- \Big) \psi_{\rm R}^{\oplus
  \para}
\; , \ls
0 \ = \ 
\Big( \Box + \frac{3}{2} \mu \, i \del_- \Big) \psi_{\rm L}^{\oplus
  \para}
\; . 
\end{align*}
Thus the zero point energies and the number of degrees of freedom of them are 
represented by 
\begin{align}
{\cal E}{}_0 (\psi_{\rm R}^{\oplus \para})
\ &= \ 
\frac{1}{2}
\; , \ls
{\cal E}{}_0 (\psi_{\rm L}^{\oplus \para})
\ = \
\frac{7}{2} 
\; , \ls
{\cal D} (\psi_{\rm R}^{\oplus \para})
\ = \ 
{\cal D} (\psi_{\rm L}^{\oplus \para})
\ = \ 8 
\; . \label{energy-dof-psi-para}
\end{align}

Now we have fully solved the field equations for fermionic
fluctuations, and have derived the spectrum of gravitino on the
plane-wave. As a result, we have found that the spectrum is splitting 
with a certain energy difference in the same manner with the spectrum of 
bosons.  
Summarizing (\ref{energy-dof-psiI-perp}),
(\ref{energy-dof-psiI'-perp}) and (\ref{energy-dof-psi-para}),
we obtain the spectrum of gravitino as in Table
\ref{fermion}:

\begin{table}[htbp]
\begin{center}
\begin{tabular}{c|@{\ls}c@{\ls}|c} \hline
energy ${\cal E}{}_0$ & {fermionic fields} 
& degrees of freedom
\\ \hline \hline
${7}/{2}$ & $\psi_{\rm L}^{\oplus \para}$ & $8$ \\ 
${5}/{2}$ & $\psi_{\wt{I} {\rm R}}^{\oplus \perp}$ \LS 
$\psi_{I' {\rm L}}^{\oplus \perp}$ & $16+40$ \\
${3}/{2}$ & $\psi_{\wt{I} {\rm L}}^{\oplus \perp}$ \LS
$\psi_{I' {\rm R}}^{\oplus \perp}$ & $16+40$ \\
${1}/{2}$ & $\psi_{\rm R}^{\oplus \para}$ & $8$ \\ \hline
\end{tabular}
\caption{\sl Zero point energy spectrum of fermionic fields
in eleven-dimensional supergravity on the plane-wave background.}
\label{fermion}
\end{center}
\end{table}


\section{Result}

Until the previous sections 
we constructed the field equations for fluctuation fields of
linearized supergravity and calculated the zero point energies of
fluctuations. 
We summarize the results of spectrum of fluctuation fields in Table
\ref{KG-spectrum}.
\begin{table}[h]
\begin{align*}
\begin{array}{c||@{\ls}c@{\ls}|c} \hline
\mbox{energy $E_0$} 
& \mbox{bosonic/fermionic fields}  
& \mbox{degrees of freedom} \\ \hline \hline
2 \mu & h & 1 \\ 
7\mu/4 & \psi_{\rm L}^{\oplus \para} & 8 \\
3\mu/2 & H_{\wt{I} J'} \LS {\cal C}{}_{I'J'K'}^{\ominus} & 18+10 \\
{5\mu}/{4} & \psi_{\wt{I} {\rm R}}^{\oplus \perp} \LS 
\psi_{I' {\rm L}}^{\oplus \perp} & 16 + 40 \\
\mu & h_{\wt{I} \wt{J}}^{\perp} \LS {\cal C}{}_{\wt{I} J'K'} \LS 
h_{I'J'}^{\perp} & 5+45+20 \\
{3\mu}/{4} & \psi_{\wt{I} {\rm L}}^{\oplus \perp} \LS
\psi_{I' {\rm R}}^{\oplus \perp} & 16+40 \\
\mu/2 & \ol{H}_{\wt{I} J'} \LS {\cal C}{}_{I'J'K'}^{\oplus} & 18+10 \\
{\mu}/{4} &  \psi_{\rm R}^{\oplus \para} & 8 \\
0 & \ol{h} & 1 \\ \hline
\end{array}
\end{align*}
\caption{\sl Zero point energy spectrum of all the physical fields of the
  linearized supergravity on the plane-wave background.}
\label{KG-spectrum}
\end{table}

The spectrum of the center of mass degrees of freedom of the Matrix
theory on the plane-wave background was discussed in the previous
chapter (see Table \ref{KP0207-1:table1}).
In that chapter, we found that the energy values of the multiplet
starts from zero (the ground state $\ket{\Lambda} = \ket{1,1}$) 
to $2 \mu$ (the highest state $\Ket{ \Yng(2,2,2,2) \; , \; \Yng(4,4) }
= \ket{1,1}$) 
at intervals of $\mu/4$ energy values.
In comparison with the result of the supergravity discussed in this
chapter,
we find that 
the $U(1)$ part spectrum of the Matrix theory on the plane-wave
background exactly corresponds to the spectrum of linearized
supergravity on the same background!


\chapter{Conclusion and Discussions} \label{summary}

\newpage

\setcounter{section}{0}
\renewcommand{\thesection}{\thechapter.\arabic{section}}

\section*{Conclusion}

In this doctoral thesis 
we have studied 
Matrix theory and eleven-dimensional supergravity on the plane-wave
background. 

First we reviewed the Matrix theory on the plane-wave background
called the BMN matrix model.
We constructed the single D0-brane effective action as a superparticle on the
plane-wave, and extended this to $N$ D0-branes' effective action as the
non-abelian $U(N)$ matrix model in terms of the Myers' proposition. 
The resulting action has one non-vanishing parameter $\mu$ with mass
dimension one. 
The BMN matrix model has also 32 local supersymmetry which decomposes
into the linearly realized supersymmetry called the kinematical
supersymmetry and the nonlinearly realized one called the dynamical
supersymmetry. 
We also wrote down the Hamiltonian and momentum operators, $SO(3)
\times SO(6)$ rotation operators and supercharges in terms of matrix
variables.
Unlike the flat space case, 
there are non-trivial commutation relations between the Hamiltonian
and supercharges on the plane-wave.
Thus the members in one supermultiplet which is generated by
such supercharges have different energies.
In this thesis we concentrated only the $U(1)$ free sector of this
matrix model, which is the center of mass degrees of freedom of $N$
D0-branes, or the superparticle.
States in this part are generated by the kinematical supercharges 
and we analyzed the energies of supermultiplet 
including the ground state.
The result is summarized in Table \ref{KP0207-1:table1}.
Here let us write down this result again:

\begin{gather*}
\begin{array}{c||@{\ls}c@{\ls}|c} \hline
\mbox{$N$-th Floor} & \mbox{$SU(4) \times SU(2)$ Representations} 
& \mbox{Energy Eigenvalues} \\ \hline \hline
8 & ({\bf 1}, {\bf 1}) & 2 \mu  \\
7 & (\ol{\bf 4}, {\bf 2}) & 7 \mu/4 \\
6 & (\ol{\bf 6}, {\bf 3}) \LS (\ol{\bf 10}, {\bf 1}) & 3 \mu / 2 \\ 
5 & ({\bf 4}, {\bf 4}) \LS (\ol{\bf 20}, {\bf 2}) & 5\mu/4 \\ 
4 & ({\bf 1}, {\bf 5}) \LS ({\bf 15} ,{\bf 3}) \LS({\bf 20'}, {\bf 1}) 
  &  \mu \\ 
3 & (\ol{\bf 4}, {\bf 4}) \LS ({\bf 20}, {\bf 2}) & 3\mu/4 \\ 
2 & ({\bf 6}, {\bf 3}) \LS ({\bf 10}, {\bf 1}) & \mu/2 \\
1 & ({\bf 4}, {\bf 2}) & \mu/4 \\
\mbox{ground state} & ({\bf 1}, {\bf 1}) & 0 \\ \hline
\end{array} \\
\text{\sl 
The ground state supermultiplet generated by kinematical supercharges.}
\end{gather*}

Next we investigated the eleven-dimensional supergravity on the same
background.
We prepared the eleven-dimensional supergravity Lagrangian and
classical field equations derived from it.
They are described up to torsion terms,
or quartic terms with respect to gravitino, which do not contribute to
the analysis of the spectrum on the plane-wave background.
Expanding fields around the plane-wave background and constraining the
light-cone gauge-fixing,
we obtained equations of motion for fluctuation fields.
At first sight these equations seemed to be complicated, 
however we could obtain the Klein-Gordon type field equations 
via field re-definitions.
From the result of this analysis 
we found that the fluctuation fields have different zero-point energies
as below (see also chapter \ref{MAIN}):

\begin{gather*}
\begin{array}{c||@{\ls}c@{\ls}|c} \hline
\mbox{Zero-point Energy $E_0$} 
& \mbox{Bosonic/fermionic Fields}  
& \mbox{Degrees of Freedom} \\ \hline \hline
2\mu & h & 1 \\ 
7\mu/4 & \psi_{\rm L}^{\oplus \para} & 8 \\
3\mu/2 & H_{\wt{I} J'} \LS {\cal C}{}_{I'J'K'}^{\ominus} & 18+10 \\
{5\mu}/{4} & \psi_{\wt{I} {\rm R}}^{\oplus \perp} \LS 
\psi_{I' {\rm L}}^{\oplus \perp} & 16 + 40 \\
\mu & h_{\wt{I} \wt{J}}^{\perp} \LS {\cal C}{}_{\wt{I} J'K'} \LS 
h_{I'J'}^{\perp} & 5+45+20 \\
{3\mu}/{4} & \psi_{\wt{I} {\rm L}}^{\oplus \perp} \LS
\psi_{I' {\rm R}}^{\oplus \perp} & 16+40 \\
\mu/2 & \ol{H}_{\wt{I} J'} \LS {\cal C}{}_{I'J'K'}^{\oplus} & 18+10 \\
{\mu}/{4} &  \psi_{\rm R}^{\oplus \para} & 8 \\
0 & \ol{h} & 1 \\ \hline
\end{array} \\
\text{\sl Zero point energy spectrum of physical degrees of freedom in
supergravity on the plane-wave.}
\end{gather*}

We obtained the energy spectra of the $U(1)$ part of Matrix theory and
eleven-dimensional supergravity.
Both spectra include the same number of bosonic and fermionic degrees
of freedom.
This result should be satisfied in all multiplets in any supersymmetric theory.
We also obtained the fact that the energies of the states in Matrix theory 
completely correspond to those of fields in supergravity.
Thus, we found that the Matrix theory on the plane-wave background
contains the zero-mode spectrum of the eleven-dimensional supergravity
completely.
We describe the image of the above result in Figure \ref{figure-spectrum}.
\begin{figure}[h]
\begin{center}
\includegraphics{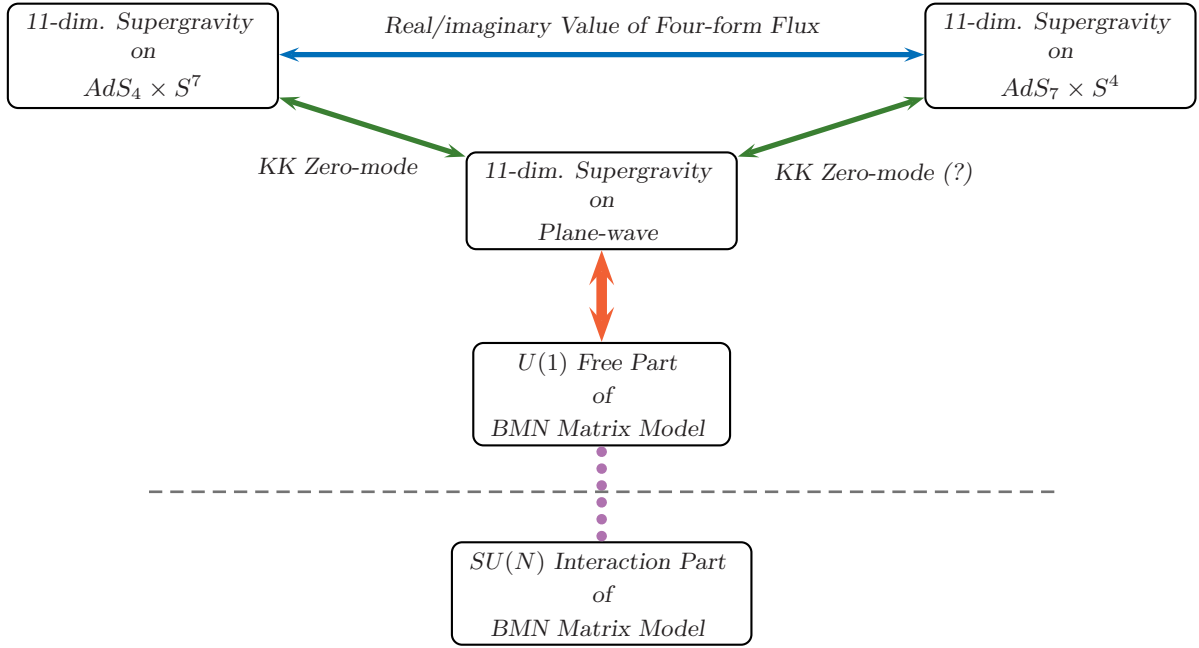}
\caption{\sl The relationships among the spectrum of the
  eleven-dimensional supergravity/Matrix theory on the maximally
  supersymmetric curved background. The $SU(N)$ interaction part in
  the BMN matrix model is independent of the $U(1)$ part which can be
  regarded as the superparticle on the plane-wave.}
\label{figure-spectrum}
\end{center}
\end{figure}

Through this result,
we can see the Matrix theory on the plane-wave background as a candidate of a
quantum extension of eleven-dimensional supergravity on the same
background, or as a candidate
of description of yet-unknown theory, i.e., M-theory, on the
plane-wave background.


\newpage

\section*{Discussions and Future Problems}

In this thesis 
we have argued the $U(1)$ free sector in BMN matrix model and
fluctuation fields in eleven-dimensional supergravity on the plane-wave.
We have found the essential evidence 
that the BMN matrix model 
also includes the supergravity on the plane-wave as in the case of the
theories on flat background.
In order to confirm this evidence more clearly,
we must study other kinds of 
correspondence between the BMN matrix model and
supergravity beyond the correspondence of the spectra between them.
The next study we should do is to compare graviton scattering
amplitudes in those models on the plane-wave background \cite{BFSS96,
  BBPT97, OY98, LW03}. 
There are still few direct discussions about interactions of 
superparticles and scattering amplitudes which should be
calculated in both models.
In order to argue this topic,
vertex operator method seems to be a useful tool
as in string theory.
There already exists the vertex operator formulation for supergravity in
light-cone gauge discussed by Green, Gutperle and Kwon \cite{GGK99},
and there also exists the vertex operator formulation for supermembrane
or Matrix theory proposed by Dasgupta, Nicolai and Plefka \cite{DNP00}.
These formulations can be organized on weakly curved background and 
we would be able to apply these methods to the analysis on the
plane-wave background \cite{TvR98}.
Shin and Yoshida studied one-loop quantum corrections of the BMN
matrix model on the classical plane-wave background in the framework
of path integration \cite{SY0309}.
This analysis would give us a helpful information for the graviton
scatterings.

There also exist many important tasks which we should work around the
physics on the eleven-dimensional plane-wave background.
As mentioned in appendix \ref{geometry},
the plane-wave background connects,
from the purely geometric point of view, 
to $AdS_4 \times S^7$ and $AdS_7 \times S^4$ via the Penrose limit. 
We should study the ``physics'' on the plane-wave also connect to 
the ones on $AdS_{4(7)} \times S^{7(4)}$ background.
We have already understood the properties of 
the linearized and nonlinear full supergravity on $AdS_{4(7)} \times
S^{7(4)}$ background \cite{DP82, D83, PvN84, dWN87, NVvN99, D99}.
In fact, Fernando, G\"{u}naydin and Pavlyk discussed in this topic via
oscillator method \cite{FGP02}
and the oscillator modes on the plane-wave connects to the ones on
$AdS_{4(7)} \times S^{7(4)}$ consistently.
Thus we can
trace how the {\sl fluctuation fields} transform and re-define
in supergravity on $AdS_{4(7)} \times S^{7(4)}$ through the Penrose limit.
When we understand these connections,
we will be able to investigate their gauge theory duals, i.e.,
AdS$_4$/CFT$_3$ and AdS$_7$/CFT$_6$ correspondences \cite{CGLP0102, 
AH9812} as the AdS$_5$/CFT$_4$ correspondence \cite{AMO} 
in type IIB superstring theory.
In particular, we would like to study the strong coupling region of
conformal field theory in six dimensions, which would be one of the
most mysterious theory in quantum field theory.

Myers' term played a central role in the BMN matrix model construction.
Because of the existence of this term,
there is one solution that the supermembrane wrapping on the fuzzy
two-sphere \cite{BMN02}.
How about in the supergravity side?
Myers' effect seems to influence the decomposition of 
three-form gauge fields into self-dual and anti-self-dual part \cite{Yo0308}.

Where is M-brane configuration?
The BMN matrix model has two classical vacua \cite{BMN02}.
One is the ``fuzzy sphere vacuum'' obtained by
\begin{gather*}
X^{\wt{I}} \ = \ \frac{\mu}{3R} J^{\wt{I}}
\; , \ls
X^{I'} \ = \ 0 \; ,
\end{gather*} 
where $J^{\wt{I}}$ form a representation
of the $SU(2)$ algebra
\begin{align*}
[ J^{\wt{I}} , J^{\wt{J}} ] \ &= \ i \eps^{\wt{I} \wt{J} \wt{K}}
J^{\wt{K}}
\;.
\end{align*} 
In the large $N$ limit 
this vacuum is related to ``giant gravitons'' in the plane-wave
background
which are M2-branes wrapping the two-sphere given by $\sum_{\wt{I}}
(x^{\wt{I}})^2 = (\text{\it constant})$ and classically sitting at a fixed
position $x^-$, but with non-zero momentum $p_-$.
The other vacuum is given by $X^I = 0$ for all $I = 1,2, \cdots, 9$,
which is called the ``trivial vacuum''.
This solution is regarded as giant gravitons which are (transverse) M5-branes
wrapping the $S^5$ given by $\sum_{I'} (x^{I'})^2 = (\text{\it constant})$ in
the large $N$ limit.
However, this does not appear as a classical solution of the BMN
matrix model.
This is partly because it is more difficult to
describe M5-brane worldvolume theory than to describe M2-brane
\cite{W9610, BLNPST9701, BLNPST9703}.  
As in the case of ${\cal N}=1^*$ super Yang-Mills theory discussed by
Polchinski and Strassler \cite{PS00},
it is natural to conjecture that the trivial vacuum in the quantum
mechanics theory corresponds to a single large M5-brane.
Further discussions are given by Maldacena, Sheikh-Jabbari and Van
Raamsdonk \cite{MSvR0211}.
M-brane configurations on the plane-wave
background are also studied by Mas and Ramallo \cite{MR0303}, etc.
Thus it is quite interesting for us to investigate how the M-brane
configurations are given in the supergravity on the plane-wave.

In the BMN matrix model, 
there exist a lot of BPS solutions generated by dynamical and kinematical
supercharges \cite{DSvR0205, KP0207-1, KP0207-2}.
Longitudinal and transverse M-branes should be also BPS states
preserving parts of supersymmetry \cite{BSS9612}. 
It is interesting to study the realization of these BPS states
in the supergravity side.

\begin{quote}
In this doctoral thesis 
I have considered the investigation 
about the eleven-dimensional theory on the plane-wave background.
I have also introduced various {\sl tales} for future works.  
The work which has been done here seems to be a small one.
But {\sl what you take} around these topics would become a {\sl giant step} in
M-theory. 
\begin{center}
{\sl I hope that someone gets my message in the thesis!} 
\end{center}
\end{quote}


\newpage

\begin{appendix}


\chapter{Convention} \label{convention}

\newpage

\setcounter{section}{0}
\renewcommand{\thesection}{\thechapter.\arabic{section}}

\section{Eleven-dimensional Spacetime}

First we should define the signature of spacetime 
in order to discuss various properties of symmetries, transformation
laws and Lagrangian of the system.
In this thesis,
we adopt the {\sl almost plus} signature 
to eleven-dimensions Minkowski spacetime: $(-,+,+,\cdots,+)$.

We describe the curved spacetime metric and the tangent space metric
as $g_{MN}$ and $\eta_{\r{A} \r{B}}$, respectively.
Notice that the capital letters which start from $M,N,P, \cdots$ refer to
eleven-dimensional world indices (curved spacetime indices) 
and the capital letters which start from $\r{A},\r{B},\r{C}, \cdots$  
denotes eleven-dimensional tangent space indices.
Note that the vielbein $e_M{}^{\r{A}}$ and its inverse vielbein $E_{\r{A}}{}^M$
are related to the curved spacetime metric $g_{MN}$ and the tangent
space metric $\eta_{\r{A} \r{B}}$ as follows:
\begin{align*}
g_{MN} \ = \ e_M{}^{\r{A}} \, e_N{}^{\r{B}} \, \eta_{\r{AB}} \; , \ls 
\eta_{\r{AB}} \ = \ E_{\r{A}}{}^M \, E_{\r{B}}{}^N \, g_{MN} 
\; .
\end{align*}
We prepare a character such as $\ve^{MNPQRSUVWXY}$ which makes the
three-form gauge field $C_{MNP}$ and its field strength $F_{MNPQ} = 4
\del_{[M} C_{NPQ]}$ couple to each other.
This character is an invariant tensor density in eleven-dimensional
spacetime (weight +1), whose normalization is 
$\ve^{012\cdots\nt} = 1$.


\section{Clifford Algebra: $SO(10,1)$ Representation}
\label{SO11}

In chapter \ref{MAIN} we will use various spinor variables.
Thus it is necessary for us to introduce the Clifford algebra and
Dirac gamma matrices in order to define various transformations.
Here let us define the Clifford algebra and gamma matrices in
eleven-dimensional Minkowski spacetime which has $SO(10,1)$ Lorentz
symmetry. In the next sections we will decompose them into various
representations.

Let us first write down the Clifford algebra and Dirac gamma matrices
in eleven-dimensional spacetime
\begin{gather*}
\{ \hG{}^{\r{A}} , \hG{}^{\r{B}} \} \ = \ 
2 \eta^{\r{AB}} \cdot {\bf 1}_{32} \; .
\end{gather*}
Note that $\eta^{\r{A} \r{B}}$ is the tangent space metric.
Hermitian conjugate of the gamma matrices is defined by
\begin{gather*}
(\hG{}^{\r{A}})^{\dagger} \ = \ \hG_{\r{A}} \ = \ 
- \hG{}^{\r{0}} \hG{}^{\r{A}} 
(\hG{}^{\r{0}})^{-1} 
\; .
\end{gather*}
Note that the gamma matrices $\hG^{\r{I}}$ are Hermitian except for
$\hG^{\r{0}}$, which is anti-Hermitian.
For the convenience we define the following anti-symmetrized products
of gamma matrices with unit weight:
\begin{gather*}
\hG{}_{\r{A_1} \r{A_2} \cdots \r{A_n}} 
\ \equiv \ 
\hG_{[\r{A_1}} \hG_{\r{A_2}} \cdots \hG_{\r{A_n}]} 
\ = \ 
\frac{1}{n!} \sum_{\sigma} {\rm sgn} (\sigma) \,
\hG{}_{\r{A_{\sigma_1}}} \hG{}_{\r{A_{\sigma_2}}} \cdots 
\hG{}_{\r{A_{\sigma_n}}} 
\; .
\end{gather*}
Utilizing this definition, we write an identity for the gamma matrices:
\begin{align}
&\wh{\Gamma}{}^{\r{A_1} \r{A_2} \cdots \r{A_p}} 
\, \wh{\Gamma}{}_{\r{B_1} \r{B_2} \cdots \r{B_q}}
\nn \\
&\LS \ = \ 
\sum_{k=0}^{\min (p,q)} 
(-1)^{\half k (2p-k-1)} \frac{p! \, q!}{(p-k)! (q-k)! k!}
\delta^{[\r{A_1}}_{[\r{B_1}} \cdots 
\delta^{\phantom{[}\r{A_k}}_{\phantom{[}\r{B_k}} 
\wh{\Gamma}{}_{\phantom{[\r{B}}}^{\r{A_{k+1}} 
\cdots \r{A_p}]}{}^{\phantom{\r{A}]}}_{\r{B_{k+1}} \cdots \r{B_q}]} \; .
\label{identity-gamma}
\end{align}
Utilizing these properties, we can define a spinor in
eleven-dimensional Minkowski spacetime.
In particular, 
we can define a {\sl Majorana spinor} $\theta$ 
as a irreducible representation of $SO(10,1)$ spinor\footnote{Notice
  that we denote the gravitino (vectorial $SO(10,1)$ Majorana
  spinor) as $\Psi_M$ in eleven-dimensional supergravity 
(see chapter \ref{MAIN}).}.
Let us define the Dirac conjugate of the spinor $\theta$ as
\begin{align*}
\ol{\theta} \ = \ i \theta^{\dagger} \hG{}^{\r{0}} 
\;.
\end{align*}
Note that the product $\ol{\theta} \theta$ is Hermitian in this definition.
In terms of this Dirac conjugate, we describe the Majorana condition of the
spinors as
\begin{align*}
\ol{\theta} \ &= \ \theta^T C \; ,
\end{align*}
where $C$ is the charge conjugation matrix.
In this thesis this charge conjugation matrix 
is defined as antisymmetric:
$C =  
- C^{-1} = - C^T$.
Under this definition, 
charge conjugations of the gamma matrices and antisymmetrized gamma matrices
are given by
\bsubeq \label{CC}
\begin{gather}
C \hG{}^{\r{A}} C^{-1} 
\ = \ 
- (\hG{}^{\r{A}})^T \; , \\
C \hG{}^{\r{A_1} \cdots \r{A_{2n}}} C^{-1} 
\ = \ 
- ( \hG{}^{\r{A_1} \cdots \r{A_{2n}}} )^T \; , \ls
C \hG{}^{\r{A_1} \cdots \r{A_{2n+1}}} C^{-1} 
\ = \ 
+ ( \hG{}^{\r{A_1} \cdots \r{A_{2n+1}}} )^T 
\; .
\end{gather}
\esubeq

\section{Lorentz Algebra}

The Lorentz symmetry on the tangent space is important to describe
vectors, tensors, and spinors in curved spacetime via vielbeins and
inverse vielbeins. 
It is also important to understand the dynamics of the
theory in the weak coupling limit of gravity.
Thus, let us define here 
the Lorentz algebra in the eleven-dimensional tangent space as
\begin{align*}
i [ \Sigma_{\r{A} \r{B}} , \Sigma_{\r{C} \r{D}} ] 
\ &= \ 
\eta_{\r{A} \r{C}} \, \Sigma_{\r{B} \r{D}} 
+ \eta_{\r{B} \r{D}} \, \Sigma_{\r{A} \r{C}}  
- \eta_{\r{A} \r{D}} \, \Sigma_{\r{B} \r{C}}
- \eta_{\r{B} \r{C}} \, \Sigma_{\r{A} \r{D}} 
\; ,
\end{align*}
where the Lorentz generators $\Sigma_{\r{A} \r{B}}$ are Hermitian 
and they are represented by
\begin{align*}
\Sigma_{\r{A} \r{B}} \ &= \ 0 && \mbox{scalar} \; , \\
(\Sigma_{\r{C} \r{D}})^{\r{A}}{}_{\r{B}} \ &= \ 
i \big( \delta^{\r{A}}_{\r{C}} \, \eta_{\r{D} \r{B}} 
- \delta^{\r{A}}_{\r{D}} \, \eta_{\r{C} \r{B}} \big) && \mbox{vector} 
\; , \\
\Sigma_{\r{A} \r{B}} \ &= \ \frac{i}{2} \hG_{\r{A} \r{B}} && \mbox{spinor}
\; .
\end{align*}


\section{$SO(9)$ Representation} \label{conv-membrane}

We defined the Dirac gamma matrices $\hG^{\r{A}}$ in eleven 
dimensions in appendix \ref{SO11}. 
Performing the fermion light-cone gauge fixing (or $\kappa$-symmetry
gauge fixing),
we decompose these $SO(10,1)$ gamma matrices
$\hG^{\r{A}}$ in terms of $16 \times 16$ unit matrix ${\bf 1}_{16}$ and the 
$SO(9)$ gamma matrices $\gamma^{\r{I}}$.
First we put the fermionic light-cone gauge fixing on the $SO(10,1)$
Majorana spinor $\theta$:
\begin{gather}
\hG^{\r{+}} \theta \ = \ 0 \; , \ls 
\ol{\theta} \hG^{\r{+}} \ = \ 0
\; . 
\end{gather}
By virtue of this constraint 16 degrees of freedom of $SO(10,1)$
Majorana spinor is gauged away and 
we write down $\theta$ by using the $SO(9)$ Majorana spinor $\Psi$ as
\begin{gather}
\theta \ = \ \frac{1}{2^{3/4}} \left(
\begin{array}{c}
0 \\
\Psi
\end{array} \right)
\; , \ls
\ol{\theta} \ = \ i \theta^{\dagger} \hG^{\r{0}} \ = \ 
\theta^T C \ \equiv \ 
\frac{1}{2^{3/4}} \Big( \, - \Psi^T \, , \,  0 \, \Big)
\; .
\end{gather}
This representation denotes that the $SO(9)$ Majorana spinor $\Psi$
satisfies the reality condition $\Psi^{\dagger} = \Psi^T$ explicitly;
the normalization of $\Psi$ is defined so as to satisfy the following:
\begin{align*}
- \ol{\theta} \hG^{\r{-}} \del \theta \ &= \ 
\frac{i}{2} \Psi^{\dagger} \del \Psi 
\; .
\end{align*}
Under this convention, the charge conjugation matrix $C$ in
eleven-dimensional Minkowski spacetime is represented by
\begin{align}
C \ &= \ \left(
\begin{array}{cc}
0 & {\bf 1}_{16} \\
-{\bf 1}_{16} & 0
\end{array} \right)
\; . \label{C}
\end{align}
Let us express the $SO(10,1)$ gamma matrices in the light-cone
directions $\hG^{\r{+}}$ and $\hG^{\r{-}}$ in terms of $16 \times 16$
matrices
\begin{align*}
\hG^{\r{0}} \ &= \ 
\left(
\begin{array}{cc}
0 & i {\bf 1}_{16} \\
i {\bf 1}_{16} & 0
\end{array} \right)
\; , \ls
\hG^{\r{10}} \ = \ 
\left(
\begin{array}{cc}
0 & -i {\bf 1}_{16} \\
i {\bf 1}_{16} & 0
\end{array} \right)
\; , \\
\hG^{\r{\pm}} \ &\equiv \ 
\frac{1}{\sqrt{2}} \big( \hG^{\r{0}} \pm \hG^{\r{10}}
\big) 
\; , \ls \{ \hG^{\r{+}} , \hG^{\r{-}} \} \ = \ -2 \cdot {\bf 1}_{32} \; , \\
\hG^{\r{+}} \ &= \ \sqrt{2} \left(
\begin{array}{cc}
0 & 0 \\
i {\bf 1}_{16} & 0
\end{array} \right)
\; , \ls
\hG^{\r{-}} \ = \ \sqrt{2} \left(
\begin{array}{cc}
0 & i {\bf 1}_{16} \\
0 & 0
\end{array} \right)
\; .
\end{align*}
Next we describe the gamma matrices in the longitudinal directions
$\hG^{\r{I}}$ in terms of the $SO(9)$
gamma matrices $\gamma^{\r{I}}$
\begin{align}
\hG^{\r{I}} \ &= \ \left(
\begin{array}{cc}
- (\gamma^{\r{I}})^T & 0 \\
0 & \gamma^{\r{I}}
\end{array} \right)
\; . \label{Gamma-I}
\end{align}
Note that $\gamma^{\r{I}}$ satisfies the Clifford algebra: 
$\{ \gamma^{\r{I}} , \gamma^{\r{J}} \} = 2 \delta^{\r{I} \r{J}}$.
Since we define the hermitian conjugation of $\hG^{\r{I}}$ as  
$(\hG^{\r{I}})^{\dagger} = \hG^{\r{I}}$,
we obtain the hermitian conjugation of the $SO(9)$ gamma matrices below:
\begin{align*}
(\gamma^{\r{I}})^{\dagger} \ &= \ \gamma^{\r{I}}
\;, \ls
(\gamma^{\r{I} \r{J}})^{\dagger} \ = \ - \gamma^{\r{I} \r{J}}
\; , \ls
(\gamma^{\r{I} \r{J} \r{K}})^{\dagger} \ = \ - \gamma^{\r{I} \r{J} \r{K}}
\; .
\end{align*}


\section{$SU(4) \times SU(2)$ Representation} \label{4-2-repre}

Let us decompose the $SO(9)$ Majorana spinors $\Psi$ and the 
gamma matrices in the $SO(9)$ representations 
into the ones in the $SU(4) \times SU(2)$ representations \cite{DSvR0205}.
The ${\bf 16}$ representation of the $SO(9)$ Majorana spinor are 
split up as
\begin{align*}
{\bf 16} \ &= \ ({\bf 4}, {\bf 2}) \oplus (\ol{\bf 4}, {\bf 2}) 
\LS
\Psi \ \to \ \{ \psi_{i \alpha} \; , \  \wt{\psi}{}^{j \beta} \} 
\; ,
\end{align*}
where ${\bf 4}$ and $\ol{\bf 4}$ are the fundamental and
anti-fundamental representations of $SU(4)$ spinor; ${\bf 2} (=
\ol{\bf 2})$ is the
fundamental representation of $SU(2)$ spinor.
We express the $SU(4) \times SU(2)$ spinor as $\psi_{i \alpha}$ 
in terms of indices $i$, the fundamental $SU(4)$ indices, and 
the fundamental $SU(2)$ indices $\alpha$.
These spinors obey a reality condition,
which in the reduced notation becomes simply as  
$\wt{\psi}{}^{j \beta} = \psi^{\dagger j \beta}$.
More concretely
we represent the $SO(9)$ Majorana spinor $\Psi$ in terms of
$SU(2) \times SU(4)$ representations $\psi_{i \alpha}$ as
\begin{align}
\Psi \ &= \ \left(
\begin{array}{c}
\psi_{i \alpha} \\
\eps_{\alpha \beta} \, \psi^{\dagger i \beta}
\end{array} \right)
\; ,
\end{align} 
and we decompose 
the $SO(9)$ gamma matrices\footnote{The $SO(9)$ gamma matrices and
  Majorana spinors are defined in appendix \ref{conv-membrane}.} 
to the direct product of $SU(4)$
and $SU(2)$ gamma matrices
\begin{align}
\gamma^{\r{\wt{I}}} \ &= \ \left(
\begin{array}{cc}
- \sigma^{\r{\wt{I}}} \otimes {\bf 1}_4 & 0 \\
0 & \sigma^{\r{\wt{I}}} \otimes {\bf 1}_4  
\end{array} \right) \; , \ls 
\gamma^{\r{I'}} \ = \ \left(
\begin{array}{cc}
0 & {\bf 1}_2 \otimes {\sf g}^{\r{I'}} \\
{\bf 1}_2 \otimes ({\sf g}^{\r{I'}})^{\dagger} & 0
\end{array} \right)
\; .
\end{align}
Note that the matrices $\sigma^{\r{\wt{I}}}$ are the ordinary Pauli matrices
and  the $SU(4)$ gamma matrices ${\sf g}^{\r{I'}}$ satisfy the Clifford algebra
\begin{align*}
\sigma^{\r{\wt{I}}} \sigma^{\r{\wt{J}}} 
+ \sigma^{\r{\wt{J}}} \sigma^{\r{\wt{I}}}
\ &= \ 2
\delta^{\r{\wt{I}} \r{\wt{J}}} \; , \ls
{\sf g}^{\r{I'}} ({\sf g}^{\r{J'}})^{\dagger} 
+ {\sf g}^{\r{J'}} ({\sf g}^{\r{I'}})^{\dagger} \ = \ 
2 \delta^{\r{I'}\r{J'}}
\; .
\end{align*}

\section{Connections and Curvature Tensors}
\label{connections}

In this appendix we define the geometrical variables such as
connections and their curvature tensors which appear 
in the eleven-dimensional supergravity Lagrangian.
Explicit expressions of these definitions are of important to calculate 
the fluctuation fields, superspace coset formalism, etc.

First let us define 
the covariant derivative $\nabla_M$ for 
general coordinate transformation by using the affine connection
$\Gamma^P_{NM}$ as
\begin{align}
\nabla_M A_N \ &= \ \del_M A_N - \Gamma^P_{NM} \, A_P \; ,
\end{align}
where $A_P$ is an arbitrary covariant vector.
Riemann curvature tensor $R^R{}_{PMN}$ for the affine connection is defined 
from the commutator of the covariant derivative as
\begin{gather}
\begin{split}
[ \nabla_M , \nabla_N ] A_P \ &= \ 
- R^R{}_{PMN} A_R + T^R{}_{MN} A_R
\; , \\
R^R{}_{PMN} \ &= \ 
\del_M \Gamma^R_{PN} - \del_N \Gamma^R_{PM}
+ \Gamma^R_{QM} \Gamma^Q_{PN} - \Gamma^R_{QN} \Gamma^Q_{PM} 
\; ,
\end{split} \label{cr-affine} 
\end{gather}
where $T^R{}_{MN}$ is a torsion coming from the antisymmetric part of
the affine connection:
\begin{align*}
\Gamma^R_{[MN]} \ &= \ \half \big( \Gamma^R_{MN} - \Gamma^R_{NM} \big)
\ \equiv \ \half T^R{}_{MN} \; .
\end{align*}
In addition, let us introduce the Christoffel symbol
$\christoffel{R}{M}{N}$ in terms of the metric and torsion:
\begin{align}
\begin{split}
\christoffel{R}{M}{N} \ &\equiv \
\half g^{RQ} \big( \del_M g_{NQ} + \del_N g_{MQ} - \del_Q g_{MN} \big)
\\
\ &= \ \Gamma^R_{(MN)} + \half T_M{}^R{}_N + \half T_N{}^R{}_M \; ,
\end{split}
\end{align}
where $\Gamma^R_{(MN)} \equiv \half (\Gamma^R_{MN} + \Gamma^R_{NM}) $ 
is the symmetric part of the affine connection.
Thus the affine connection is written in terms of the Christoffel
symbol and torsion:
\bsubeq
\begin{align}
\Gamma^R_{MN} \ &= \ \Gamma^R_{(MN)} + \Gamma^R_{[MN]}
\ = \ \christoffel{R}{M}{N} + K^R{}_{MN} \; , \\
K^R{}_{MN} \ &\equiv \ \half \big( 
T^R{}_{MN} + T_{MN}{}^R - T_{NM}{}^R \big) \; .
\end{align}
\esubeq
Note that the tensor $K^R{}_{MN}$ is called the contorsion.
When the torsion vanishes, the affine connection is equal to the
Christoffel symbol.

Similarly, we define the covariant derivative $D_M$ for the local Lorentz
transformations by using the spin connection $\omega_M{}^{\r{A} \r{B}}$ as
\begin{align}
D_M \phi \ &= \ 
\del_M \phi - \frac{i}{2} \omega_M{}^{\r{A} \r{B}} \, 
\Sigma_{\r{A} \r{B}} \, \phi \; , 
\end{align}
where $\phi$ is an arbitrary field and 
we write the Lorentz generators as $\Sigma_{\r{A} \r{B}}$ whose
representations are described in appendix \ref{SO11}.
The curvature tensor $\wt{R}^{\r{A} \r{B}}{}_{MN}$ for the spin
connection is defined from the
commutator of the covariant derivative as
\begin{gather}
\begin{split}
[ D_M , D_N ] \phi \ &= \ 
- \frac{i}{2} \wt{R}^{\r{A} \r{B}}{}_{MN} \, \Sigma_{\r{A} \r{B}} \, \phi 
\; , \\
\wt{R}^{\r{A} \r{B}}{}_{MN} \ &= \ 
\del_M \omega_N{}^{\r{A} \r{B}} - \del_N \omega_M{}^{\r{A} \r{B}} 
+ \omega_M{}^{\r{A}}{}_{\r{C}} \, \omega_N{}^{\r{C} \r{B}}
- \omega_N{}^{\r{A}}{}_{\r{C}} \, \omega_M{}^{\r{C} \r{B}} 
\; .
\end{split} \label{cr-spin} 
\end{gather}
We can obtain the curvature tensor $\wt{R}^{\r{A} \r{B}}{}_{MN}$ in
terms of vielbein $e_M{}^{\r{A}}$ and spin connection $\omega_M{}^{\r{A}
  \r{B}}$. 
As in the case of the covariant derivative for the
affine connection (\ref{cr-affine}), 
the torsion term also appears if we write the above 
commutation relation (\ref{cr-spin}) in terms
of the covariant derivative on the tangent space as $D_{\r{A}} =
E_{\r{A}}{}^M D_M$.

We also define the total covariant derivative $\wt{D}_M \equiv
\nabla_M - \frac{i}{2} \omega^{\r{A} \r{B}} \Sigma_{\r{A} \r{B}}$
which contains
both the affine connection and the spin connection.
By virtue of the total covariant derivative
we can simply consider the {\sl vielbein postulate} 
\begin{align}
0 \ &= \ \wt{D}_P \, e_M{}^{\r{A}} 
\ = \ \del_P e_M{}^{\r{A}} - \Gamma^R_{MP} \, e_R{}^{\r{A}} +
\omega_P{}^{\r{A}}{}_{\r{B}} \, e_M{}^{\r{B}}
\; ,
\end{align}
which is equivalent to the equivalent principle for the metric
$\nabla_P g_{MN} = 0$.
Under this postulate
the curvature tensor for the spin connection $\wt{R}^{\r{A} \r{B}}{}_{MN}$
is associated with the curvature for the affine connection 
$R^R{}_{PMN}$:
\begin{align}
R^R{}_{PMN} \ &= \ \eta_{\r{B} \r{C}} \, E_{\r{A}}{}^{R} \,
e_P{}^{\r{C}}
\, \wt{R}^{\r{A} \r{B}}{}_{MN}
\; .
\end{align}
Ricci tensor and scalar curvature are defined below:
\begin{align*}
{\cal R}^M{}_N \ &= \ g^{PQ} \, R^M{}_{PNQ}
\; , \ls
{\cal R} \ = \ {\cal R}^M{}_M
\; .
\end{align*}

Finally let us introduce the ``Cartan's structure equations'' from the
viewpoint of differential geometry
\bsubeq 
\begin{gather*}
\d s^2 \ = \ g_{MN} \, \d x^M \, \d x^N \ = \ \eta_{\r{A} \r{B}}
\, e^{\r{A}} \, e^{\r{B}} \; , \\
T^{\r{A}} \ = \ \d e^{\r{A}} + \omega^{\r{A}}{}_{\r{B}} \w e^{\r{B}}
\; , \ls
\wt{R}^{\r{A} \r{B}} \ = \ 
\d \omega^{\r{A} \r{B}} + \omega^{\r{A}}{}_{\r{C}} \w \omega^{\r{C}
  \r{B}}
\; ,  
\end{gather*}
\esubeq
which can be written more explicitly as
\begin{align*}
T^{\r{A}}{}_{MN} \ &= \ 
\del_M e_N{}^{\r{A}} - \del_N e_M{}^{\r{A}} 
+ \omega_M{}^{\r{A}}{}_{\r{B}} \, e_N{}^{\r{B}}
- \omega_M{}^{\r{A}}{}_{\r{B}} \, e_N{}^{\r{B}} 
\; , \\
\wt{R}^{\r{A} \r{B}}{}_{MN} \ &= \ 
\del_M \omega_N{}^{\r{A} \r{B}} - \del_N \omega_M{}^{\r{A} \r{B}} 
+ \omega_M{}^{\r{A}}{}_{\r{C}} \, \omega_N{}^{\r{C} \r{B}}
- \omega_N{}^{\r{A}}{}_{\r{C}} \, \omega_M{}^{\r{C} \r{B}}
\; .
\end{align*}
Note that $T^{\r{A}}$ is a torsion two-form which vanishes on coset spaces.


\chapter{Lagrangians} \label{Lagrangians}

\newpage


\section{Matrix Theory Lagrangian: Super Yang-Mills Action} \label{SYM}

Matrix theory Lagrangian suggested by Banks, Fischler, Shenker and Susskind
\cite{BFSS96} is described by $N$ D0-branes' effective action
, i.e., the dimensionally reduced model of 
ten-dimensional $U(N)$ super Yang-Mills theory.
Thus, in this appendix
we derive the Matrix theory Lagrangian on the flat background.

We have not completely understood how to describe the $N$ coincident
D-branes' effective action yet.
But, in the low energy region,
we now believe that the effective action should be described by the
dimensionally reduced model of ten-dimensional $U(N)$ super Yang-Mills
theory.
We will also introduce another derivation of the low energy effective
theory of $N$ coincident D-branes' system with/without non-vanishing
background fields in appendix \ref{nonabelian}.

Here we consider the low energy region of the D$p$-branes system.
In low energy region, or weak string coupling region,
the transverse fluctuations of D$p$-branes would freeze.
Thus D$p$-branes appear as heavily massive solitons in string theory.
The resulting modes in D$p$-branes are the
longitudinal modes moving in $(p+1)$-dimensional hypersurface of
D$p$-branes,
which are the massless excitation modes of open strings.
These dynamics could be described by the $(p+1)$-dimensional super
Yang-Mills theory with $U(N)$ gauge symmetry.
We can obtain this theory Lagrangian from the dimensional reduction to
$(p+1)$-dimensions of the ten-dimensional $U(N)$ super Yang-Mills.
In this context, 
we first introduce the ten-dimensional $U(N)$ super Yang-Mills
Lagrangian 
and perform its dimensional reduction procedure.

The Lagrangian of ten-dimensional $U(N)$ super Yang-Mills is described
in terms of the $N \times N$ matrix valued gauge fields and
$SO(9,1)$ Majorana-Weyl spinors as follows:
\begin{gather*}
S \ = \ \int \! \d^{9+1} x \, {\cal L}_{9+1} \ = \ 
\int \! \d^{9+1} x \Big\{ 
- \frac{1}{4} \Tr \big( F_{\mu \nu} F^{\mu \nu} \big)
- \Tr \big( \ol{\theta} \Gamma^{\mu} D_{\mu} \theta \big) \Big\}
\; ,  \\
F_{\mu \nu} \ = \ 
\del_{\mu} A_{\nu} - \del_{\nu} A_{\mu} + i g [ A_{\mu} , A_{\nu} ] 
\; , \ls
D_{\mu} \theta \ = \ 
\del_{\mu} \theta + i g [ A_{\mu} , \theta ] 
\; ,
\end{gather*}
where $\mu, \nu = 0,1,2, \cdots, 9$.
Note that $A_{\mu}$ and $\theta$ are the $N \times N$ matrix valued
dynamical fields whose mass dimensions are $4$ and $9/2$, respectively. 
The Yang-Mills coupling constant is denoted by $g$ of mass dimensions $-3$.
They are the $U(N)$ gauge potential and the $SO(9,1)$ spinor
(Dirac representation), repetitively.
In particular, the spinor $\theta$ can be expressed by the irreducible
Majorana-Weyl spinor $\chi$ (real 16 components) as
\begin{align}
\theta \ &= \ \frac{1}{\sqrt{2}} \left(
\begin{array}{c}
0 \\
\chi
\end{array} \right) \; , \ls 
\chi^* \ = \ \chi \; . \label{MW}
\end{align}
The overall factor is a convention.

It is useful to write down the $SO(9,1)$ Clifford algebra.
This algebra corresponds to the one in eleven-dimensions\footnote{Strictly
speaking, the Clifford algebra in eleven-dimensions is defined in the
same way as the one in ten dimensions.}.
So the Dirac matrices can be described by the same form of
eleven-dimensional ones\footnote{In this section we discuss the
  Lagrangian in the flat Minkowski spacetime. Thus we do not
  distinguish the curved spacetime indices $M$ and tangent space
  indices $\r{A}$ which are described in the other chapters.}:
\bsubeq \label{10-clifford}
\begin{gather}
\{ \G^{\mu} , \G^{\nu} \} \ = \ 2 \eta^{\mu \nu} \; , 
\ls
\{ \gamma^{I} , \gamma^{I} \} \ = \ 2 \delta^{IJ} \; , \\
\G^{0} \ = \ 
\left(
\begin{array}{cc}
0 & i {\bf 1}_{16} \\
i {\bf 1}_{16} & 0 
\end{array} \right) \; , \ls
\G^{I} \ = \ \half \left(
\begin{array}{cc}
- (\gamma^I)^T + \gamma^I & - i (\gamma^I)^T - i \gamma^I \\
i (\gamma^I)^T + i \gamma^I & - (\gamma^I)^T + \gamma^I
\end{array} \right) \; , \\
\G \ = \ 
\left(
\begin{array}{cc}
{\bf 1}_{16} & 0 \\
0 & - {\bf 1}_{16}
\end{array} \right) \; , 
\end{gather}
\esubeq
where $\G$ is the chirality matrix.
The Lorentz indices $I$ runs from 1 to 9, which denotes the spatial
directions of ten-dimensional Minkowski spacetime.
The gamma matrices $\gamma^I$ are defined as the generators of $SO(9)$
Clifford algebra.
Although the above descriptions (\ref{10-clifford}) seems somewhat 
complicated,
these are useful expressions to perform the chiral decomposition of $SO(9,1)$
Dirac spinors $\theta$ 
and to construct the Majorana-Weyl spinors $\chi$ (\ref{MW}).
Note that we can easily connect the above gamma matrices 
to the ones described in
appendix \ref{conv-membrane} in terms of the following
unitary rotation
\begin{gather*}
\G^{\mu} \ = \ U \hG^{\mu} U^{-1} \; , \ls
\G \ = \ U \hG^{10} U^{-1} \; , \ls
U \ = \ \frac{1}{\sqrt{2}} \left(
\begin{array}{cc}
1 & -i \\
-i & 1
\end{array} \right) \; .
\end{gather*}

Utilizing the above descriptions (\ref{10-clifford}),
we can easily write down the Lagrangian of ten-dimensional super Yang-Mills 
in terms of the Majorana-Weyl spinors $\chi$ as follows:
\begin{align*}
S \ &= \ \int \! \d^{9+1} x \, {\cal L}_{9+1}
\ = \ 
\int \! \d^{9+1} x \Big\{ 
- \frac{1}{4} \Tr \big( F_{\mu \nu} F^{\mu \nu} \big)
+ \frac{i}{2} \Tr \big( \chi^T D_0 \chi \big)
- \frac{i}{2} \Tr \big( \chi^T \gamma^I D_I \chi \big) 
\Big\}
\; . 
\end{align*}
Now let us perform the dimensional reduction to
$(0+1)$-dimensional ``spacetime''. 
Under this reduction,
the gauge potential $A_I$ becomes an $U(N)$ adjoint scalar fields 
denoted by $X^I$.
The field strength $F_{\mu \nu}$ and the covariant derivative $D_I
\chi$ also reduce to 
\begin{align*}
F_{0I} \ &= \ 
\del_0 X^I + i g [ A_0 , X^I ] \ \equiv \ D_0 X^I 
\; , \ls
F_{IJ} \ = \ i g [ X^I , X^J ] \; , \\
D_0 \chi \ &= \ \del_0 \chi + i g [ A_0 , \chi ] \; , \ls
D_I \chi \ = \ i g [ X^I , \chi ] 
\; .
\end{align*}
Note that the bosonic fields $A_0$ and $X^I$, the fermionic fields $\chi$, 
and the Yang-Mills coupling $g$ are appropriately 
rescaled by the volume factor of the reduced nine-dimensional space. 
Thus the dimensionally reduced Lagrangian is obtained as
\begin{align}
\begin{split}
S \ &= \ \int \! \d^{0+1} x \, {\cal L}_{0+1} 
\\
\ &= \
\int \! \d^{0+1} x \,
\Tr \Big\{ \half D_0 X'{}^I D_0 X'{}^I 
+ \frac{1}{4} g'{}^2 \, [ X'{}^I , X'{}^J ]^2 
+ \frac{i}{2} \chi'{}^T D_0 \chi' + \half g' \, \chi'{}^T 
\gamma^I [ X'{}^I , \chi' ] \Big\}
\; . 
\end{split} \label{1-SYM}
\end{align}
We denote the dimensionally reduced variables to $X' = \sqrt{L^9} X$,
$\chi' = \sqrt{L^9} \chi$ and $g' = g/\sqrt{L^9}$, 
where $\dps \int \! \d^9 x = L^9$ is a reduced volume.
This is the effective Lagrangian of $N$ D0-branes system.
In chapter \ref{BMN} we use this action 
in order to write down the action (\ref{S-4'}).

We of course start the ten-dimensional Yang-Mills action with rescaled
field variables as
\begin{align}
\begin{split}
S \ &= \ \frac{1}{g^2} \int \! \d^{9+1} x \, \Tr \Big\{
- \frac{1}{4} \wt{F}_{\mu \nu} \wt{F}^{\mu \nu} 
- \ol{\wt{\theta}} \Gamma^{\mu} \wt{D}_{\mu} \wt{\theta} \Big\}
\\
\ &= \ 
\int \! \d^{9+1} x \, \Tr \Big\{
- \frac{1}{4} \wt{F}_{\mu \nu} \wt{F}^{\mu \nu} 
+ \frac{i}{2} \wt{\chi}^T D_0 \wt{\chi} - \frac{i}{2} \chi^T \gamma^I
D_I \wt{\chi} \Big\}
\; , 
\end{split} \label{rescaled-10SYM}
\end{align}
where we performed the following field re-definitions
\begin{align*}
g A_{\mu} \ &\equiv \ \wt{A}_{\mu} \; , \ls
F_{\mu \nu} \ = \ \frac{1}{g} \wt{F}_{\mu \nu} \;, \ls
g \chi \ = \ \wt{\chi} \; , \ls
D_{\mu} \chi \ = \ \frac{1}{g} \wt{D}_{\mu} \wt{\chi} \; .
\end{align*}
Under the above rescaling, 
the mass dimensions of rescaled variables in ten-dimensions are 
$[\wt{A}_{\mu}] = 1$ and $[ \wt{\chi} ] = 3/2$.
Now let us perform the dimensional reduction to $(0+1)$-dimensional
spacetime:
\begin{align}
S \ &= \ \frac{1}{g'{}^2} \int \! \d^{0+1} x \, 
\Tr \Big\{
\half \wt{D}_0 \wt{X}^I D_0 \wt{X}^I 
+ \frac{1}{4} [ \wt{X}^I , \wt{X}^J ]^2 
+ \frac{i}{2} \wt{\chi}^T D_0 \wt{\chi} - \frac{i}{2} \chi^T \gamma^I
D_I \wt{\chi} \Big\}
\; . \label{res-1-SYM}
\end{align}
We find that only the Yang-Mills coupling $g' \equiv g / \sqrt{L^9}$ 
changes the mass dimensions to $3/2$, 
and that the mass dimensions of fields $\wt{A}_{\mu} =
(\wt{A}_0, \wt{X}^I)$ and $\wt{\chi}$ remain $1$ and $3/2$, respectively.
Since this phenomenon also occurs in the dimensional reduction to any
dimensional spacetime,
we sometimes write down the field theory Lagrangian in the same
description as (\ref{rescaled-10SYM}).
The representation of (\ref{S-4}) is also this type.

These Lagrangians (\ref{1-SYM}) or (\ref{res-1-SYM}) are 
represented the low energy region of $N$ D0-branes system in flat background.
They could be deformed 
when the non-vanishing background fields turn on.
In chapter \ref{BMN},
we will construct an effective theory Lagrangian of $N$ D0-branes
system on such a non-trivial background.


\section{Matrix Theory Lagrangian: 
Dirac-Born-Infeld Type Action} \label{nonabelian}

In appendix \ref{SYM}
we discussed the effective action of $N$ coincident D0-branes system
without background fields.
Here we consider the effective action for $N$ D-branes with non-vanishing
massless Ramond-Ramond background field strength.
The argument presented in this appendix is given by Myers' lecture
\cite{M0303}.

As discussed in the previous appendix,
we have not completely understood the microscopic description of
Dirichlet $p$-brane(s) (or simply D$p$-brane(s)) action yet.
But we now believe that the D$p$-branes' effective action can be described by
the Dirac-Born-Infeld (DBI) type Lagrangian 
at least in the low energy region \cite{DLP89, L89, Pol95}.
Furthermore, Tseytlin, Myers, and a lot of other people have found that
the effective theory of $N$ coincident D-branes system should be added 
the Chern-Simons term from the viewpoint of T-duality \cite{T9701,
  T9908, M99, M0303}. 
Thus we introduce a short review of single D-brane effective action
and $N$ coincident D-branes' effective action.
We sometimes call the latter action the {\sl nonabelian D-branes'} action.

First let us construct a single D$p$-brane action.
As you know, 
a D$p$-brane is a $(p+1)$-dimensional extended hypersurface in
spacetime which supports the endpoints of open string within the
framework of perturbative string theory.
The massless modes of the open string theory form a supersymmetric
$U(1)$ gauge theory with a gauge potential $A_{a}$ ($a = 0,1, \cdots,
p$), $9-p$ real scalars $X^i$ ($i=p+1,\cdots, 9$) and their
superpartner fermions.
As discussed in the previous appendix,
the low energy effective action corresponds to the dimensional
reduction of the ten-dimensional $U(1)$ super Yang-Mills theory.
However, as usual in string theory,
there are higher order $\alpha' = \l_s^2$ corrections, i.e., the stringy
corrections (where $\l_s$ is the string length).
Due to this stringy corrections,
the effective action of D$p$-brane is deformed to the
DBI form\footnote{In this appendix we ignore contributions from the fermionic
  fields for simplicity.}
\begin{align}
S_{\rm DBI} \ &= \ 
- T_p \int \! \d^{p+1} \sigma
\, \Big( \e^{-\phi} 
\sqrt{- \det \big\{ 
{\rm P} [ G + B ]_{ab} + \lambda F_{ab} \big\}} \Big) 
\; , \label{DBI-U1}
\end{align}
where $T_p$ is the D$p$-brane tension and $\lambda$ denotes the
inverse of the string tension, i.e., $\lambda = 2 \pi \alpha'$.
the action (\ref{DBI-U1}) contains the field strength of the gauge
potential $F_{ab}$ with dimensions of $({\rm mass})^2$.
This DBI action describes the couplings of the D$p$-brane to the {\sl
  massless} Neveu-Schwarz (NS) fields of the bulk closed string as the
metric $G_{\mu \nu}$ ($\mu\, \nu = 0,1,\cdots, 9$), the dilaton
$\phi$, and the Kalb-Ramond field $B_{\mu \nu}$.
They are all dimensionless fields.
The interactions with the {\sl massless} Ramond-Ramond (RR) fields are
described by the Wess-Zumino term as
\begin{align}
S_{\rm WZ} \ &= \ \mu_p \int \! {\rm P} \Big[ \sum C^{(n)} \, \e^{B}
  \Big] \, \e^{\lambda F}
\; . \label{WZ-U1}
\end{align}
Note that $C^{(n)}$ is the $n$-form RR potential defined as
\begin{align*}
C^{(n)} \ &= \ \frac{1}{n!} C_{\mu_1 \mu_2 \cdots \mu_n} \, \d
x^{\mu_1} \w \d x^{\mu_2} \w \cdots \w \d x^{\mu_n}
\; .
\end{align*}
The Wess-Zumino term (\ref{WZ-U1}) shows that a D$p$-brane is
naturally charged under the $(p+1)$-form RR potential with charge
$\mu_p$, which relates to the D$p$-brane tension as $\mu_p = \pm T_p$
due to the spacetime supersymmetry.
Summarizing (\ref{DBI-U1}) and (\ref{WZ-U1}),
we describe the single D$p$-brane effective action as
\begin{align}
S_{\text{D$p$}} \ &= \ S_{\rm DBI} + S_{\rm WZ} \; .
\label{a-Dp}
\end{align}
On the flat spacetime without nontrivial constant background fields
(i.e., $G_{\mu \nu} = \eta_{\mu \nu}$ and $B = F = 0$),
the leading order of the 
action (\ref{a-Dp}) reduces to the $(p+1)$-dimensional $U(1)$
gauge theory action.
In the case of $p=0$,
the Yang-Mills coupling $g'$ in (\ref{res-1-SYM}) is represented in
terms of the D0-brane tension $T_0$ and the Regge parameter $\alpha'$ 
\begin{align*}
\frac{1}{g'{}^2} \ &= \ (2 \pi \alpha')^2 T_0 \; .
\end{align*}

The symbol ${\rm P}[\cdots]$ in (\ref{DBI-U1}) and (\ref{WZ-U1}) 
denotes the pull back of the bulk
spacetime tensors to the D$p$-brane worldvolume.
The DBI action (\ref{DBI-U1}) expresses that the D$p$-brane moves
dynamically in the spacetime.
This dynamics becomes more evident with an explanation of the static gauge.
To begin, we employ the spacetime diffeomorphism to set the position of
the worldvolume  $x^i = 0$.
With the worldvolume diffeomorphism,
we can match the worldvolume coordinates with the remaining spacetime
coordinates as $x^a = \sigma^a$.
Then the worldvolume scalar fields $X^i$ play the role of describing
the transverse displacements of the D$p$-brane through the following
identification
\begin{align}
x^i (\sigma) \ &= \ \lambda X^i (\sigma) \; .
\label{id-scalar-coord}
\end{align}
With this identification,
the general formula for the pullback is written by
\begin{align}
{\rm P} [E]_{ab} \ &= \ 
E_{\mu \nu} \frac{\del x^{\mu}}{\del \sigma^a} \frac{\del
  x^{\nu}}{\del \sigma^b}
\ = \ 
E_{ab} + \lambda E_{i b} \del_a X^i + \lambda E_{a j} \del_b X^j 
+ \lambda^2 E_{ij} \del_a X^i \del_b X^j
\; . 
\label{pullback}
\end{align}

Now let us generalize the above effective action for the single
D$p$-brane (\ref{a-Dp}) to the $N$ coincident D$p$-branes system.
As $N$ parallel D$p$-branes approach each other,
the ground state modes of strings stretching between the different
D$p$-branes become {\sl massless}.
These extra massless states carry the appropriate charges to fill out
representations under a $U(N)$ symmetry.
Thus the $U(1)^N$ symmetry of the individual D$p$-branes enhances to
the nonabelian $U(N)$ group for the coincident D$p$-branes.
The vector $A_a$ becomes a nonabelian gauge potential
\begin{align}
A_a \ &= \ A_a^k \, T_k \; , \ls
F_{ab} \ = \ \del_a A_b - \del_b A_a + i [ A_a , A_b ] \; ,
\end{align}
where $T_k$ are $N^2$ Hermitian generators of $U(N)$ group with $\Tr
(T_k T_l) = N \delta_{kl}$.
The scalar fields $X^i$ become also matrix valued transforming in the
adjoint of $U(N)$.
The covariant derivative of the scalar fields is given by
$D_a X^i = \del_a X^i + i [ A_a , X^i ]$. 

Under the above extension,
the DBI action (\ref{DBI-U1}) is generalized to
\begin{align}
S_{\rm DBI} \ &= \ 
- T_p \int \! \d^{p+1} \sigma \, {\rm STr} \Big(
\e^{- \phi} \sqrt{ \det (Q^i{}_j) }
\cdot \sqrt{- \det \big\{ {\rm P} 
[ E_{ab}
+ E_{ai} (Q^{-1} - \delta)^{ij} E_{jb}] + \lambda F_{ab} \big\}
}
\Big) \; , \label{DBI-UN}
\end{align}
where $E_{\mu \nu} = G_{\mu \nu} + B_{\mu \nu}$ and 
$Q^i{}_j = \delta^i_j + i \lambda \, [ X^i , X^k ] \, E_{kj}$.
We also generalize the Wess-Zumino term (\ref{WZ-U1}) to
\begin{align}
S_{\rm WZ} \ &= \ 
\mu_p \int \! {\rm STr} \Big( {\rm P} \Big[ 
\e^{i \lambda {\rm i}_{X} {\rm i}_{X}} 
\big( \sum C^{(n)} \, \e^{B} \big) \Big] \, \e^{\lambda F} \Big)
\; . \label{WZ-UN}
\end{align}
The symbol ${\rm STr}$ in (\ref{DBI-UN}) and (\ref{WZ-UN}) 
denotes the maximally symmetrized trace in which we average over all
possible orderings of the matrices in the trace.
Furthermore, 
the symbol ${\rm i}_X$ in the Wess-Zumino term (\ref{WZ-UN}) denotes
the interior product with $X^i$ defined as
\begin{align*}
{\rm i}_{X} {\rm i}_{X} C^{(n)}
\ &= \ 
\frac{1}{2(n-2)!}
\, [ X^i , X^j ] \, C_{ji \mu_3 \cdots \mu_n} \, 
\d x^{\mu_3} \w \cdots \w \d x^{\mu_n}
\; . 
\end{align*}
Note that acting on forms, the interior product is an anticommuting
operator.
Thus if the scalar fields $X^i$ are ordinary vector fields $v^i$,
the above equation vanishes: ${\rm i}_v {\rm i}_v C^{(n)} = 0$.

Now let us consider a specific situation for $N$ coincident D0-branes ($p=0$).
If there is a non-vanishing RR four-form field strength 
$F^{(4)} = \d C^{(3)}$
in the background, and if the other background fields vanish 
($B= C^{(1)} = C^{(5)} = C^{(7)} = C^{(9)} = 0$), 
the Wess-Zumino term (\ref{WZ-UN}) reduces to 
\begin{align}
\begin{split}
S_{\rm WZ}
\ &= \ 
i \lambda \mu_0 \int \! {\rm Tr} \Big( {\rm P} \Big[ 
({\rm i}_{X} {\rm i}_{X}) \, C^{(3)}
\Big] \Big)
\\
\ &= \ 
\frac{i \lambda}{2} \mu_0 \int \! \d t \, {\rm Tr} \Big( 
C_{t ij} \, [X^i , X^j]
+ \lambda C_{ijk} \, D_t X^i \, [ X^{k} , X^{j} ] \Big) 
\; .
\end{split} \label{red-WZ-UN}
\end{align}
Now we assume that the four-form field strength $F^{(4)}$ can be
written in terms of a constant $f$ with dimension of mass:
\begin{align*}
F_{t \wt{I} \wt{J} \wt{K}} \ &= \ - f \, \eps_{\wt{I} \wt{J} \wt{K}} 
\; ,
\end{align*}
where $\eps_{\wt{I} \wt{J} \wt{K}}$ denotes the $SO(3)$ Levi-Civita
tensor whose normalization is $\eps_{123} = 1$.
This assumption is regarded as the Freund-Rubin ansatz.
Under this ansatz,
we can write the reduced Wess-Zumino term (\ref{red-WZ-UN}) more
simply as
\begin{align*}
S_{\rm WZ} \ &= \ \frac{i}{3} \lambda^2 \mu_0 \int \! \d t \,
\Tr (X^{\wt{I}} X^{\wt{J}} X^{\wt{K}}) \, F_{t \wt{I} \wt{J} \wt{K}}
\; .
\end{align*}
This term appears in the Lagrangian of the BMN matrix model (\ref{S-3}).


\section{Supergravity Lagrangian with Full Interactions} \label{11SUGRA}

In this appendix we introduce the eleven-dimensional supergravity
Lagrangian which was described by Cremmer, Julia and Scherk \cite{CJS78}.
It is not so difficult to describe the Lagrangian 
with full interactions including torsion and quartic
terms with respect to the gravitino.
Here we write the full supergravity Lagrangian below:
\begin{align}
\begin{split}
{\cal L} \ &= \ 
e \, {\cal R} (e , \omega) 
- \half \ol{\Psi}_M \hG^{MNP} D_N [{\textstyle \half} 
(\omega + \wh{\omega})] \Psi_P
- \frac{1}{48} e \, F_{MNPQ} \, F^{MNPQ}
\\
\ & \ \ \ \ 
- \frac{1}{192} e \, \ol{\Psi}_M \wt{\Gamma}^{MNPQRS} \Psi_N 
\cdot \half (F + \wh{F})_{PQRS}
\\
\ & \ \ \ \ 
- \frac{1}{(144)^2} \ve^{MNPQRSUVWXY} \, F_{MNPQ} \, F_{RSUV} \,
C_{WXY}
\; , 
\end{split} \label{11-SUGRA-complete}
\end{align}
where definitions of $\wh{\omega}$ and $\wh{F}_{MNPQ}$ are 
\begin{gather*}
D_{[M} (\wh{\omega}) e_{N]}{}^{\r{A}} \ = \ \frac{1}{8} \ol{\Psi}_M
\hG^{\r{A}} \Psi_N
\; , \ls
\wh{F}_{MNPQ} \ = \ F_{MNPQ} + \frac{3}{2} \ol{\Psi}_{[M} \hG_{NP}
  \Psi_{Q]}
\; .
\end{gather*}
Note that the definition of the covariant derivative $D_M$ is
described in appendix \ref{connections}.
We also the full supersymmetry transformation rules for the vielbein,
three-form gauge field and gravitino:
\begin{gather*}
\delta e_M{}^{\r{A}} \ = \ \half \ol{\ve} \hG^{\r{A}} \Psi_M
\; , \ls
\delta C_{MNP} \ = \ - \frac{3}{2} \ol{\ve} \hG_{[MN} \Psi_{P]}
\; , \\
\delta \Psi_M \ = \ 
2 D_M (\wh{\omega}) \ve + 2 T_M{}^{MNPQ} \ve \wh{F}_{MNPQ}
\; , \ls
T_M{}^{NPQR} \ = \ \frac{1}{288} \Big( \hG_M{}^{NPQR}
- 8 \delta_M^{[N} \hG_{\phantom{M}}^{PQR]} \Big)
\; .
\end{gather*}
We can see these descriptions in various lectures with respect to
higher-dimensional supergravities
(for instance, see \cite{PvN81, D83, DNP86, dW99}).


\chapter{Background Geometry} \label{geometry}

\newpage

\setcounter{section}{0}
\renewcommand{\thesection}{\thechapter.\arabic{section}}

\section{Anti-de Sitter Spaces} \label{ADS}

Anti-de Sitter spaces emerge in the spontaneous compactifications of
higher-dimensional supergravities via Kaluza-Klein mechanism.
Since the anti-de Sitter space is a maximally symmetric space with negative
cosmological constant, we can study supergravity in this
spacetime background.
In the case of eleven-dimensions, for example,
the four- and seven-dimensional anti-de Sitter spaces are derived from the
existence of the constant four-form flux and some constraints: 
the seven- and four-dimensional compactified space should be Einstein
spaces and the uncompactified spacetime should be maximally symmetric
\cite{D83, DNPW84, DNP86}. 
In ten-dimensions, $AdS_5 \times S^5$ geometry also appears in type
IIB supergravity \cite{KRN85} and type IIB superstring in the near
horizon limit of D3-brane via constant self-dual five-form
Ramond-Ramond flux \cite{M97}.

Representations of supersymmetry in anti-de Sitter spaces are
discussed by Nicolai \cite{Nic84} and de Wit and Herger \cite{dWH99}.
The superalgebra and its unitary representation \cite{Nic84, dWH99} 
are discussed in terms of the oscillator method, which is
one of the powerful tool to investigate the supermultiplets and their
dynamics on the anti-de Sitter space \cite{GvNW85, GM98, GMZ98, FGP02}.

In this thesis we will argue the (linearized) supergravity on the
plane-wave background, which is a specific limit of the product space
of anti-de Sitter space and Einstein space, called the ``Penrose limit''.
Next we will discuss the definition of the Penrose
limit of the product spaces and will explain a physical meaning.


\subsection*{Penrose Limit of Maximally Supersymmetric Spaces} \label{PL}

Let us consider the Penrose limit of the product spaces of anti-de
Sitter space and higher-dimensional sphere in eleven-dimensions.
Since we would like to consider the maximally supersymmetric
spacetime,
we concentrate the discussions of 
the Penrose limit of $AdS_4 \times S^7$ and $AdS_7 \times S^4$
\cite{BFHP0201}. 

Since the $AdS_4 \times S^7$ geometry \cite{DP82} has 32 Killing spinors 
this spacetime is maximally supersymmetric.
This appears as a geometry of the near horizon limit of
M2-brane, whose line element is described by the global coordinates 
\begin{align*}
\d s^2 \ &= \ 
R_A^2 \big\{ - \cosh^2 \rho \cdot \d t^2 + \d \rho^2 
+ \sinh^2 \rho \cdot \d \Omega_2^2 \big\}
+ R_S^2 \big\{ \cos^2 \theta \d \varphi^2 
+ \d \theta^2 + \sin^2 \theta \cdot \d \Omega_5'{}^2 \big\}
\; .
\end{align*}
We introduce the following coordinates around a null geodesic 
$\gamma = \{ R_S = 2 R_A, t = 2 \varphi, \rho = \theta = 0\}$:
\begin{gather*}
\alpha \ = \ \frac{R_S}{R_A} \ = \ 2
\; ,
\ls 
x^+ \ = \ \half (t + 2 \varphi) \cdot \frac{3}{\mu}
\; , 
\ls
x^- \ = \ R_A^2 (t - 2 \varphi) \cdot \frac{\mu}{3}
\; , 
\\
x \ = \ R_A \rho \; , \ls 
y \ = \ 2 R_A \theta \; .
\end{gather*}
Notice that we added the rescaling factor $3/\mu$ for
later convenience.
Performing the large $R_A$ limit and retaining $x$ and $y$ to be
finite (the Penrose limit),
we obtain the following simple line element
\begin{align}
\d s^2 \ &= \ 
-2 \d x^+ \d x^- 
- \Big( \frac{\mu}{3} \Big)^2 
\Big\{ x^2 + \frac{1}{4} y^2 \Big\} (\d x^+)^2 
+ \big\{ \d x^2 + x^2 \d \Omega_2^2 \big\} 
+ \big\{ \d y^2 + y^2 \d \Omega_5'{}^2 \big\}
\; . 
\label{KG-solution}
\end{align}
This metric was constructed by Kowalski-Glikman \cite{KG8401}.
Thus this spacetime metric is sometimes called ``Kowalski-Glikman (KG)
solution'' and this spacetime is a maximally supersymmetric 
solution of the eleven-dimensional supergravity with constant flux
$F_{123+} = \mu$ because of the existence of 32 Killing
spinors \cite{KG8402}.

Let us consider the Penrose limit of another spacetime, i.e., the
Penrose limit of the $AdS_7 \times S^4$ spacetime.
Since the $AdS_7 \times S^4$ spacetime \cite{PvNT84} also has 32
Killing spinors, 
this is maximally supersymmetric in eleven-dimensions.
Moreover it is known that 
this spacetime appears in the near horizon limit of M5-branes.
Now we write down the global coordinates of $AdS_7 \times S^4$
\begin{align*}
\d s^2 \ &= \ 
R_A^2 \big\{ - \cosh^2 \rho \cdot \d t^2 + \d \rho^2 
+ \sinh^2 \rho \cdot \d \Omega_5^2 \big\}
+ R_S^2 \big\{ \cos^2 \theta \d \varphi^2 
+ \d \theta^2 + \sin^2 \theta \cdot \d \Omega_2'{}^2 \big\}
\; ,
\end{align*}
where $R_A$ and $R_S$ are the radius of $AdS_7$ and $S^4$,
respectively.
In order to perform the Penrose limit we take the following constraints:
\begin{gather*}
\alpha \ = \ \frac{R_S}{R_A} \ = \ \half 
\; , 
\ls 
x^+ \ = \ \half \Big(t + \half \varphi \Big) \cdot \frac{6}{\mu} 
\; , \ls
x^- \ = \ R_A^2 \Big( t - \half \varphi \Big) \cdot \frac{\mu}{6}
\; , \\
x \ = \ R_A \rho \; , \ls 
y \ = \ \half R_A \theta
\end{gather*}
around a null geodesic $\gamma = \{R_A = 2 R_S, 
t = \half \varphi$, $\theta = \rho = 0\}$.
Taking $R_A \to \infty$ and remaining the coordinates ($x$, $y$) to be finite, 
and exchanging the coordinate labels between $x$ and $y$,
we obtain the same line element as the Penrose limit of $AdS_4 \times S^7$
\begin{align*}
\d s^2 \ &= \ 
-2 \d x^+ \d x^- 
- \Big( \frac{\mu}{3} \Big)^2 
\Big\{ x^2 + \frac{1}{4} y^2 \Big\} (\d x^+)^2 
+ \big\{ \d x^2 + x^2 \d \Omega_2^2 \big\} 
+ \big\{ \d y^2 + y^2 \d \Omega_5'{}^2 \big\} 
\; .
\end{align*}

We can easily find that the vanishing limit of the parameter $\mu$ of the
Penrose limit of $AdS_{4(7)} \times S^{7(4)}$ is the flat Minkowski
spacetime, which is of course maximally supersymmetric.
Here we draw the relations of the four maximally supersymmetric spacetime
in eleven-dimensions in Figure \ref{4-BG}:
\begin{figure}[h]
\begin{center}
\includegraphics{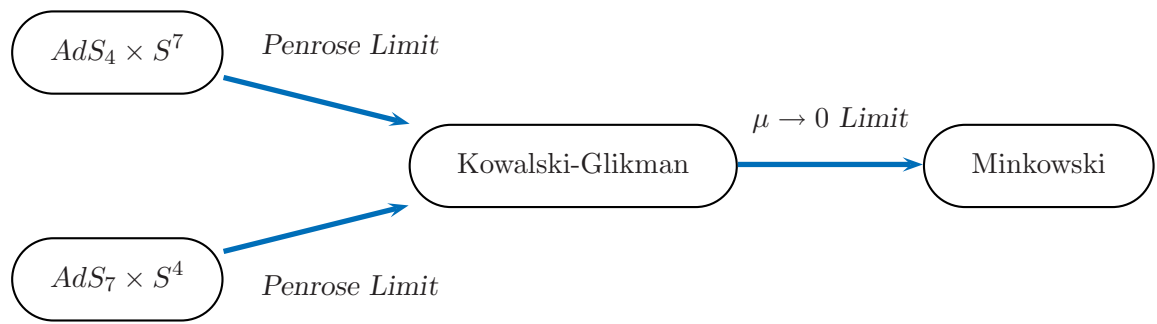}
\caption{\sl The relationships among the maximally supersymmetric spacetimes
  in eleven-dimensions.}
\label{4-BG}
\end{center}
\end{figure}

So far we discussed the procedure of the Penrose limit of the product
spaces such as $AdS_4 \times S^7$ and $AdS_7 \times S^4$.
We can also argue the Penrose limit of other spaces, for example,
the $AdS_4 \times \text{squashed $S^7$}$, 
the $AdS_4 \times Q^{1,1,1}$, the $AdS_4 \times N^{0,1,0}$, and so on,
which also reduce to the above Kowalski-Glikman solution
(\ref{KG-solution}) via Penrose limit \cite{GNS0203}.
Now it is important to explain the physical meaning of the procedure
of the Penrose limit. Let us quote a famous paragraph from the
Penrose's lecture \cite{Penrose} (see also \cite{BFP02}):
\begin{quote}
{\sl There is a `physical' interpretation of the above mathematical
    procedure, which is the following. 
    We envisage a succession of
    observers travelling in the space-time $M$ whose world lines
    approach the null geodesic $\gamma$ more and more closely; so we
    picture these observers as travelling with greater and greater
    speeds, approaching that of light. As their speeds increase they
    must correspondingly recalibrate their clocks to run faster and
    faster (assuming that all space-time measurements are referred to
    clock measurements in the standard way), so that in the limit the
    clocks measure the affine parameter $x^0$ along $\gamma$. (Without
    clock recalibration a degenerate space-time metric would result.)
    In the limit the observers measure the space-time to have the
    plane wave structure $W_\gamma$.}
\end{quote}
In other words, the Penrose limit can be understood as a boost
followed by a commensurate uniform rescaling of the coordinates in
such a way that the affine parameter along the null geodesic remains
invariant.
The obtained spacetime backgrounds are called the
``plane-wave'' or ``pp-wave'' backgrounds, where 
the term ``pp-wave'' is the abbreviation of ``{\sl plane fronted
gravitational wave with parallel rays}''.


\subsection*{Geometrical Variables on the Plane-wave}

Here we discuss several properties of 
the maximally supersymmetric plane-wave background. 
As discussed above,
this solution was found by Kowalski-Glikman \cite{KG8401, KG8402} and 
often called the KG solution. This is the unique plane-wave type 
solution preserving maximal 32 supersymmetries in eleven dimensions. 
The metric of this background is given by (\ref{KG-solution}) as
\begin{align}
\begin{split}
\d s^2 \ &= \ - 2 \d x^+ \d x^- + G_{++} \cdot (\d x^+)^2 + \sum_{I=1}^9 (
\d x^{I})^2 \; , 
\\
G_{++} \ &= \ - \Big[ \Big( \frac{\mu}{3} \Big)^2 \sum_{\wt{I}=1}^3
  (x^{\wt{I}})^2 + \Big( \frac{\mu}{6} \Big)^2 \sum_{I'=4}^9
  (x^{I'})^2 \Big] \; , 
\end{split} \label{KG-BG1-app}
\end{align}
which is equipped with the constant four-form flux 
\begin{align*}
F_{123+} \ &= \ \mu \ \neq \ 0 \; . 
\end{align*}
In our consideration the contribution from torsion is not included, i.e.,
affine connection is symmetric under lower indices: $\Gamma^P_{MN} =
\Gamma^P_{NM}$. 
For the metric on the KG solution (\ref{KG-BG1-app}), 
we obtain the following variables:
\begin{gather}
e_+{}^{\r{+}} \ = \ e_-{}^{\r{-}} \ = \ 1
\; , \ls
e_+{}^{\r{-}} \ = \ - \half G_{++} 
\; , \nn \\
E_{\r{+}}{}^+ \ = \ E_{\r{-}}{}^- \ = \ 1
\; , \ls
E_{\r{+}}{}^- \ = \ \half G_{++} 
\; , \nn \\
\omega_+{}^{\r{I} \r{-}} 
\ = \ 
\half \del_I G_{++} \; , \label{KG-BG2-app}
\\
\Gamma^{I}_{++} \ = \ 
\Gamma^-_{+ I} \ = \ 
- \half \del_{I} G_{++}
\; , \nn \\
R^{I}{}_{+J+} \ = \ 
- \half \del_I \del_J G_{++}
\; ,
\ls
{\cal R}_{++} \ = \ 
\half \mu^2 
\;, \ls 
{\cal R} \ = \  0 
\; . \nn
\end{gather}


\section{Coset Construction} \label{coset}

Here
we discuss the coset construction of product spaces of anti-de Sitter
and sphere, in particular $AdS_{4} \times S^{7}$ and $AdS_{7} \times S^{4}$
spacetimes, which lead to the KG solution in the Penrose limit.
In this construction we define supervielbeins to all order in
$\theta$, the superspace coordinates ($SO(10,1)$ Majorana
spinor coordinates) in eleven dimensions \cite{dWPP98, dWPPS98, dW99}.

\subsection*{Superalgebra}

Let us consider the superalgebra of the plane-wave in terms of the
Penrose limit of $AdS_4 \times S^7$ spacetime superalgebra.
Thus we first prepare the superalgebras of $AdS_{4(7)} \times S^{7(4)}$. 
This spacetime is solutions of the eleven-dimensional supergravity
with a constant four-form flux given by
\begin{align} 
F_{\wt{M} \wt{N} \wt{P} \wt{Q}} \ &= \ 
f \, e \, E^{-1}_{\wt{M} \wt{N} \wt{P} \wt{Q}}
\; .
\label{TK-FR-4-7}
\end{align}
Note that the indices $\wt{M}, \wt{N}, \cdots$ are the indices
expanded to the four-dimensional spacetime directions and 
the variable $e = \sqrt{|\det g_{MN}|}$ denotes to the square root of
the determinant of the metric in the eleven-dimensional curved spacetime.
The $E_{\wt{M} \wt{N} \wt{P} \wt{Q}}^{-1}$ is an
invariant tensor density in four dimensions (weight $-1$) whose
normalization is defined by $E^{-1}_{0123} = 1$.
The constant $f$ decides the property of spacetime: 
if $f$ is real and non-vanishing, 
we can obtain the $AdS_4 \times S^7$ spacetime and 
if $f$ is non-zero pure imaginary, $AdS_7 \times S^4$ spacetime appears.
Of course we obtain the flat spacetime when we choose
$f=0$\footnote{We assume that the spacetime is (maximally) symmetric.}. 
Under this setup\footnote{This setup is called the ``Freund-Rubin
  ansatz'' \cite{FR80}. 
This situation can be derived under some simple assumption \cite{FP0211}.},
the Riemann tensors of four- and seven-dimensional spaces are given
by equations of motion of eleven-dimensional supergravity
  (\ref{eom-class-g-2-2}): 
\bsubeq \label{TK-4-7-curv}
\begin{align}
R_{\wt{M} \wt{N} \wt{P} \wt{Q}} \ &= \ 
- \frac{1}{9} f^2 \big( g_{\wt{M} \wt{P}} \, g_{\wt{N} \wt{Q}} 
- g_{\wt{M} \wt{Q}} \, g_{\wt{N} \wt{P}} \big)
& & \text{four-dimensional space} \; , \\
R_{M'N'P'Q'} \ &= \ 
\frac{1}{36} f^2 \big( g_{M'P'} \, g_{N'Q'} - g_{M'Q'} \, g_{N'P'} \big)
& & \text{seven-dimensional space} \; , 
\end{align} 
\esubeq
where $M', N', \cdots$ denotes the seven-dimensional space indices.
In this configuration the superalgebra of $AdS_{4(7)} \times S^{7(4)}$
can be written down in terms of Hermitian generators $\{ P_{\r{A}} ,
\Sigma_{\r{A} \r{B}} \}$ and fermionic generators $Q_{a a'}$ 
below (see the lecture note written by de Wit \cite{dW99}):
\bsubeq \label{TK-coset-algebra}
\begin{align}
[ P_{\r{\wt{A}}} , P_{\r{\wt{B}}} ] \ &= \ 
\frac{i}{9} f^2 \, \Sigma_{\r{\wt{A}} \r{\wt{B}}} 
\; , &
[ P_{\r{A'}} , P_{\r{B'}} ] \ &= \ 
- \frac{i}{36} f^2 \, \Sigma_{\r{A'} \r{B'}}
\; , \\
[ P_{\r{\wt{A}}} , \Sigma_{\r{\wt{B}} \r{\wt{C}}} ] \ &= \ 
i \big( \eta_{\r{\wt{A}} \r{\wt{B}}} \, P_{\r{\wt{C}}} 
- \eta_{\r{\wt{A}} \r{\wt{C}}} \, P_{\r{\wt{B}}} \big)
\; , &
[ P_{\r{A'}} , \Sigma_{\r{B'} \r{C'}} ] \ &= \ 
i \big( 
\eta_{\r{A'} \r{B'}} \, P_{\r{C'}} - \eta_{\r{A'} \r{C'}} \, P_{\r{B'}} 
\big)
\; , 
\end{align}
\begin{align}
i [ \Sigma_{\r{\wt{A}} \r{\wt{B}}} , \Sigma_{\r{\wt{C}} \r{\wt{D}}} ] \ &= \ 
  \eta_{\r{\wt{A}} \r{\wt{C}}} \, \Sigma_{\r{\wt{B}} \r{\wt{D}}}
+ \eta_{\r{\wt{B}} \r{\wt{D}}} \, \Sigma_{\r{\wt{A}} \r{\wt{C}}}
- \eta_{\r{\wt{A}} \r{\wt{D}}} \, \Sigma_{\r{\wt{B}} \r{\wt{C}}}
- \eta_{\r{\wt{B}} \r{\wt{C}}} \, \Sigma_{\r{\wt{A}} \r{\wt{D}}}
\; , 
\\
i [ \Sigma_{\r{A'} \r{B'}} , \Sigma_{\r{C'} \r{D'}} ] \ &= \ 
  \eta_{\r{A'} \r{C'}} \, \Sigma_{\r{B'} \r{D'}}
+ \eta_{\r{B'} \r{D'}} \, \Sigma_{\r{A'} \r{C'}}
- \eta_{\r{A'} \r{D'}} \, \Sigma_{\r{B'} \r{C'}}
- \eta_{\r{B'} \r{C'}} \, \Sigma_{\r{A'} \r{D'}}
\; , 
\end{align}
\begin{align}
[ P_{\r{\wt{A}}} , Q_{a a'} ] \ &= \ 
- \frac{i}{6} f \, (\gamma_{\r{\wt{A}}} \gamma_5)_a{}^b \, Q_{b a'}
\; , &
[ P_{\r{A'}} , Q_{a a'} ] \ &= \ 
- \frac{i}{12} f \, (\Gamma_{\r{A'}})_{a'}{}^{b'} \, Q_{a b'}
\; , \\
[ \Sigma_{\r{\wt{A}} \r{\wt{B}}} , Q_{a a'} ] \ &= \ 
- \frac{i}{2} (\gamma_{\r{\wt{A}} \r{\wt{B}}})_a{}^b \, Q_{b a'}
\;, &
[ \Sigma_{\r{A'} \r{B'}} , Q_{aa'} ] \ &= \ 
- \frac{i}{2} (\Gamma_{\r{A'} \r{B'}})_{a'}{}^{b'} \, Q_{a b'}
\; , 
\end{align}
\begin{align}
\{ Q_{aa'} , Q_{bb'} \} 
\ &= \ 
- C'_{a'b'} 
\Big\{ - 2 i \, (\gamma_{\r{\wt{A}}} C)_{ab} \, P^{\r{\wt{A}}}
+ \frac{i}{6} f \, 
(\gamma_{\r{\wt{A}} \r{\wt{B}}} \gamma_5 C)_{ab} \,
\Sigma^{\r{\wt{A}} \r{\wt{B}}} 
\Big\}
\nn \\
\ & \ \ \ \ 
- (\gamma_5 C)_{ab} 
\Big\{ - 2 i \, (\Gamma_{\r{A'}} C')_{a'b'} \, P^{\r{A'}}
- \frac{i}{3} f \, 
(\Gamma_{\r{A'} \r{B'}} C')_{a'b'} \, M^{\r{A'} \r{B'}} 
\Big\}
\; .
\end{align}
\esubeq
Notice that the indices $\r{\wt{A}}$ and $\r{B'}$ are the indices of four- and
seven-dimensional tangent spaces;
indices $(a,b)$ are the spinor indices in four-dimensional space 
and $(a',b')$ are the spinor indices in seven-dimensional space.
The matrices 
$\gamma_{\r{\wt{A}}}$, $\gamma_5$ and $C_{ab}$ are Dirac gamma matrices
and charge conjugation matrix in four-dimensional space
and $\Gamma_{\r{A'}}$, $C'_{a'b'}$ are gamma
matrices and charge conjugation matrix in seven-dimensional space.
We can rewrite the above superalgebra (\ref{TK-coset-algebra}) in the
language of eleven-dimensional spacetime
\begin{gather}
[ P_{\r{\wt{A}}} , P_{\r{\wt{B}}} ] \ = \ 
\frac{i}{9} f^2 \, \Sigma_{\r{\wt{A}} \r{\wt{B}}} 
\; , \ls
[ P_{\r{A'}} , P_{\r{B'}} ] \ = \ 
- \frac{i}{36} f^2 \, \Sigma_{\r{A'} \r{B'}}
\; , 
\nn \\
[ P_{\r{\wt{A}}} , \Sigma_{\r{\wt{B}} \r{\wt{C}}} ] \ = \ 
i \big( \eta_{\r{\wt{A}} \r{\wt{B}}} \, P_{\r{\wt{C}}} 
- \eta_{\r{\wt{A}} \r{\wt{C}}} \, P_{\r{\wt{B}}} \big)
\; , 
\nn \\
[ P_{\r{A'}} , \Sigma_{\r{B'} \r{C'}} ] \ = \ 
i \big( 
\eta_{\r{A'} \r{B'}} \, P_{\r{C'}} - \eta_{\r{A'} \r{C'}} \, P_{\r{B'}} 
\big)
\; , 
\nn \\
i [ \Sigma_{\r{\wt{A}} \r{\wt{B}}} , \Sigma_{\r{\wt{C}} \r{\wt{D}}} ] \ = \ 
  \eta_{\r{\wt{A}} \r{\wt{C}}} \, \Sigma_{\r{\wt{B}} \r{\wt{D}}}
+ \eta_{\r{\wt{B}} \r{\wt{D}}} \, \Sigma_{\r{\wt{A}} \r{\wt{C}}}
- \eta_{\r{\wt{A}} \r{\wt{D}}} \, \Sigma_{\r{\wt{B}} \r{\wt{C}}}
- \eta_{\r{\wt{B}} \r{\wt{C}}} \, \Sigma_{\r{\wt{A}} \r{\wt{D}}}
\; , \label{TK-coset-algebra-11}
\\
i [ \Sigma_{\r{A'} \r{B'}} , \Sigma_{\r{C'} \r{D'}} ] \ = \ 
  \eta_{\r{A'} \r{C'}} \, \Sigma_{\r{B'} \r{D'}}
+ \eta_{\r{B'} \r{D'}} \, \Sigma_{\r{A'} \r{C'}}
- \eta_{\r{A'} \r{D'}} \, \Sigma_{\r{B'} \r{C'}}
- \eta_{\r{B'} \r{C'}} \, \Sigma_{\r{A'} \r{D'}}
\; , 
\nn \\
[ P_{\r{A}} , \ol{Q} ] \ = \ 
i \ol{Q} \, T_{\r{A}}{}^{\r{B} \r{C} \r{D} \r{E}} \, 
F_{\r{B} \r{C} \r{D} \r{E}} 
\; , \ls
[ \Sigma_{\r{A} \r{B}} , \ol{Q} ] \ = \ 
\frac{i}{2} \ol{Q} \, \hG_{\r{A} \r{B}} 
\; , 
\nn \\
\{ Q, \ol{Q} \} \ = \ 
2 i \, \hG_{\r{A}} P^{\r{A}}
- \frac{i}{144} \big\{ 
\hG^{\r{A} \r{B} \r{C} \r{D} \r{E} \r{F}} \, F_{\r{C} \r{D} \r{E} \r{F}}
+ 24 \, \hG_{\r{C} \r{D}} \, F^{\r{A} \r{B} \r{C} \r{D}}
\big\} \Sigma_{\r{A} \r{B}}
\; , \nn
\end{gather}
where $P_{\r{A}}$ and $\Sigma_{\r{A} \r{B}}$ are bosonic Hermitian
generators 
and fermionic generators $Q$ are $SO(10,1)$ Majorana
spinors\footnote{We define the Dirac conjugate of the Majorana spinor
  as $\ol{Q}
= i Q^{\dagger} \hG^{\r{0}} = Q^T C$.
Thus the product of two Majorana spinors has the following properties:
$\ol{\theta} Q = - Q^T 
  C^T \theta = Q^T C \theta = \ol{Q} \theta$ and
$(\ol{\theta} Q)^{\dagger} = - i
  Q^{\dagger} (\hG^{\r{0}})^{\dagger} \theta = i Q^{\dagger}
  \hG^{\r{0}} \theta = \ol{Q} \theta = \ol{\theta} Q$.}.
The symbol $T_{\r{A}}{}^{\r{B} \r{C} \r{D} \r{E}}$ is described by the
gamma matrices as
\begin{align*}
T_{\r{A}}{}^{\r{B} \r{C} \r{D} \r{E}} 
\ &= \ 
\frac{1}{288} \Big( \hG_{\r{A}}{}^{\r{B} \r{C} \r{D} \r{E}}
- 8 \delta_{\r{A}}^{[\r{B}} \hG_{\phantom{\r{A}}}^{\r{C} \r{D} \r{E}]}
\Big) \; .
\end{align*}
On the geometry of the plane-wave background 
this superalgebra is also satisfied
because the plane-wave is continuously connected to $AdS_{4(7)}
\times S^{7(4)}$ geometries.

We will construct supervielbeins on the $AdS_{4(7)} \times S^{7(4)}$
and on the plane-wave by utilizing this superalgebra
(\ref{TK-coset-algebra-11}).
The supervielbeins are important to construct Lagrangians of
supermembranes and Matrix theory on the $AdS$ background and the
plane-wave background, which are discussed in chapter \ref{BMN}. 


\subsection*{Coset Space Representatives and Supervielbeins}

Here we construct the supervielbeins on the $AdS_{4(7)} \times
S^{7(4)}$ background and the plane-wave background of them.
First we define a (super)representative $L(Z)$ on the backgrounds
\begin{align*}
L (Z) \ &= \ \l (x) \cdot \wh{L} (\theta)
\; , \ls
\l (x) \ = \ \exp (i \, x^{\r{A}} P_{\r{A}})
\ls \mbox{and} \ls
\wh{L} (\theta) \ = \ \exp (i \, \ol{\theta} Q) 
\; ,
\end{align*}
where $Z = (x^{\r{A}}, \theta)$ are the tangent space coordinates of
eleven-dimensional curved spacetime; the bosonic generators
$P_{\r{A}}$ are Hermitian and the fermionic generators $Q$ are the
$SO(10,1)$ Majorana spinors. In this definition the representative is
unitary. 
Utilizing this representative we define a ``super'' Maurer-Cartan
one-form $\alpha$ in the same way as bosonic one-form
\begin{align}
\alpha \ &= \ i^{-1} \, L^{-1} \d L \ = \ \wt{E} + \wt{\Omega}
\; .
\label{TK-MC}
\end{align}
Here we introduce a supervielbein $\wt{E}$ and super $H$-connection
$\wt{\Omega}$ which are expanded by the bosonic and fermionic generators
\begin{align}
\wt{E} \ &= \ \wh{E}^{\r{A}} \, P_{\r{A}} + \ol{Q} \wh{E}
\; , \ls
\wt{\Omega} \ = \ 
\frac{1}{2} \wh{\Omega}{}^{\r{A} \r{B}} \, \Sigma_{\r{A} \r{B}}
\; .
\label{TK-supervielbein}
\end{align}
Note that we also refer the components $\wh{E}^{\r{A}}$, $\wh{E}$ 
and $\wh{\Omega}^{\r{A} \r{B}}$ to supervielbeins and super $H$-connections.
Maurer-Cartan one-form (\ref{TK-MC}) satisfies the following relation
\begin{align}
\d \alpha + i \, \alpha \w \alpha \ &= \ 0
\; .
\label{TK-MC-eq}
\end{align}
Utilizing the equations (\ref{TK-MC}) and (\ref{TK-supervielbein}),
we obtain the ``super'' Cartan's structure equations
\begin{align}
\begin{split}
0 \ &= \ 
\d \wt{\Omega} + i \, \wt{\Omega} \w \wt{\Omega} 
+ \frac{i}{2} \wh{E}{}^{\r{A}} \w \wh{E}{}^{\r{B}} [ P_{\r{A}} , P_{\r{B}} ] 
\\
\ & \ \ \ \ 
+ \frac{1}{288} \ol{\wh{E}} \, \Big\{ 
\hG^{\r{A} \r{B} \r{C} \r{D} \r{E} \r{F}} \, 
F_{\r{C} \r{D} \r{E} \r{F}}
+ 24 \, 
\hG_{\r{C} \r{D}} \, F^{\r{A} \r{B} \r{C} \r{D}} \Big\}
\wh{E} \, \Sigma_{\r{A} \r{B}}
\; , \\
0 \ &= \ 
\d \wh{E}{}^{\r{A}} 
- \wh{\Omega}{}^{\r{A}}{}_{\r{B}} \w \wh{E}{}^{\r{B}}
- \ol{\wh{E}} \, \hG^{\r{A}} \w \wh{E} 
\; , \\
0 \ &= \ 
\d \wh{E} 
- \wh{E}{}^{\r{A}} \w T_{\r{A}}{}^{\r{B} \r{C} \r{D} \r{E}} 
\, \wh{E} \, F_{\r{B} \r{C} \r{D} \r{E}}
- \frac{1}{4} \wh{\Omega}{}^{\r{A} \r{B}} \w \hG_{\r{A} \r{B}} \, \wh{E}
\; .
\end{split} \label{TK-coset-str}
\end{align}
Substituting the superalgebra (\ref{TK-coset-algebra-11}) into the
above super Cartan's structure equations (\ref{TK-coset-str}),
we solve the supervielbeins and $H$-connections
\begin{align}
\wh{E}^{\r{A}} \ &= \ 
e^{\r{A}} + {\cal O} (\theta^2) \; , \ls
\wh{\Omega}^{\r{A} \r{B}} \ = \ 
- \omega^{\r{A} \r{B}} + {\cal O} (\theta^2) \; .
\label{TK-bosonic-sol}
\end{align}
Here we wrote down the solutions up to fermionic contributions.
The vielbeins $e^{\r{A}}$ and the spin connections $\omega^{\r{A}
  \r{B}}$ are obtained 
such that they satisfy the Riemann tensors (\ref{TK-4-7-curv}).
It is somewhat difficult to solve the equations (\ref{TK-coset-str})
with all the fermionic contributions. 
Thus we introduce a trick proposed by Kallosh, Rahmfeld and Rajaraman
\cite{KRR98}.
We rescale the fermionic coordinates $\theta$ to $t \theta$ with one
arbitrary parameter $t \in [0,1]$ which we put to unity at the end.
Taking the derivative with respect to this parameter $t$ of 
the Maurer-Cartan one-form (\ref{TK-MC}) leads to first order
differential equations for supervielbeins $\wh{E}$ and $H$-connection 
$\wh{\Omega}$:
\begin{align}
\frac{\d}{\d t} (\wt{E} + \wt{\Omega}) \ &= \ 
\d \ol{\theta} Q 
+ i \, (\wt{E} + \wt{\Omega}) \ol{\theta} Q
- i \, \ol{\theta} Q (\wt{E} + \wt{\Omega}) 
\; .
\end{align}
The left-hand side and right-hand side of this equation is calculated
respectively: 
\begin{align*}
\frac{\d}{\d t} (\wt{E} + \wt{\Omega}) 
\ &= \ 
\frac{\d}{\d t} \wh{E}^{\r{A}} P_{\r{A}}
+ \ol{Q} \frac{\d}{\d t} \wh{E} 
+ \frac{1}{2} \frac{\d}{\d t} \wh{\Omega}^{\r{A} \r{B}} 
\Sigma_{\r{A} \r{B}}
\; , \\
(\wt{E} + \wt{\Omega}) \ol{\theta} Q - \ol{\theta} Q (\wt{E}+ \wt{\Omega})
\ &= \ 
i \ol{Q} T_{\r{A}}{}^{\r{B} \r{C} \r{D} \r{E}} \theta \, F_{\r{B}
  \r{C} \r{D} \r{E}} \, \wh{E}^{\r{A}}
+ \frac{i}{4} \wh{\Omega}^{\r{A} \r{B}} \ol{Q} \hG_{\r{A} \r{B}} \theta
\nn \\
\ & \ \ \ \ 
- \ol{\theta} \Big[ 2 i \hG^{\r{A}} P_{\r{A}}
- \frac{i}{144} \big\{ \hG^{\r{A} \r{B} \r{C} \r{D} \r{E} \r{F}}
F_{\r{C} \r{D} \r{E} \r{F}} + 24 \, \hG_{\r{C} \r{D}} 
\, F^{\r{A} \r{B} \r{C} \r{D}} \big\} \Sigma_{\r{A} \r{B}} \Big] \wh{E}
\; .
\end{align*}
Note that we substituted the superalgebra (\ref{TK-coset-algebra-11}) 
into the above equations. 
Summarizing the equations in terms of the supersymmetry generators
$P_{\r{A}}$, $\Sigma_{\r{A} \r{B}}$ and $Q$, we find that a couple of
first-order differential equations
\begin{align*}
\frac{\d}{\d t} \wh{E}{}^{\r{A}} \ &= \ 
2 \ol{\theta} \hG^{\r{A}} \wh{E} 
\; , \\
\frac{\d}{\d t} \wh{E} \ &= \
\d \theta - \wh{E}{}^{\r{A}} \, T_{\r{A}}{}^{\r{B} \r{C} \r{D} \r{E}}
\theta \, F_{\r{B} \r{C} \r{D} \r{E}}
- \frac{1}{4} \wh{\Omega}{}^{\r{A} \r{B}} \, \hG_{\r{A} \r{B}} \theta 
\; , \\
\frac{\d}{\d t} \wh{\Omega}{}^{\r{A} \r{B}} \ &= \ 
- \frac{1}{72} \ol{\theta}
\Big\{
\hG^{\r{A} \r{B} \r{C} \r{D} \r{E} \r{F}} \, F_{\r{C} \r{D} \r{E} \r{F}}
+ 24 \, \hG_{\r{C} \r{D}} \, F^{\r{A} \r{B} \r{C} \r{D}} \Big\}
\wh{E}
\; .
\end{align*}
Since these equations have a structure of coupled harmonic oscillators
with  respect to $\wh{E}^{\r{A}}$ and $\wh{\Omega}^{\r{A} \r{B}}$,
we can solve these completely as
\begin{align}
\begin{split}
\wh{E}^{\r{A}} (x, \theta) \ &= \ 
e^{\r{A}} + \ol{\theta} \hG^{\r{A}} D \theta
+ 2 \sum_{n=1}^{15} \frac{1}{(2n+2)!} 
\ol{\theta} \, \hG^{\r{A}} {\cal M}^{2n} \, D \theta
\; , \\
\wh{E} (x, \theta) \ &= \ 
D \theta + \sum_{n=1}^{16} \frac{1}{(2n +1)!} {\cal M}^{2n} D \theta
\; , \\
\wh{\Omega}{}^{\r{A} \r{B}} (x, \theta) \ &= \ 
- \omega^{\r{A} \r{B}}
- \frac{1}{72}
\sum_{n=0}^{15}
\frac{1}{(2n+2)!} \ol{\theta} \big\{ 
\hG^{\r{A} \r{B} \r{C} \r{D} \r{E} \r{F}} \, F_{\r{C} \r{D} \r{E} \r{F}}
+ 24 \, \hG_{\r{C} \r{D}} \, F^{\r{A} \r{B} \r{C} \r{D}} \big\} 
{\cal M}^{2n} D \theta
\; , \\
D \theta \ &= \ \frac{\d}{\d t} \wh{E} \Big|_{t=0}
\ = \ \d \theta - e^{\r{A}} T_{\r{A}}{}^{\r{B} \r{C} \r{D} \r{E}}
\theta \, F_{\r{B} \r{C} \r{D} \r{E}} + \frac{1}{4} \omega^{\r{A} \r{B}}
\hG_{\r{A} \r{B}} \theta
\; , \\
{\cal M}^2 
\ &= \ 
- 2 \big( T_{\r{A}}{}^{\r{B} \r{C} \r{D} \r{E}} \, \theta \big) \, 
F_{\r{B} \r{C} \r{D} \r{E}} \, 
\big( \ol{\theta} \hG^{\r{A}} \big)
\\
\ & \ \ \ \ 
+ \frac{1}{288} \big( \hG_{\r{A} \r{B}} \, \theta \big) 
\Big( \ol{\theta} \Big[ \hG^{\r{A} \r{B} \r{C} \r{D} \r{E} \r{F}} \, 
F_{\r{C} \r{D} \r{E} \r{F}} 
+ 24 \, \hG_{\r{C} \r{D}} \, F^{\r{A} \r{B} \r{C} \r{D}} \Big] \Big)
\; .
\end{split} \label{TK-supervielbein-sol}
\end{align}
Notice that the coordinate $\theta$ is the anticommuting 
$SO(10,1)$ Majorana spinor and we put the free parameter $t$ to unity.
These variables correctly represents the superspaces of the $AdS_{4(7)} \times
S^{7(4)}$.
In chapter \ref{BMN} we use these variables on the
plane-wave which is continuously related to $AdS_{4(7)} \times
S^{7(4)}$ as discussed in appendix \ref{PL}.


\end{appendix}



}
\end{document}